\documentclass{aa}
 
\usepackage[varg]{txfonts}
 
\usepackage[]{hyphenat}
\usepackage[]{microtype}
\usepackage{amssymb}
\usepackage{float}
\usepackage{bm}
\usepackage{epsfig}
\usepackage{soul}
\usepackage{subcaption}
\usepackage{multirow,balance}
\usepackage{amsfonts}
\usepackage{amsmath}
\usepackage{aas_macros}
\usepackage{graphicx}
\usepackage{natbib}
\usepackage{color}
\usepackage{listings}
\usepackage{hyperref}  % use [draft] mode if you get a \pdfendlink error and find the citation which is split between two pages and surround with a \mbox{}.
\usepackage{url}
\usepackage{cases}
\usepackage{tikz}
\usetikzlibrary{decorations.pathreplacing}

\hypersetup{dvips, colorlinks=true, linkcolor=cyan, citecolor=black, filecolor=black, urlcolor=black}

\definecolor{grey}{rgb}{0.75,0.75,0.75}
\definecolor{Orange}{rgb}{1.0,0.5,0.15}
\definecolor{brown}{rgb}{0.7,0.25,0.0}
\definecolor{pink}{rgb}{1.0,0.5,0.5}
\definecolor{darkerred}{rgb}{0.8,0,0}
\definecolor{darkerblue}{rgb}{0,0,0.8}
\definecolor{Blue}{rgb}{0,0.08,0.65}
\definecolor{Red}{rgb}{0.65,0.08,0.05}
\definecolor{Green}{rgb}{0.15,0.45,0.25}

\begin{document}

\setcounter{tocdepth}{3}

\title
{
Secular diffusion in discrete self-gravitating tepid discs \\
II: accounting for swing amplification via the matrix method 
}

\titlerunning{Secular diffusion in discrete self-gravitating tepid discs}

\author{J.~B. Fouvry
\inst{\ref{inst1}}
\and
C. Pichon
\inst{\ref{inst1},\ref{inst2}}
\and
J. Magorrian
\inst{\ref{inst3},\ref{inst1}}
\and
P.~H. Chavanis
\inst{\ref{inst4}}
}

\institute{
Institut d'Astrophysique de Paris and UPMC, CNRS (UMR 7095), 98 bis Boulevard Arago, 75014, Paris, France\\
\email{fouvry@iap.fr}\label{inst1}
%\email{fouvry@iap.fr; pichon@iap.fr}\label{inst1}
\and
Institute of Astronomy \& KICC, University of Cambridge, Madingley Road, Cambridge, CB3 0HA, United Kingdom
\label{inst2}
\and
Rudolf Peierls Centre for Theoretical Physics, University of Oxford, Keble Road, Oxford OX1 3RH, United Kingdom 
%\email{john.magorrian@physics.ox.ac.uk}
\label{inst3}
\and
Laboratoire de Physique Th\'eorique (IRSAMC), CNRS and UPS, Univ. de Toulouse, F-31062 Toulouse, France
%\email{chavanis@irsamc.ups-tlse.fr}
\label{inst4}
}

\date{Received \today /
Accepted --}
\abstract{
The secular evolution of an infinitely thin tepid isolated galactic disc  made of a finite number of particles
is investigated using the  inhomogeneous Balescu-Lenard equation expressed in terms of  angle-action variables.
The matrix method is 
implemented numerically in order to model the induced gravitational polarization.
Special care is taken to account for the  amplification of potential fluctuations of mutually resonant orbits
and the unwinding of the induced swing amplified transients.  
Quantitative comparisons with ${N-}$body simulations yield consistent scalings with the number of particles and with the self-gravity of the disc:  the fewer particles and
the colder the disc, the faster the secular evolution.  
Secular evolution is driven by resonances, but does \textit{not} depend on the 
initial phases of the disc. 
For a Mestel disc with ${ Q \!\sim\! 1.5 }$, 
the polarization cloud around each star boosts up its secular effect by a factor of the order of a thousand or more, promoting accordingly
the dynamical relevance of self-induced collisional secular evolution.
The position and shape of the induced resonant ridge
are found to be in very good agreement with 
the prediction of the Balescu-Lenard equation, which  scales with the \textit{square} of the susceptibility of the disc. 
\\
In astrophysics, the inhomogeneous Balescu-Lenard equation may  describe the secular diffusion of giant molecular clouds in galactic discs, the secular migration  and segregation of planetesimals in proto-planetary discs, or even the long-term evolution of  population of stars within the Galactic centre.
It  could be used as a valuable check of the accuracy of ${N-}$body integrators  over secular timescales.
}


\maketitle

\section{Introduction}
\label{sec:introduction}
Galactic astronomy has striven to 
understand  the dynamical  evolution of discs  over cosmic times. 
For these self-gravitating systems, fluctuations of the potential  induced by discrete   encounters may be strongly amplified \citep{Kalnajs1972}, 
 while resonances tend to confine and localise their dissipation: such small stimuli can lead to  long-term spontaneous  evolution towards distinct  galactic equilibria.
The effect of this susceptibility on secular timescales will be addressed here in the context of an extended kinetic theory which takes explicitly into account such interactions.
  
The kinetic theory of stellar systems  was initiated by~\cite{Jeans1929} and 
\cite{Chandrasekhar1942}  in the context of hot spherical stellar systems such as elliptical galaxies and globular clusters
for which the gravitational susceptibility can safely be neglected. 
In contrast, self-gravitating galactic discs are cold  dynamical systems, for which rotation represents an important reservoir of free energy. 
More generally, in astrophysics, the secular diffusion of giant molecular clouds in galactic discs, the secular migration  and segregation of planetesimals in proto-planetary or debris discs,   or even the long-term evolution of  population of stars within the Galactic centre are all processes for which it is of interest to  quantify the 
dynamical effect of gravitationally amplified potential fluctuations induced by the finite number of stars involved.
 
 More than fifty five years ago, \cite{Balescu1960} and
\cite{Lenard1960} developed a rigorous kinetic theory  taking
collective effects into account, and obtained the corresponding kinetic equation for plasmas, the
Balescu-Lenard equation.  More recently \cite{Heyvaerts2010} and \cite{Chavanis2012} have transposed the corresponding non-linear
kinetic equation to the angle-action variables that are the appropriate variables
to describe \textit{spatially inhomogeneous} multi-periodic systems. 
The corresponding inhomogeneous Balescu-Lenard  equation 
 accounts for self-driven orbital secular diffusion of a self-gravitating system induced by the intrinsic shot noise due to its discreteness. 
Note that the formal transposition  from position-velocity to angle-action
implies that the secular interaction need not be local in space: they only need to correspond to gravitationally amplified long range correlations and resonances,
which are indeed the driving mechanism for the secular evolution of isolated astrophysical discs via angular momentum redistribution~\citep{LyndenBell1972}. 

 The Balescu-Lenard equation is valid
at the order ${ 1/N }$ in an expansion of the dynamics in terms of this
small parameter, where ${ N \!\gg\! 1 }$ is the number of stars. Therefore, it takes
finite${-N}$ effects into account and describes the evolution of
the system on a timescale of the order ${ N t_D }$, where $t_D$ is the dynamical
time.
For self-gravitating systems, the collective effects are responsible
for an  anti-shielding which tends to  increase the effective mass of the
stars, hence reducing the relaxation time.  
 When the system is  cold, each particle is dressed by the very strong gravitational polarization it induces, hence
the secular effects may occur on much shorter timescales than one would naively think,
so that, say ${ N_{\rm eff} \!\sim\! N/10^{\rm few}}$. 
The purpose of this paper is to quantify this effect for stable but strongly susceptible 
 galactic discs.

   The Balescu-Lenard  formalism has
seldomly been applied in its prime context,  but only in various limits
where it reduces to simpler kinetic equations \citep{Landau1936,Vlasov1938,Chandrasekhar1942,Rosenbluth1957}.
\cite{Weinberg1993} presents an interesting first implementation, though
 in a  somewhat over-simplified cartesian geometry.  Yet, this formalism
 is quite unique in accounting for the \textit{non-linear} evolution of discs and galaxies over secular timescales.
${N-}$body simulations, while potentially probing similar processes, should be scrutinized in such regime, as shadowing may, over many orbital times
impact resonant interactions. ${N-}$body simulations have been shown to more or less reproduce growth rates of discs on \textit{dynamical} timescales \citep[see, e.g.][and references therein, together with Appendix~\ref{sec:MatrixOK}]{SellwoodEvans2001}; qualifying them quantitatively  over \textit{secular} timescales
  is now within reach of the Balescu-Lenard  formalism.

 The companion paper, \cite{Fouvry2015}, hereafter paper~I, 
presented a simple and tractable quadrature for the Balescu-Lenard  drift and diffusion coefficients while assuming that the transient response of the disc was  described by
tightly wound spirals. Paper I  applied the corresponding WKB approximation, while assuming that the disc was tepid and that the epicyclic approximation held.
These simple expressions provided insight into the physical processes at work during the secular diffusion of  self-gravitating discrete discs. 
When applied to the secular evolution of an isolated stationary self-gravitating Mestel disc, it identified  the  importance of the corotation resonance in the inner regions of the disc leading to a regime with both radial migration and heating, in qualitative agreement with  numerical simulations.

Yet, the tightly wound approximation is quantitatively questionable when transient spirals unwind. 
Indeed paper I found a timescale discrepancy between the predicted secular evolution timescale and the measured one, which
might be driven by the incompleteness of the WKB basis.  Such   basis 
can only correctly represent tightly wound spirals. 
It also enforced \textit{local resonances}, and did not allow for remote orbits to resonate, or wave packets to propagate between such non-local resonances. 
Yet, the seminal works from \cite{GoldreichLyndenBell1965a,JulianToomre1966,Toomre1981} showed that any leading spiral wave undergoes  significant amplification during its unwinding to a trailing wave. 
Because it involves unwinding spirals this mechanism is not captured by the WKB formalism of paper I.

In this paper, we will make no such  approximations and will therefore compute numerically  the corresponding 
diffusion and drift  coefficients while relying on the matrix method \citep{Kalnajs2}
to estimate the gravitational amplification of the secular response. It will allow us to assess the amplitude of the cross-talk between non-local resonances.  
We will then compare those predictions to   crafted sets of numerical experiments, allowing us to estimate ensemble averaged secular responses
of a sizable number of simulations 
 as a function of the total number of particles $N$. Such ensemble average will allow us to make robust predictions for the $N$-scaling of the 
 secular response and its dependence on halo to disc mass fraction, hence probing the secular importance of gravitational polarization.

The paper is organized as follows. Section~\ref{sec:inhomogeneousBL} briefly presents the content of the inhomogeneous Balescu-Lenard equation.
Section~\ref{sec:disccase} presents our implementation of the matrix method to compute the diffusion equation for an isolated self-gravitating tapered Mestel disc.
Section~\ref{sec:application} computes numerically the exact drift and diffusion coefficient in action space 
for such a truncated Mestel disc, and compares the divergence of the corresponding flux density to the initial measured rate of change  of the distribution function.
Section~\ref{sec:NB} presents our ${N-}$body simulations and compares scaling of the flux with the number of particles and the fraction of mass in the disc.
Finally, section~\ref{sec:conclusion} wraps up. 
Appendix~\ref{sec:Kalnajsbasis}  presents the relevant bi-orthogonal basis function.
Appendix~\ref{sec:MatrixOK} validates the response matrix method and the ${N-}$body integrator while matching growth rates and pattern speeds of unstable Mestel discs.
Appendix~\ref{sec:swing-test} investigates the roles of self-gravity and basis completeness.
Appendix~\ref{sec:sampling} describes the sampling strategy for the initial distribution. 
Appendix~\ref{sec:codes} presents briefly the available online codes.

\section{The inhomogeneous Balescu-Lenard equation}
\label{sec:inhomogeneousBL}

We intend to describe the long-term evolution of a system made of $N$ particles. We assume that the gravitational background $\psi_{0}$ of the system is stationary and integrable, and associated with the Hamiltonian $H_{0}$. As a consequence, one can always remap the physical space-coordinates ${ (\bm{x} , \bm{v}) }$ to the angle-action coordinates ${(\bm{\theta}, \bm{J}) }$ \citep{Goldstein,born1960mechanics,BinneyTremaine2008}. We define the intrinsic frequencies of motions along the action torus as
\begin{equation}
\bm{\Omega} (\bm{J}) = \dot{\bm{\theta}} = \frac{\partial H_{0}}{\partial \bm{J}} \, .
\label{definition_Omega}
\end{equation}
Within these new coordinates, one has that along the unperturbed trajectories the angles $\bm{\theta}$ are ${2 \pi-}$periodic, evolving with the frequencies $\bm{\Omega}$, whereas the actions $\bm{J}$ are conserved. We assume that the system is always in a virialised state, so that its distribution function (DF) can be written as a quasi-stationary DF of the form ${ F \!=\! F (\bm{J} , t) }$, satisfying the normalization constraint ${ \int \!\! \mathrm{d} \bm{x} \mathrm{d} \bm{v} \, F \!=\! M_{\rm tot} }$, where ${ M_{\rm tot} }$ is the total mass of the system. On secular timescales, this isolated DF evolves under the effect of stellar \textit{encounters} (finite${-N}$ effects). Such a collisional long-term evolution is descrided by the inhomogeneous Balescu-Lenard equation \citep{Heyvaerts2010,Chavanis2012} which reads
\begin{align}
\frac{\partial F}{\partial t} = \pi (2 \pi)^{d} \, \mu \, \frac{\partial}{\partial \bm{J}_{1}} \!\cdot\! \bigg[ \!\! \sum_{\bm{m}_{1} , \bm{m}_{2}} \!\! \bm{m}_{1} \!\! \int \!\! \mathrm{d} \bm{J}_{2} \, \frac{\delta_{\rm D} (\bm{m}_{1} \!\cdot\! \bm{\Omega}_{1} \!-\! \bm{m}_{2} \!\cdot\! \bm{\Omega}_{2})}{|\mathcal{D}_{\bm{m}_{1} , \bm{m}_{2}} (\bm{J}_{1} , \bm{J}_{2} , \bm{m}_{1} \!\cdot\! \bm{\Omega}_{1}) |^{2}}  \nonumber
\\
\times \, \bigg( \bm{m}_{1} \!\cdot\! \frac{\partial }{\partial \bm{J}_{1}} \!-\! \bm{m}_{2} \!\cdot\! \frac{\partial }{\partial \bm{J}_{2}} \bigg) \, F (\bm{J}_{1} , t) \, F(\bm{J}_{2} , t) \bigg] \, ,
\label{definition_BL}
\end{align}
where ${1 / \mathcal{D}_{\bm{m}_{1} , \bm{m}_{2}} (\bm{J}_{1} , \bm{J}_{2} , \omega) }$ are the dressed susceptibility coefficients, 
 $d$ is the dimension of the physical space, ${ \mu \!=\! M_{\rm tot}
   / N }$ is the mass of the individual particles, and where we used
 the shortened notation ${\bm{\Omega}_{i} \!=\! \bm{\Omega}
   (\bm{J}_{i})}$. Since it is written as the divergence of a flux,
 this diffusion equation conserves the number of stars. One should
 also note the  resonance condition encapsulated in the Dirac delta ${
   \delta_{\rm D} (\bm{m}_{1} \!\cdot\! \bm{\Omega}_{1} \!-\!
   \bm{m}_{2} \!\cdot\! \bm{\Omega}_{2}) }$, with the integration over
 the dummy variable $\bm{J}_{2}$ scanning for points where the resonance condition is satisfied.
Note importantly right away that equation~\eqref{definition_BL} scales like ${ 1/(N \mathcal{D}^{2}) }$ (since ${ \mu \!\propto\! 1/N }$), so that increasing $N$ or increasing the heat content of the disc have the same effect.
For a more detailed discussion on the content of the Balescu-Lenard equation, see paper I.
 
In order to solve the \textit{non-local} Poisson equation, we follow Kalnajs' matrix method \citep{Kalnajs2}, so that we introduce a complete biorthonormal basis of potentials and densities ${ \psi^{(p)} (\bm{x}) }$ and ${ \rho^{(p)} (\bm{x}) }$ such that
\begin{equation}
\Delta \psi^{(p)} \!= 4 \pi G \rho^{(p)} \, , \quad
\int \!\! \mathrm{d} \bm{x} \, [ \psi^{(p)} (\bm{x}) ]^{*} \, \rho^{(q)} (\bm{x}) = - \, \delta_{p}^{q} \, .
\label{definition_basis}
\end{equation}
The dressed susceptibility coefficients appearing in equation~\eqref{definition_BL} are then given by
\begin{equation}
\frac{1}{\mathcal{D}_{\bm{m}_{1} , \bm{m}_{2}} (\bm{J}_{1} , \bm{J}_{2} , \omega)} =  \sum_{p , q}  \psi_{\bm{m}_{1}}^{(p)} \!(\bm{J}_{1}) \, [ \mathbf{I} \!-\! \widehat{\mathbf{M}} (\omega) ]^{-1}_{p q} \, [ \psi_{\bm{m}_{2}}^{(q)} \!(\bm{J}_{2}) ]^{*} \, ,
\label{definition_1/D}
\end{equation}
where $\mathbf{I}$ is the identity matrix and ${ \widehat{\mathbf{M}} }$ is the response matrix defined as
\begin{equation}
\widehat{\mathbf{M}}_{pq} (\omega) = (2 \pi)^{d} \! \sum_{\bm{m}} \!\! \int \!\! \mathrm{d} \bm{J} \, \frac{\bm{m} \!\cdot\! \partial F / \partial \bm{J} }{\omega \!-\! \bm{m} \!\cdot\! \bm{\Omega}} [\psi_{\bm{m}}^{(p)} \!(\bm{J})]^{*} \psi_{\bm{m}}^{(q)} \!(\bm{J}) \, .
\label{Fourier_M}
\end{equation}
In the previous expression, we introduced as ${ \psi_{\bm{m}}^{(p)} (\bm{J}) }$ the Fourier transform in angles of the basis elements ${ \psi^{(p)} (\bm{x}) }$ using the convention that the Fourier transform of a function ${ X (\bm{\theta} , \bm{J}) }$ is given by
\begin{equation}
\begin{cases}
\displaystyle X (\bm{\theta} , \bm{J}) = \!\! \sum_{\bm{m} \in \mathbb{Z}^{d}} X_{\bm{m}} (\bm{J}) \, e^{i \bm{m} \cdot \bm{\theta}} \, ,
\\
\displaystyle X_{\bm{m}} (\bm{J}) = \frac{1}{(2 \pi)^{d}} \!\! \int \!\! \mathrm{d} \bm{\theta} \, X (\bm{\theta} , \bm{J}) \, e^{- i \bm{m} \cdot \bm{\theta}} \, .
\end{cases}
\label{definition_Fourier_angles}
\end{equation}

In order to ease the understanding of the Balescu-Lenard equation~\eqref{definition_BL}, one may rewrite it under the form of an anisotropic Fokker-Planck equation, by introducing the relevant drift and diffusion coefficients. Indeed, equation~\eqref{definition_BL} can be put under the form
\begin{equation}
  \frac{\partial F}{\partial t} =  \sum_{\bm{m}_{1}} \frac{\partial }{\partial \bm{J}_{1}} \!\cdot\! \left[  \bm{m}_{1} \left( A_{\bm{m}_{1}} ( \bm{J}_{1} ) \, F ( \bm{J}_{1} ) +  D_{\bm{m}_{1}}  ( \bm{J}_{1} ) \, \bm{m}_{1}  \!\cdot\!  \frac{\partial F}{\partial \bm{J}_{1}}  \right)  \right] ,
\label{initial_BL_rewrite}
\end{equation}
where ${ A_{\bm{m}_{1}} (\bm{J}_{1}) }$ and ${ D_{\bm{m}_{1}} (\bm{J}_{1}) }$ are respectively the drift and diffusion coefficients associated with a given resonance $\bm{m}$. One should note that they both depend secularly on the distribution function, but this dependence was not exactly written out in order to shorten the notations. The drift coefficients are given by
\begin{equation}
A_{\bm{m}_{1}} (\bm{J}) \! = \! - \pi (2 \pi)^{d} \mu \!\!  \sum_{\bm{m}_{2}}  \!\!  \int  \!\!\!  \mathrm{d} \bm{J}_{2}  \frac{\delta_{\rm D} (\bm{m}_{1}  \!\cdot\!  \bm{\Omega}_{1}  \!-\!  \bm{m}_{2}  \!\cdot\!  \bm{\Omega}_{2})}{|\mathcal{D}_{\bm{m}_{1} , \bm{m}_{2}} (\bm{J}_{1} , \!\bm{J}_{2}, \!\bm{m}_{1}  \!\cdot\!  \bm{\Omega}_{1} \!) |^{2}} \bm{m}_{2}  \!\cdot\!  \frac{\partial F}{\partial \bm{J}_{2}},
\label{initial_drift}
\end{equation}
and the diffusion coefficients are given by
\begin{equation}
D_{\bm{m}_{1}} (\bm{J}_{1}) \! =\! \pi (2 \pi)^{d} \mu \!\! \sum_{\bm{m}_{2}} \!\! \int \!\! \mathrm{d} \bm{J}_{2} \frac{\delta_{\rm D} (\bm{m}_{1} \!\cdot\! \bm{\Omega}_{1} \!-\! \bm{m}_{2} \!\cdot\! \bm{\Omega}_{2})}{|\mathcal{D}_{\bm{m}_{1} , \bm{m}_{2}} (\bm{J}_{1} , \bm{J}_{2},\bm{m}_{1} \!\cdot\! \bm{\Omega}_{1}) |^{2}} F (\bm{J}_{2}) \, .
\label{initial_diff}
\end{equation}
One can also introduce the total flux of diffusion ${ \bm{\mathcal{F}}_{\rm tot} }$ as
\begin{equation}
\bm{\mathcal{F}}_{\rm tot} =  \sum_{\bm{m}} \bm{m} \left( A_{\bm{m}} (\bm{J}) \, F (\bm{J}) \!+\! D_{\bm{m}} (\bm{J}) \, \bm{m} \!\cdot\! \frac{\partial F}{\partial \bm{J}} \right) \, ,
\label{definition_F_tot}
\end{equation}
so that the Balescu-Lenard equation from equations~\eqref{definition_BL} and~\eqref{initial_BL_rewrite} takes the explicitly conservative form
\begin{equation}
\frac{\partial F}{\partial t} = \text{div} \left( \bm{\mathcal{F}}_{\rm tot} \right) \, .
\label{BL_div_Ftot}
\end{equation}

\section{The Matrix diffusion equation}
\label{sec:disccase}

When computing the Balescu-Lenard diffusion and drift coefficients, three main difficulties  have to be addressed.  First, one must build  the mapping ${ (\bm{x} , \bm{v}) \!\mapsto\! (\bm{\theta} , \bm{J}) }$, because the drift and diffusion coefficients are associated with a diffusion in action space. The second difficulty follows from the non-locality of Poisson's equation and the estimation of the response matrix $\widehat{\mathbf{M}}$ from equation~\eqref{Fourier_M}. Indeed, as noted in equation~\eqref{definition_basis}, the matrix relies on  potential basis elements $\psi^{(p)}$ which must be 
integrated over the whole action space with functions possessing a pole ${  1 / (\omega \!-\! \bm{m} \!\cdot\! \bm{\Omega}) }$. This cumbersome and difficult evaluation has to be performed numerically, along with the matrix inversion needed to estimate the susceptibility coefficients from equation~\eqref{definition_1/D}. Finally, the third difficulty arises from the resonance condition ${ \delta_{\rm D} (\bm{m}_{1} \!\cdot\! \bm{\Omega}_{1} \!-\! \bm{m}_{2} \!\cdot\! \bm{\Omega}_{2}) }$, which requires to determine how orbits may resonate one with another. In contrast, in paper I, we relied on the epicyclic approximation to build  the angle-action mapping and on a WKB basis to treat gravity locally in order to solve these issues while obtaining tractable though approximate expressions. 

For a $2D$ axisymmetric potential, one can define explicitly  the actions of the system. Following \cite{LyndenBell1972,TremaineWeinberg1984}, the two natural actions of the system are given by a quadrature and an identity
\begin{equation}
\begin{cases}
\displaystyle J_{1} = J_{r} = \frac{1}{\pi} \!\! \int_{r_{p}}^{r_{a}} \!\!\!\! \mathrm{d} r \, \sqrt{2(E \!-\! \psi_{0} (r)) \!-\! L^{2} / r^{2}} \, ,
\\
\displaystyle J_{2} = J_{\phi} = L \, ,
\end{cases}
\label{definition_actions}
\end{equation}
where $r_{p}$ and $r_{a}$ are respectively the pericentre and the apocentre of the trajectory, while $E$ and $L$ are the energy and angular momentum of the star. The first action $J_{1}$ encodes the amount of \textit{radial energy} of the star, so that ${ J_{r} \!=\! 0 }$ corresponds to circular orbits. The second action $J_{2}$ is the angular momentum $L$ of the star. One can then define the two intrinsic frequencies of motion ${ \Omega_{1} \!=\! \kappa }$ associated with the radial oscillations and ${ \Omega_{2} \!=\! \Omega_{\phi} }$ associated with the azimuthal oscillations. Indeed, one has
\begin{equation}
\frac{2 \pi}{\Omega_{1}} = 2 \!\! \int_{r_{p}}^{r_{a}} \!\!\!\! \frac{\mathrm{d} r}{\sqrt{2 (E \!-\! \psi_{0} (r)) \!-\! J_{2}^{2} / r^{2}}}  \, ,
\label{definition_Omega1}
\end{equation}
while the azimuthal frequency $\Omega_{2}$ can then be determined via the relation
\begin{equation}
\frac{\Omega_{2}}{\Omega_{1}} = \frac{J_{2}}{\pi} \!\! \int_{r_{p}}^{r_{a}} \!\!\!\! \frac{\mathrm{d} r}{r^{2} \sqrt{2 (E \!-\! \psi_{0}(r)) \!-\! J_{2}^{2} / r^{2}}} \, .
\label{definition_Omega2}
\end{equation}

At this stage, one should note that various coordinates can be used to represent the ${ 2D }$ action space. Indeed, once the background potential $\psi_{0}$ is known, one has the bijections ${ (r_{p}, r_{a}) \leftrightarrow (E,L) \leftrightarrow (J_{r}, J_{\phi}) }$. As a consequence, any orbit can equivalently be represented by the set of the pericentre and apocentre ${ (r_{p} , r_{a}) }$ or by its actions ${ (J_{1} , J_{2}) }$. However, determining the actions associated with one set ${ (r_{p},r_{a}) }$ only requires the computation of a ${ 1D }$ integral as in equation~\eqref{definition_actions}, whereas determing the pericentre and apocentre associated with a set of actions ${ (J_{1} , J_{2}) }$ requires the inversion of the same non-trivial relation. Because the peri/apocentres are the two roots of the equation ${ 2 (E \!-\! \psi_{0}(r)) \!-\! L^{2}/r^{2} \!=\! 0 }$, one also immediately obtains that for a given value of $r_{p}$ and $r_{a}$, the energy $E$ and the angular momentum $L$ of the orbit are immediately given by
\begin{equation}
E = \frac{r_{a}^{2} \, \psi_{a} \!-\! r_{p}^{2} \, \psi_{p}}{r_{a}^{2} \!-\! r_{p}^{2}} \;\;\; ; \;\;\; L = \sqrt{\frac{2 (\psi_{a} \!-\! \psi_{p})}{r_{p}^{-2} \!-\! r_{a}^{-2}}} \, ,
\label{link_rpra_EL}
\end{equation}
where we used the shortening notations ${ \psi_{p/a} \!=\! \psi_{0} (r_{p/a}) }$. Therefore, in the upcoming calculations, we will use ${ (r_{p},r_{a}) }$ as the representative variables of the ${ 2D }$ action space.

\subsection{The basis elements}

The expressions~\eqref{definition_1/D} and~\eqref{Fourier_M} of the susceptibility coefficients and the response matrix require the introduction of ${ 2D }$ potential-density basis elements. The ${ 2D }$ potential basis elements ${ \psi^{(p)} }$ that we will consider will depend on two indices spanning the two degrees of freedom so that one has
\begin{equation}
\psi^{(p)} (R , \phi) = \psi_{n}^{\ell} (R, \phi) = e^{i \ell \phi} \, \mathcal{U}_{n}^{\ell} (R , \phi) \, ,
\label{definition_psi_p}
\end{equation}
where ${ \mathcal{U}_{n}^{\ell} }$ is a real radial function and ${ (R , \phi) }$ are the usual polar coordinates. The associated surface densities elements will be of the form
\begin{equation}
\Sigma^{(p)} (R , \phi) = \Sigma_{n}^{\ell} (R , \phi) = e^{i \ell \phi} \, \mathcal{D}_{n}^{\ell} (R , \phi) \, ,
\label{definition_Sigma_p}
\end{equation}
where ${ \mathcal{D}_{n}^{\ell} }$ is a real radial function. The basis elements therefore depend on two indices ${ \ell \!\geq\! 0 }$ and ${ n \!\geq\! 0 }$. In all the numerical calculations, we used the radial functions from \cite{Kalnajs2}, which are recalled in Appendix~\ref{sec:Kalnajsbasis}.

The next step is then to determine the Fourier transform with respect to the angles $\bm{\theta}$ of the basis elements. Indeed, to compute the response matrix ${ \widehat{\mathbf{M}} }$ from equation~\eqref{Fourier_M}, one has to compute ${ \psi_{\bm{m}}^{(p)} (\bm{J}) }$, where the resonance vector is given by ${ \bm{m} \!=\! (m_{1} , m_{2}) }$. Following equation~\eqref{definition_Fourier_angles}, it is given by
\begin{equation}
\psi_{\bm{m}}^{(p)} (\bm{J}) = \frac{1}{(2 \pi)^{2}} \!\! \int \!\!\mathrm{d} \theta_{1} \mathrm{d} \theta_{2} \, \psi^{(p)} (R , \phi) \, e^{- i m_{1} \theta_{1}} \, e^{- i m_{2} \theta_{2}} \, .
\label{calculation_psi_m_p}
\end{equation}
From \cite{LyndenBell1972}, the angles $\theta_{1}$ and $\theta_{2}$ associated with the actions from equation~\eqref{definition_actions} are given by
\begin{equation}
\begin{cases}
\displaystyle \theta_{1} = \Omega_{1} \int_{\mathcal{C}_{1}} \!\! \mathrm{d} r \, \frac{1}{\sqrt{2 (E \!-\! \psi_{0}(r)) \!-\! J_{2}^{2}/r^{2}}} \, ,
\\
\displaystyle \theta_{2} = \phi \!+\! \!\! \int_{\mathcal{C}_{1}} \!\! \mathrm{d} r \, \frac{\Omega_{2} \!-\! J_{2} / r^{2}}{ \sqrt{2(E \!-\! \psi_{0} (r)) \!-\! J_{2}^{2} / r^{2}}} \, ,
\end{cases}
\label{angles_2D}
\end{equation}
where $\mathcal{C}_{1}$ is a contour starting from the pericentre $r_{p}$ and going up to the current position ${ r \!=\! r (\theta_{1}) }$ along the radial oscillation. Following the notations from \cite{TremaineWeinberg1984}, one can straightforwardly show that equation~\eqref{calculation_psi_m_p} takes the form
\begin{equation}
\psi_{\bm{m}}^{(p)} (\bm{J}) = \delta_{m_{2}}^{\ell^{p}} \, \mathcal{W}^{m_{1}}_{\ell^{p} m_{2} n^{p}} (\bm{J}) \, ,
\label{result_psi_m_p}
\end{equation}
where ${ \mathcal{W}^{m_{1}}_{\ell^{p} m_{2} n^{p}} (\bm{J}) }$ is given by
\begin{equation}
\mathcal{W}^{m_{1}}_{\ell^{p} m_{2} n^{p}} (\bm{J}) \!=\! \frac{1}{\pi} \!\! \int_{r_{p}}^{r_{a}} \!\!\!\!\!\! \mathrm{d} r \, \frac{\mathrm{d} \theta_{1}}{\mathrm{d} r}  \mathcal{U}_{n^{p}}^{\ell^{p}} (r) \cos [ m_{1} \theta_{1} [r] \!+\! m_{2} (\theta_{2} \!-\! \phi) [r] ] \, .
\label{expression_W}
\end{equation}
In equation~\eqref{expression_W}, the boundaries of the integral are given by the pericentre $r_{p}$ and apocentre $r_{a}$ associated with the action $\bm{J}$. Such an expression underlines the reason why ${ (r_{p} , r_{a}) }$ appear naturally as \textit{good} coordinates to describe the ${ 2D }$ action space. One can note that equation~\eqref{expression_W} involves an integral over ${ r }$ thanks to the change of variables ${ \theta_{1} \!\to\! r }$, which satisfies
\begin{equation}
\frac{\mathrm{d} \theta_{1}}{\mathrm{d} r} = \frac{\Omega_{1}}{\sqrt{2 (E \!-\! \psi_{0} (r)) \!-\! J_{2}^{2} / r^{2}}} \, .
\label{dtheta1_dr}
\end{equation}
In equation~\eqref{expression_W}, ${ \theta_{1} [r] }$ and ${ (\theta_{2} \!-\! \phi) [r] }$ only depend on $r$ via the mappings from equation~\eqref{angles_2D}. Provided that ${ \mathcal{U}_{n^{p}}^{\ell^{p}} }$ is a real function, the coefficients ${ \mathcal{W}^{m_{1}}_{\ell^{p} m_{2} n^{p}} }$ are always real. Because these coefficients involve two intricated integrals, they are numerically expensive to compute. However by parity, they obey ${ \mathcal{W}^{(-m_{1})}_{\ell^{p} (- m_{2}) n^{p}} \!\!=\! \mathcal{W}^{m_{1}}_{\ell^{p} m_{2} n^{p}} }$, which allows a significant reduction of the number of coefficients to compute.

\subsection{Computation of the response matrix}

We now have all the elements required to compute the response matrix from equation~\eqref{Fourier_M}. In its definition, one should note the presence of an integral over the mute variable $\bm{J}$, which, as discussed previously, will be performed in the ${ 2D }$ ${ (r_{p},r_{a})-}$space. The first step is to go from ${ \bm{J} \!=\! (J_{1} , J_{2}) }$ to ${ (E,L) }$. The Jacobian of this transformation is given by
\begin{equation}
\frac{\partial (E,L)}{\partial (J_{1} , J_{2})} =
\begin{vmatrix}
\displaystyle \frac{\partial E}{\partial J_{1}} &\displaystyle\frac{\partial E}{\partial J_{2}} 
\\
\displaystyle \frac{\partial L}{\partial J_{1}} & \displaystyle\frac{\partial L}{\partial J_{2}}
\end{vmatrix}
 = 
\begin{vmatrix}
\Omega_{1} & \Omega_{2}
\\
0 & 1
\end{vmatrix}
= \Omega_{1} \, ,
\label{Jacobian_J_EL}
\end{equation}
so that one immediately has ${ \mathrm{d} J_{1} \mathrm{d} J_{2} \!=\! \mathrm{d} E \mathrm{d} L \, /{\Omega_{1}} }$. Given the expression~\eqref{result_psi_m_p} of the ${ 2D }$ Fourier transformed basis elements, the response matrix may be written under the form
\begin{align}
\widehat{\mathbf{M}}_{pq} (\omega) = (2 \pi )^{2} \delta_{\ell^{p}}^{\ell^{q}} \! \sum_{m_{1}} \!\! \int\!\! & \, \mathrm{d} E \, \mathrm{d} L \, \frac{1}{\Omega_{1}} \, \frac{(m_{1} , \ell^{p}) \!\cdot\! \partial F_{0} / \partial \bm{J}  }{\omega \!-\! (m_{1} , \ell^{p}) \!\cdot\! \bm{\Omega}} \nonumber
\\
& \times \mathcal{W}^{m_{1}}_{\ell^{p} \ell^{p} n^{p}} (\bm{J}) \, \mathcal{W}^{m_{1}}_{\ell^{p} \ell^{p} n^{q}} (\bm{J}) \, ,
\label{response_M_EL}
\end{align}
where the sum on $m_{2}$ has been dropped. Moreover, we dropped the conjugate over $ \mathcal{W}^{m_{1}}_{\ell^{p} \ell^{p} n^{p}} $ since they are always real. We may now perform the change of variables ${ (E,L) \!\to\! (r_{p} , r_{a}) }$, so as to rewrite equation~\eqref{response_M_EL} under the form
\begin{equation}
\widehat{\mathbf{M}}_{pq} (\omega) = \delta_{\ell^{p}}^{\ell^{q}} \! \sum_{m_{1}} \!\! \int \!\! \mathrm{d} r_{p} \mathrm{d} r_{a} \, \frac{g_{m_{1}}^{\ell^{p} n^{p} n^{q}} (r_{p} , r_{a})}{h_{m_{1} \ell^{p}}^{\omega} (r_{p} , r_{a})} \, ,
\label{response_M_rpra}
\end{equation}
where the functions ${ g_{m_{1}}^{\ell^{p} n^{p} n^{q}} (r_{p} , r_{a})}$ and ${ h_{m_{1} , \ell^{p}}^{\omega} (r_{p} , r_{a}) }$ are respectively given by
\begin{align}
g_{m_{1}}^{\ell^{p} n^{p} n^{q}} (r_{p} , r_{a}) = & \, (2 \pi)^{2} \bigg| \frac{\partial (E,L)}{\partial (r_{p} , r_{a})} \bigg| \, \frac{1}{\Omega_{1}} \bigg[ (m_{1} , \ell^{p}) \!\cdot\! \frac{\partial F_{0}}{\partial \bm{J}} \bigg] \nonumber
\\
& \times \mathcal{W}^{m_{1}}_{\ell^{p} \ell^{p} n^{p}} (\bm{J}) \, \mathcal{W}^{m_{1}}_{\ell^{p} \ell^{p} n^{q}} (\bm{J}) \, ,
\label{definition_g}
\end{align}
and
\begin{equation}
h_{m_{1} \ell^{p}}^{\omega} (r_{p} , r_{a}) = \omega \!-\! (m_{1} , \ell^{p}) \!\cdot\! \bm{\Omega} \, .
\label{definition_h}
\end{equation}
It is important to note here that the response matrix is diagonal with respect to the $\ell^{p}$ and $\ell^{q}$ indices so that each ${ \ell}$ may be treated independently. The definition of the function $g$ from equation~\eqref{definition_g} involves the Jacobian ${ \partial (E,L)/ \partial (r_{p} , r_{a} ) }$ of the transformation ${ (E,L) \!\to\! (r_{p} , r_{a}) }$ which can be immediately computed from the expressions~\eqref{link_rpra_EL} of ${ E \!=\! E (r_{p} , r_{a}) }$ and ${ L \!=\! L (r_{p} , r_{a}) }$. Moreover, in some situations, the DF ${ F \!=\! F (\bm{J}) }$, may also rather be defined as ${ F \!=\! F (E , L) }$. It is straightforward to show that one has
\begin{equation}
\bm{m} \!\cdot\! \frac{\partial F}{\partial \bm{J}} = m_{1} \Omega_{1} \frac{\partial F}{\partial E} \bigg|_{L} \!+ m_{2} \left[ \Omega_{2} \frac{\partial F}{\partial E} \bigg|_{L} \!+ \frac{\partial F}{\partial L} \bigg|_{E} \right] \, .
\label{calculation_m.dF/dJ}
\end{equation}

\subsection{Sub-region integration}

The next step of the calculation is then to perform the remaining integration over ${ (r_{p} , r_{a}) }$ from equation~\eqref{response_M_rpra}. Because of the presence of the resonant pole ${ 1 / h_{m_{1} , \ell^{p}}^{\omega} }$, such a numerical integration has to be performed carefully. We cut out the integration domain ${ (r_{p} , r_{a}) }$ in various subregions indexed by $i$. The $i^{\rm th}-$region will be centred around the position ${ (r_{p}^{i},r_{a}^{i}) }$ and will correspond to the square domain such that ${ r_{p} \!\in\! [r_{p}^{i} \!-\! \Delta r / 2 \,;\, r_{p}^{i} \!+\! \Delta r / 2 ]}$ and ${ r_{a} \!\in\! [r_{a}^{i} \!-\! \Delta r / 2 \,;\, r_{a}^{i} \!+\! \Delta r / 2 ] }$, where $\Delta r$ corresponds to the size of the subregions. The smaller $\Delta r$, the better will be the approximated estimations of the response matrix. Within the ${i^{\rm th}-}$ region, one can write first-order Taylor expansions of the functions $g$ and $h$ from equations~\eqref{definition_g} and~\eqref{definition_h} around the centre ${ (r_{p}^{i} , r_{a}^{i}) }$ of the region such that
\begin{equation}
\begin{cases}
\displaystyle g (r_{p}^{i} \!+\! \Delta r_{p} , r_{a}^{i} \!+\! \Delta r_{a}) \simeq a_{g}^{i} \!+\! b_{g}^{i} \Delta r_{p} \!+\! c_{g}^{i} \Delta r_{a} \, ,
\\
\displaystyle h (r_{p}^{i} \!+\! \Delta r_{p} , r_{a}^{i} \!+\! \Delta r_{a}) \simeq a_{h}^{i} \!+\! b_{h}^{i} \Delta r_{p} \!+\! c_{h}^{i} \Delta r_{a} \, ,
\end{cases}
\label{DL_g_h}
\end{equation}
where for convenience we shortened the index dependences from equations~\eqref{definition_g} and~\eqref{definition_h}. The coefficients $a_{g}^{i}$, $b_{g}^{i}$ and $c_{g}^{i}$ (similarly for $h$) are given by
\begin{equation}
a_{g}^{i} = g (r_{p}^{i} , r_{a}^{i}) \; ; \; b_{g}^{i} = \frac{\partial g}{\partial r_{p}} \bigg|_{(r_{p}^{i} , r_{a}^{i})} \; ; \; c_{g}^{i} = \frac{\partial g}{\partial r_{a}} \bigg|_{(r_{p}^{i} , r_{a}^{i})} \, ,
\label{definitions_a_b_c}
\end{equation}
where it is important to note that these coefficients are only functions of the central coordinates ${ (r_{p}^{i} , r_{a}^{i} ) }$ and will be treated as constants on each sub-region. In the numerical implementation, the coefficients involving partial derivatives will be estimated by finite differences, so that one will have for instance
\begin{equation}
b_{g} (r_{p}^{i} , r_{a}^{i}) = \frac{g (r_{p}^{i} \!+\! \Delta r , r_{a}^{i}) \!-\! g (r_{p}^{i} \!-\! \Delta r , r_{a}^{i})}{2 \Delta r} \, ,
\label{finite_differences_b}
\end{equation}
which allows to minimize the number of evaluations of $g$ required. The approximated integration on each sub-region can then be performed and takes the form
\begin{align}
\int \!\! \int_{i} \!\! \mathrm{d} r_{p} \mathrm{d} r_{a} \, \frac{g (r_{p} , r_{a})}{h (r_{p} , r_{a})} & \, \simeq \!\! \int_{- \frac{\Delta r}{2}}^{\frac{\Delta r}{2}} \!\! \int_{- \frac{\Delta r}{2}}^{\frac{\Delta r}{2}} \!\!\! \mathrm{d} x_{p} \mathrm{d} x_{a} \, \frac{a_{g}^{i} \!+\! b_{g}^{i} x_{p} \!+\! c_{g}^{i} x_{a}}{a_{h}^{i} \!+\! b_{h}^{i} x_{p} \!+\! c_{h}^{i} x_{a} \!+\! i \eta} \nonumber
\\
& = \aleph (a_{g}^{i} , b_{g}^{i} , c_{g}^{i} , a_{h}^{i} , b_{h}^{i} , c_{h}^{i} , \eta , \Delta r) \, ,
\label{integration_subregion}
\end{align}
where $\aleph$ is an analytical function which only depends on the coefficients obtained in the limited developments from equation~\eqref{DL_g_h}. In order to have a well-defined integral, we added an imaginary part ${ \eta \!>\! 0 }$ to the temporal frequency $\omega$, so that ${ \omega \!=\! \omega_{0} \!+\! i \eta }$. When looking for unstable modes in a disc, this imaginary part $\eta$ corresponds to the growth rate of the mode. It is also crucial to note here that one always has $a_{g}^{i}$, $b_{g}^{i}$, ${ c_{g}^{i} \!\in\! \mathbb{R} }$ and similarly $a_{h}^{i}$, $b_{h}^{i}$, ${ c_{h}^{i} \!\in\! \mathbb{R} }$. The effective computation of the function $\aleph$ is presented in Appendix~\ref{sec:aleph}. Thanks to equation~\eqref{integration_subregion}, the expression~\eqref{response_M_rpra} becomes
\begin{equation}
\widehat{\mathbf{M}}_{pq} (\omega) = \delta_{\ell^{p}}^{\ell^{q}} \! \sum_{m_{1}}  \sum_{i}  \aleph (a_{g}^{i} , b_{g}^{i} , c_{g}^{i}, a_{h}^{i} , b_{h}^{i} , c_{h}^{i} , \eta , \Delta r) \, .
\label{response_M_sum}
\end{equation}
In the previous expression, in order to effectively compute numerically the sum on $m_{1}$, we introduce a bound $m_{1}^{\rm max}$, so that the sum is only reduced to ${ |m_{1}| \!\leq\! m_{1}^{\rm max} }$.
Because of the requirement to truncate the action space in various subregions as in equation~\eqref{response_M_sum}, the computation of the response matrix still remains a daunting task, to ensure appropriate numerical convergence. In Appendix~\ref{sec:MatrixOK}, we detail the validation of our implementation of the response matrix calculation, by recovering known unstable modes of truncated Mestel discs~\citep{Zang1976,EvansRead1998II,SellwoodEvans2001}. Once the response matrix ${ \widehat{\mathbf{M}} }$ is known, the determination of the dressed susceptibility coefficients ${ 1/ |\mathcal{D}|^{2} }$ from equation~\eqref{definition_1/D} involves a straightforward summation\footnote{One could if needed regularize the inversion of ${ \mathbf{I} \!-\! \widehat{\mathbf{M}} }$, to avoid Gibbs rigging,
 since our basis is significantly truncated;
 this has proven not necessary here.}.

\subsection{Critical resonant line}

The  resonance condition encapsulated in the Dirac delta ${ \delta_{\rm D} (\bm{m}_{1} \!\cdot\! \bm{\Omega}_{1} \!-\! \bm{m}_{2} \!\cdot\! \bm{\Omega}_{2}) }$ generates an additional difficulty in the calculation of the Balescu-Lenard drift and diffusion coefficients from equations~\eqref{initial_drift} and~\eqref{initial_diff}. Recall the definition of the composition of a Dirac delta and a function~\citep{Hormander2003}, which in a ${d-}$dimensional setup takes the form
\begin{equation}
\int_{\mathbb{R}^{d}} \!\! \mathrm{d} \bm{x} \, f (\bm{x}) \, \delta_{\rm D} (g (\bm{x})) = \!\! \int_{g^{-1} (0)} \!\!\!\!\!\! \mathrm{d} \sigma (\bm{x}) \, \frac{f(\bm{x})}{|\nabla g (\bm{x})|} \, ,
\label{composition_Dirac}
\end{equation}
where ${ g^{-1} (0) \!=\! \{ \bm{x} \, | \, g (\bm{x}) \!=\! 0 \} }$ is the  hyper-surface of dimension (generically) ${ (d \!-\! 1) }$ defined by the constraint ${ g(\bm{x}) \!=\! 0 }$, and ${ \mathrm{d} \sigma (\bm{x}) }$ is the surface measure on ${ g^{-1} (0) }$. We have also defined ${ |\nabla g| }$ as the euclidean norm of the gradient of $g$, so that one has
\begin{equation}
| \nabla g  (\bm{x})| = \sqrt{ \left|\! \frac{\partial g}{\partial x_{1}} \!\right|^{2} \!\!+\! ... \!+\! \left|\! \frac{\partial g}{\partial x_{d}} \!\right|^{2} } \, .
\label{definition_nabla_g}
\end{equation}
Here we have assumed that the resonance condition associated with the function ${ g(\bm{J}_{2}) \!=\! \bm{m}_{1} \!\cdot\! \bm{\Omega}_{1} \!-\! \bm{m}_{2} \!\cdot\! \bm{\Omega}_{2} }$ is \textit{non-degenerate}, so that ${ \forall \bm{x} \!\in\! g^{-1} (0) \, , \,  | \nabla g (\bm{x}) | \!>\! 0  }$, which also ensures that the dimension of ${ g^{-1} (0) }$ is ${ ( d \!-\! 1 ) }$. One should note that this degeneracy condition is not satisfied by the Keplerian or harmonic potentials. Because we are considering an infinitely thin disc, the dimension of the physical space is given by ${ d \!=\! 2 }$, so that the set ${ g^{-1} (0) }$ will take the form of a curve $\gamma$, that we will call the critical resonant curve. Generically, it will take the form of an application of the type
\begin{equation}
\gamma \, : \, u \!\in\! [0 \, ;  1]  \mapsto \gamma (u) = (\gamma_{1} (u) , \gamma_{2} (u)) \, .
\label{definition_critical_line}
\end{equation}
One can then immediately rewrite the r.h.s of equation~\eqref{composition_Dirac} under the form
\begin{equation}
\int_{\gamma} \!\! \mathrm{d} \sigma (\bm{x}) \, \frac{f(\bm{x})}{|\nabla g (\bm{x})|} = \!\! \int_{0}^{1} \!\! \mathrm{d} u \, \frac{f (\gamma (u))}{|\nabla g (\gamma (u))|} |\gamma ' (u)| \, ,
\label{rhs_composition_delta}
\end{equation}
where we have naturally defined ${ | \gamma ' (u) | }$ as
\begin{equation}
|\gamma ' (u)| = \sqrt{ \left| \frac{\mathrm{d} \gamma_{1}}{\mathrm{d} u} \right|^{2} \!+\! \left| \frac{\mathrm{d} \gamma_{2}}{\mathrm{d} u} \right|^{2} } \, ,
\label{definition_gradient_gamma}
\end{equation}
Therefore, as soon as the critical resonant curve $\gamma$ has been identified, the integration from equation~\eqref{composition_Dirac} can be computed.

As  noted in equation~\eqref{link_rpra_EL}, using the pericentres and apocentres ${ (r_{p} , r_{a}) }$, given the Jacobian from equation~\eqref{Jacobian_J_EL} and proceeding in the same way as in equation~\eqref{response_M_rpra} for the response matrix, one may rewrite the drift and diffusion coefficients from equations~\eqref{initial_drift} and~\eqref{initial_diff} under the form
\begin{equation}
A_{\bm{m}_{1}} (\bm{J}_{1}) = \!\! \sum_{\bm{m}_{2}} \!\! \int \!\! \mathrm{d} r_{p} \mathrm{d} r_{a} \, \delta_{\rm D} (\bm{m}_{1} \!\cdot\! \bm{\Omega}_{1} \!-\! \bm{m}_{2} \!\cdot\! \bm{\Omega}_{2}) \, G_{\bm{m}_{1} , \bm{m}_{2}}^{A} (r_{p} , r_{a}) \, ,
\label{rewriting_drift}
\end{equation}
and
\begin{equation}
D_{\bm{m}_{1}} (\bm{J}_{1}) = \!\! \sum_{\bm{m}_{2}} \!\! \int \!\! \mathrm{d} r_{p} \mathrm{d} r_{a} \, \delta_{\rm D} (\bm{m}_{1} \!\cdot\! \bm{\Omega}_{1} \!-\! \bm{m}_{2} \!\cdot\! \bm{\Omega}_{2}) \, G_{\bm{m}_{1} , \bm{m}_{2}}^{D} (r_{p} , r_{a}) \, .
\label{rewriting_diff}
\end{equation}
where the functions $G_{\bm{m}_{1} , \bm{m}_{2}}^{\rm A}$ and $G_{\bm{m}_{1} , \bm{m}_{2}}^{\rm D}$ are respectively defined as
\begin{equation}
G_{\bm{m}_{1} , \bm{m}_{2}}^{\rm A} (r_{p} , r_{a}) =   -  \frac{1}{\Omega_{1}} \left| \frac{\partial (E,L)}{\partial (r_{p} , r_{a})} \right|
 \frac{4 \pi^{3} \, \mu \,\bm{m}_{2} \!\cdot\! \partial F / \partial \bm{J}_{2}}{| \mathcal{D}_{\bm{m}_{1} , \bm{m}_{2}} (\bm{J}_{1} , \bm{J}_{2} , \bm{m}_{1} \!\cdot\! \bm{\Omega}_{1}) |^{2}} \, ,
\label{definition_G_A}
\end{equation}
and 
\begin{equation}
G_{\bm{m}_{1} , \bm{m}_{2}}^{\rm D} (r_{p} , r_{a}) = \frac{1}{\Omega_{1}} \left| \frac{\partial (E , L)}{\partial (r_{p} , r_{a})} \right|  \frac{ \, 4 \pi^{3} \, \mu \,  F (\bm{J}_{2})}{| \mathcal{D}_{\bm{m}_{1} , \bm{m}_{2}} (\bm{J}_{1} , \bm{J}_{2} , \bm{m}_{1} \!\cdot\! \bm{\Omega}_{1}) |^{2}} \, .
\label{definition_G_D}
\end{equation}
For a given value of $\bm{J}_{1}$, $\bm{m}_{1}$ and $\bm{m}_{2}$, and introducing ${ \omega_{1} \!=\! \bm{m}_{1} \!\cdot\! \bm{\Omega}_{1} }$, we define the critical curve ${ \gamma_{\bm{m}_{2}} (\omega_{1}) }$ as
\begin{equation}
\gamma_{\bm{m}_{2}} (\omega_{1}) = \bigg\{ (r_{p} , r_{a}) \, \big| \, \bm{m}_{2} \!\cdot\! \bm{\Omega} (r_{p} , r_{a}) \!=\! \omega_{1}  \bigg\} \, .
\label{definition_gamma_m2}
\end{equation}
The expressions~\eqref{rewriting_drift} and~\eqref{rewriting_diff} of the drift and diffusion coefficients immediately become
\begin{equation}
\begin{cases}
\displaystyle A_{\bm{m}_{1}} (\bm{J}_{1}) = \!\! \sum_{\bm{m}_{2}} \!\! \int_{\gamma_{\bm{m}_{2}} (\omega_{1})} \!\!\!\!\!\!\!\! \mathrm{d} \sigma \, \frac{G_{\bm{m}_{1} , \bm{m}_{2}}^{\rm A}}{| \nabla (\bm{m}_{2} \!\cdot\! \bm{\Omega}_{2}) |} \, ,
\\
\displaystyle D_{\bm{m}_{1}} (\bm{J}_{1}) = \!\! \sum_{\bm{m}_{2}} \!\! \int_{\gamma_{\bm{m}_{2}} (\omega_{1})} \!\!\!\!\!\!\!\! \mathrm{d} \sigma \, \frac{G_{\bm{m}_{1} , \bm{m}_{2}}^{\rm D}}{| \nabla (\bm{m}_{2} \!\cdot\! \bm{\Omega}_{2}) |} \, ,
\end{cases}
\label{rewriting_drift_diff_II}
\end{equation}
where the resonant contribution ${ | \nabla (\bm{m}_{2} \!\cdot\! \bm{\Omega}_{2}) | }$ is defined as
\begin{equation}
| \nabla (\bm{m}_{2} \!\cdot\! \bm{\Omega}_{2}) | = \sqrt{ \bigg[ \bm{m}_{2} \!\cdot\! \frac{\partial \bm{\Omega}_{2}}{\partial r_{p}} \bigg]^{2} \!+\! \bigg[ \bm{m}_{2} \!\cdot\! \frac{\partial \bm{\Omega}_{2}}{\partial r_{a}} \bigg]^{2} } \, .
\label{definition_nabla_Omega}
\end{equation}
The derivatives of the intrinsic frequencies with respect to $r_{p}$ and $r_{a}$ appearing in equation~\eqref{definition_nabla_Omega} will be computed as in equation~\eqref{finite_differences_b} using finite differences. Once the critical lines of resonances have been determined, the computation of the drift and diffusion coefficients from equation~\eqref{rewriting_drift_diff_II} is straightforward, so that the full secular diffusion flux ${ \bm{\mathcal{F}}_{\rm tot} }$ from equation~\eqref{definition_F_tot} may be determined.

\section{Predicting Balescu-Lenard flux divergences}
\label{sec:application}

We may now illustrate how the previous computations of the response matrix and the Balescu-Lenard drift and diffusion coefficients can be used to recover some results obtained in well-crafted numerical simulations of galactic discs.
Indeed, \cite{Sellwood2012} (hereafter S12), studied the long-term evolution of an isolated stable truncated Mestel disc~\citep{Mestel1963}. After letting the disc evolve for hundreds of dynamical times, S12 observed a secular diffusion of the disc DF in action space, through the spontaneous generation of transient spiral waves. The most striking result of this evolution is given in figure 7 of S12, which exhibits the late time formation of a resonant ridge in action space along a specific resonant direction. Such diffusion features observed in the late evolution of an isolated stable and discrete system are thought to be signatures of a secular evolution induced by finite${-N}$ effects, as described by the Balescu-Lenard formalism. Because the system is made of a finite number $N$ of pointwise particles, it undergoes (long range) \textit{resonant encounters} leading to an irreversible secular evolution. In order to investigate such a collisional evolution, paper I applied the WKB limit of the Balescu-Lenard formalism to S12 simulation. While most of the secular diffusion was qualitatively recovered, there remained a significant timescale discrepancy, since the typical timescale diffusion predicted by this approach was typically $10^{3}$ times too slow compared to the observations made in S12. 
The use of a non-local basis such as equation~\eqref{definition_psi_p} and the numerical computation of the response matrix from equation~\eqref{response_M_EL} allows to incorporate in the present paper
 these previously ignored contributions from the WKB approach. 
 In the upcoming sections, we therefore present briefly the disc considered by S12 and our determination of the secular diffusion flux predicted by the Balescu-Lenard formalism.

\subsection{Initial setup}
\label{sec:initialsetup}

\begin{figure}[!htbp]
\begin{center}
\epsfig{file=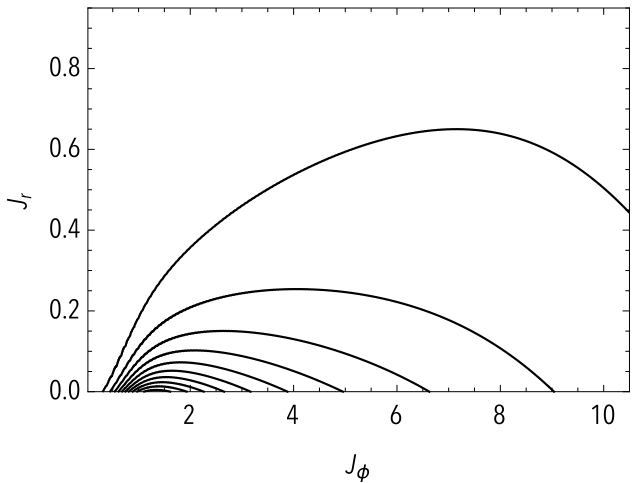,angle=-00,width=0.45\textwidth}
\caption{\small{Contours of the initial distribution function $F_{\rm star}$ from equation~\eqref{definition_Fstar}, in action space $(J_{\phi},J_{r})$. The contours are spaced linearly between 95\% and 5\% of the distribution function maximum.
}}
\label{figcontourDF}
\end{center}
\end{figure}

We consider the same disc as considered in~\cite{Sellwood2012}. It is an infinitely thin Mestel disc for which the circular speed $v_{\phi}$ is a constant $V_{0}$ independent of the radius. The stationary background potential $\psi_{\rm M}$ and its associated surface density $\Sigma_{\rm M}$ are given by
\begin{equation}
\psi_{\rm M} (R) = V_{0}^{2} \log \!\left[\! \frac{R}{R_{\rm max}} \!\right] \;\;\; ; \;\;\; \Sigma_{\rm M} (R) = \frac{V_{0}^{2}}{2 \pi G R} \, .
\label{psi_Mestel_Sigma_Mestel}
\end{equation}
where $R_{\rm max}$ is a scale parameter of the disc. Following~\cite{Toomre1977,BinneyTremaine2008}, a self-consistent DF for this system is given by
\begin{equation}
F_{\rm M} (E , J_{\phi}) = C_{\rm M} \, J_{\phi}^{q} \, \exp [ - E / \sigma_{r}^{2}] \, ,
\label{DF_Toomre}
\end{equation}
where the exponent $q$ is given by
\begin{equation}
q = \frac{V_{0}^{2}}{\sigma_{r}^{2}} \!-\! 1 \, ,
\label{definition_q}
\end{equation}
with $\sigma_{r}$ being the constant radial velocities spread within the disc. In equation~\eqref{DF_Toomre}, $C_{\rm M}$ is a normalization constant given by
\begin{equation}
C_{\rm M} = \frac{V_{0}^{2}}{2^{1 + q/2} \, \pi^{3/2} \, G \, \sigma_{r}^{q+2} \,  \Gamma \big[ \frac{1}{2} \!+\! \frac{q}{2} \big] \, R_{\rm max}^{q+1} } \, .
\label{definition_C_Toomre}
\end{equation}
In order to deal with the central singularity of the Mestel disc along with its infinite extent, we introduce two tapering functions
\begin{equation}
\begin{cases}
\displaystyle T_{\rm inner} (J_{\phi}) = \frac{J_{\phi}^{\nu_{\rm t}}}{(R_{\rm i} V_{0})^{\nu_{\rm t}} \!+\! J_{\phi}^{\nu_{\rm t}} } \, ,
\\
\displaystyle T_{\rm outer} (J_{\phi}) = \bigg[ 1 \!+\! \bigg[ \frac{J_{\phi}}{R_{0} V_{0}} \bigg]^{\mu_{\rm t}} \bigg]^{-1} \, ,
\end{cases}
\label{definition_tapering}
\end{equation}
where the indices $\nu_{\rm t}$ and $\mu_{\rm t}$ control the sharpness of the two tapers, while the radii $R_{\rm i}$ and $R_{0}$ are two scale parameters. These tapers $T_{\rm inner}$ and $T_{\rm outer}$ respectively represent the bulge and the outer truncation of the disc. In addition to these taperings, we also suppose that only a fraction $\xi$ of the stellar disc is self-gravitating, with ${ 0 \!\leq\! \xi \!\leq\! 1 }$, while the rest of the gravitational potential is provided by the static halo. As a consequence, the active distribution function $F_{\rm star}$ is given by
\begin{equation}
F_{\rm star} (E , J_{\phi}) = \xi \, F_{\rm M} (E , J_{\phi}) \, T_{\rm inner} (J_{\phi}) \, T_{\rm outer} (J_{\phi}) \, .
\label{definition_Fstar}
\end{equation}
We place ourselves in the same units system as in S12, so that we have ${ V_{0} \!=\! G \!=\! R_{\rm i} \!=\! 1 }$. The other numerical factors are given by ${ q \!=\! 11.4 }$, ${ \nu_{\rm t} \!=\! 4 }$, ${ \mu_{\rm t} \!=\! 5 }$, ${ \xi \!=\! 0.5 }$, ${ R_{0} \!=\! 11.5 }$ and ${ R_{\rm max} \!=\! 20 }$. The contours of the tapered DF $F_{\rm star}$ are illustrated in figure~\ref{figcontourDF}.
At this stage, it is important to note that S12 restricted the perturbations forces to the harmonic sector ${ m_{\phi} \!=\! 2 }$, so that we may consider the same restriction on the considered azimuthal number $m_{\phi}$. As a consequence, in the double resonance sum on $\bm{m}_{1}$ and $\bm{m}_{2}$ present in the Balescu-Lenard flux from equation~\eqref{definition_BL}, we will assume that $\bm{m}_{1}$ and $\bm{m}_{2}$ belong to the restricted set ${ \{ m_{r} , m_{\phi}\} \!\in\! \{ (-1,2),(0,2),(1,2) \} }$, where ${ (m_{r} , m_{\phi}) \!=\! (-1,2) }$ corresponds to the Inner Lindblad resonance (ILR), ${ (m_{r} , m_{\phi}) \!=\! (0,2) }$ to the Corotation resonance (COR), and ${ (m_{r} , m_{\phi}) \!=\! (1,2) }$ to the Outer Lindblad Resonance (OLR). All the upcoming calculations have also been performed while taking into account the contributions from the resonances associated with ${ m_{r} \!=\! \pm 2 }$, which were checked to be largely subdominant.

\subsection{Initial drift and diffusion}
\label{sec:initialD}

As detailed in equation~\eqref{response_M_sum}, the computation of the response matrix requires to consider a grid in the ${ (r_{p} , r_{a})-}$space. We considered a grid such that ${ r_{p}^{\rm min} \!=\! 0.08 }$, ${ r_{a}^{\rm max} \!=\! 4.92 }$ and ${ \Delta r \!=\! 0.05 }$. The sum on $m_{1}$ appearing in equation~\eqref{response_M_sum} was reduced to ${ | m_{1} | \!\leq\! m_{1}^{\rm max} \!=\! 7 }$. The basis considered was Kalnajs ${ 2D }$ basis~\citep{Kalnajs2} with the parameters ${ k_{\rm Ka} \!=\! 7 }$ and ${ R_{\rm Ka} \!=\! 5 }$. One should note that despite having a disc which extends up to ${ R_{\rm max} \!\!=\! 20 }$, one can still consider a basis truncated at such a small $R_{\rm Ka}$, so as to be able to efficiently capture the diffusion properties of the system in its inner regions, from where the secular diffusion is known to start. The radial basis elements were restricted to ${ 0 \!\leq\! n \!\leq\! 8 }$. When evaluating the response matrix, as in equation~\eqref{integration_subregion}, one has to add a small imaginary part $\eta$ to the frequency so as to regularize the resonant denominator. Throughout the calculations presented below, we considered ${ \eta \!=\! 10^{-4} }$
and  checked that this choice had no impact on our results.

\begin{figure}[!htbp]
\begin{center}
\epsfig{file=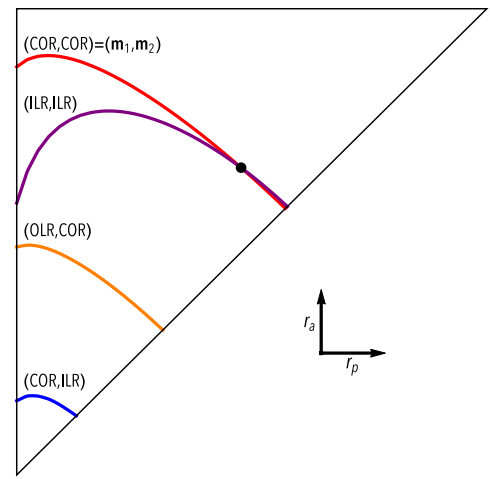,angle=-00,width=0.35\textwidth}
\caption{\small{Illustration of $4$ different critical resonant lines in the ${ (r_{p},r_{a})-}$space. As defined in equation~\eqref{definition_gamma_m2}, a critical line is characterized by the resonant vectors $\bm{m}_{1}$, $\bm{m}_{2}$ and a location ${ \bm{J}_{1} \!\leftrightarrow\! (r_{p}^{1} , r_{a}^{1}) }$ in action space. Each of the $4$ plotted critical lines are associated with the same location ${ (r_{p}^{1} , r_{a}^{1}) }$, represented by the black dot. The critical lines correspond to various choices of resonant vectors $\bm{m}_{1}$ and $\bm{m}_{2}$ among the three inner outer and corotation Lindblad resonances, respectively  ILR, OLR and COR. One should also note that for ${ \bm{m}_{1} \!=\! \bm{m}_{2} }$, the critical lines go through the point ${ (r_{p}^{1} , r_{a}^{1}) }$.
}}
\label{figCriticalLines}
\end{center}
\end{figure}

Since the total potential $\psi_{\rm M}$ is known via equation~\eqref{psi_Mestel_Sigma_Mestel}, the mapping to the angle-action coordinates is completely determined. The two intrinsic frequencies of the system can then be computed on the ${ (r_{p} , r_{a})-}$grid via equations~\eqref{definition_Omega1} and~\eqref{definition_Omega2}. Once these frequencies are known, the critical resonant lines introduced in equation~\eqref{definition_gamma_m2} can be determined and are illustrated in figure~\ref{figCriticalLines}. 
It is along these lines that one will have to perform the integration present in the definitions of the drift and diffusion coefficients from equation~\eqref{rewriting_drift_diff_II}.

\begin{figure}[!htbp]
\begin{center}
\epsfig{file=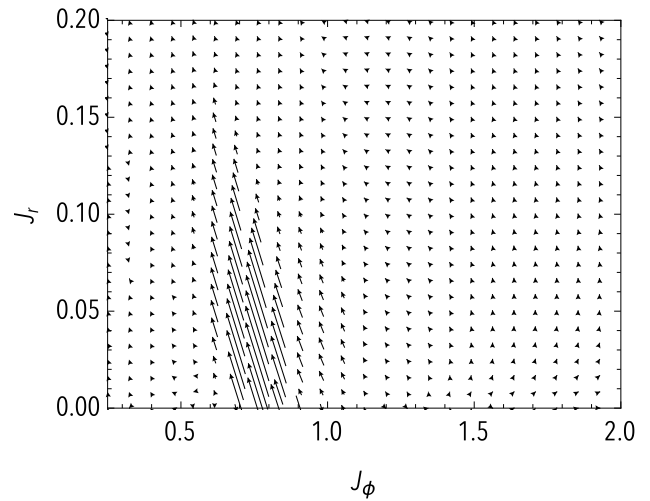,angle=-00,width=0.45\textwidth}
\caption{\small{Map of ${-  N \bm{\mathcal{F}}_{\rm tot} }$, where the flux has been computed with ${ \bm{m}_{1} , \bm{m}_{2} \!\in\! \left\{ \bm{m}_{\rm ILR}, \, \bm{m}_{\rm COR} , \, \bm{m}_{\rm OLR} \right\} }$. Following equation~\eqref{BL_div_Ftot}, ${ - N \bm{\mathcal{F}}_{\rm tot} }$ corresponds to the direction of diffusion of individual particles in action-space.
}}
\label{figPlotFlux}
\end{center}
\end{figure}

Thanks to this expression, one can then compute the secular diffusion flux ${ \bm{\mathcal{F}}_{\rm tot} }$ defined in equation~\eqref{BL_div_Ftot}. Because the mass of the particles is given by ${ \mu \!=\! M_{\rm tot} / N }$, it is natural to consider the quantity ${ N \bm{\mathcal{F}}_{\rm tot} }$ which is independent of $N$. The vector field ${ - N \bm{\mathcal{F}}_{\rm tot} \!=\! - (N \mathcal{F}_{J_{\phi}} , N \mathcal{F}_{J_{r}}) }$, which represents the direction of diffusion of individual particles, is illustrated in figure~\ref{figPlotFlux}.
One can already note in figure~\ref{figPlotFlux} that the diffusion vector field is along a narrow resonant direction. Along this ridge, one typically has ${ \mathcal{F}_{J_{\phi}} \!\simeq\! - 2 \mathcal{F}_{J_{r}} }$, so that the diffusion appears as aligned with the direction of the ILR resonance given by ${\bm{m}_{\rm ILR} \!=\! (2, -1)}$.
If one considers only the curl free part of this vector field, a sink and a source can be easily identified within that flow. 

\begin{figure*}[!htbp]
\centering
\begin{tabular}{@{}c@{\hskip 0.3in}c@{}}
{\epsfig{file=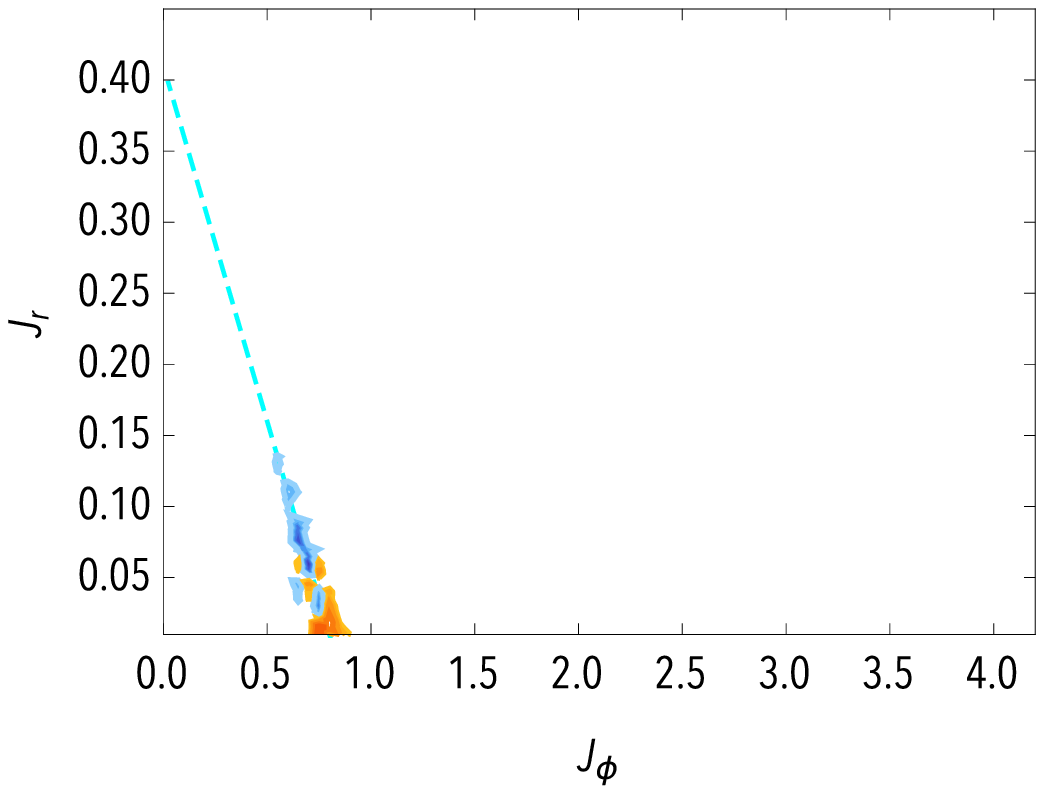,angle=-00,width=0.45\textwidth}}
&
{\epsfig{file=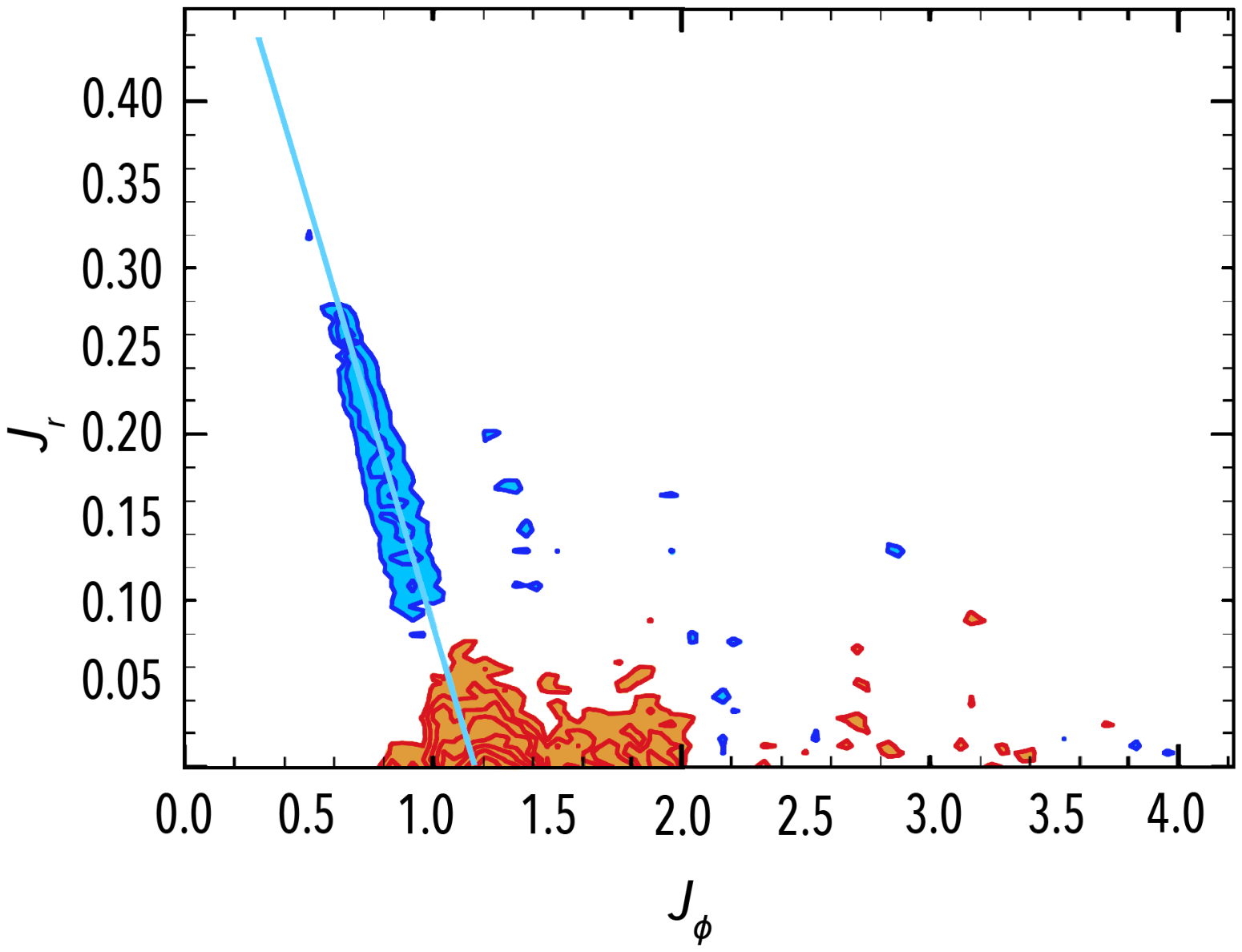,angle=-00,width=0.475\textwidth} }
\end{tabular}
\caption{\small{\textit{Left panel}: Map of ${ N \text{div} (\bm{\mathcal{F}}_{\rm tot}) }$, where the total flux has been computed with ${ \bm{m}_{1} , \, \bm{m}_{2} \!\in\! \left\{ \bm{m}_{\rm ILR}, \, \bm{m}_{\rm COR} , \, \bm{m}_{\rm OLR} \right\} }$. Red contours, for which ${ N \text{div} ( \bm{\mathcal{F}}_{\rm tot} ) \!<\! 0 }$ are associated with regions from which the orbits will be depleted, whereas blue contours, for which ${ N \text{div} (\bm{\mathcal{F}}_{\rm tot} ) \!>\! 0 }$ correspond to regions where the value of the DF will be increased during the secular diffusion. The contours are spaced linearly between the minimum and the maximum of ${ N \text{div} (\bm{\mathcal{F}}_{\rm tot}) }$. The maximum value for the positive blue contours corresponds to ${ N \text{div} (\bm{\mathcal{F}}_{\rm tot} ) \!\simeq\! 350 }$, while the mininum value for the negative red contours is associated with ${ N \text{div} (\bm{\mathcal{F}}_{\rm tot} ) \!\simeq\! -250 }$.
\textit{Right panel}: From~\cite{Sellwood2012} -- figure $7$, contours of the change in the DF between the time ${ t_{\rm S12} \!=\! 1400 }$ and ${ t_{\rm S12} \!=\! 0 }$, for a run with ${ 50M }$ particles. Similarly to the left panel, red contours correspond to negative differences, i.e. regions emptied from their orbits, while blue contours correspond to positive differences, i.e. regions where the DF has increased during the diffusion. Both of these contours are aligned with the ILR direction of ${ \bm{m}_{\rm ILR} \!=\! (2 , -1) }$ in the ${ (J_{\phi} , J_{r})-}$plane, corresponding to the cyan line.
}}
\label{figContoursDressed}
\end{figure*}

Once the diffusion flux ${ N \bm{\mathcal{F}}_{\rm tot} }$ has been determined, one can compute the divergence of this flux, so as to determine the regions for which the DF is expected to change during the secular diffusion. Figure~\ref{figContoursDressed} illustrates the contours of ${ N \text{div} (\bm{\mathcal{F}}_{\rm tot}) }$.
In figure~\ref{figContoursDressed}, we obtained that the Balescu-Lenard formalism indubitably predicts the formation of a narrow resonant ridge aligned with the ILR-direction, as was observed in S12 simulation.
 One also recovers that the stars which will populate the resonant ridge originate from the basis of the ridge and diffuse along the direction associated with the ILR resonance.
 It is most likely that the slight shift in the position of the ridge is due to the fact that the Balescu-Lenard prediction was carried at ${ t \!=\! 0^{+} }$, while S12's measurements are at ${ t \!=\! 1400 }$, so that we do not expect a perfect match.
Other sources of discrepancies might be the use of a softening length in numerical simulations, which modifies the two-body interaction potential, or the difference between an ensemble average (as predicted by the secular formalism) and one specific realization -- our own simulations suggest that there is some variation in the position of the ridge between one run and another.
Because we explicitly determined the value of ${ N \text{div} ( \bm{\mathcal{F}}_{\rm tot} ) }$, we may now study the typical timescale of collisional relaxation predicted by this Balescu-Lenard estimation as detailed in section~\ref{sec:timescale}. One may also investigate the respective roles of the self-gravitating amplification and the limitation to the tightly-wound basis elements as presented in Appendices~\ref{sec:noamplification} and~\ref{sec:backtoWKB}.

\subsection{Timescale of diffusion}
\label{sec:timescale}

The most significant disagreement found in paper I, while applying the WKB approximation of the Balescu-Lenard equation to S12's simulation was a timescale discrepancy between the time required to observe the resonant ridge in S12 simulation and the collisional timescale 
for which the finite${-N}$ effects come into play. As already noted in paper I, since the Balescu-Lenard equation~\eqref{definition_BL} only depends on $N$ through the mass of the individual particles ${ \mu \!=\! M_{\rm tot} / N }$, we may rewrite it under the form
\begin{equation}
\frac{\partial F}{\partial t} = \frac{1}{N} C_{\rm BL} [F] \, ,
\label{BL_with_N}
\end{equation}
where ${ C_{\rm BL} [F] \!=\! N \text{div} (\bm{\mathcal{F}}_{\rm tot}) }$ is the ${N-}$independent Balescu-Lenard collisional operator, i.e. the r.h.s of equation~\eqref{definition_BL} multiplied by ${ N \!=\! M_{\rm tot} / \mu }$. As expected, the larger the number of particles, the slower the secular evolution. This also illustrates the fact that the Balescu-Lenard equation comes from a kinetic Taylor expansion in the small parameter ${ \varepsilon \!=\! 1 /N \!\ll\! 1 }$. Introducing the rescaled time ${ \tau \!=\! {t}/{N} }$,
so that equation~\eqref{BL_with_N} reads
\begin{equation}
\frac{\partial F}{\partial \tau} = C_{\rm BL} [ F ] \, ,
\label{BL_with_tau}
\end{equation}
letting us express the Balescu-Lenard equation without any explicit appearance of $N$. In paper I, we estimated the time ${ \Delta \tau_{\rm S12} }$ required to observe the ridge as ${ \Delta \tau_{\rm S12} \!\simeq\! 3 \!\times\! 10^{-5} }$. When performing the same measurement thanks to the contours of the diffusion flux ${ \text{div} (\bm{\mathcal{F}}_{\rm tot}) }$ computed within the WKB approximation, we obtained ${ \Delta \tau_{\rm WKB} \!\simeq\! 3 \!\times\! 10^{-2} }$, so that paper I obtained the ratio ${ \Delta \tau_{\rm S12} / \Delta \tau_{\rm WKB} \!\simeq\! 10^{-3} }$. 
This discrepancy was due to the limitation to tightly wound spirals.
Because the estimation of the secular diffusion flux $\bm{\mathcal{F}}_{\rm tot}$ presented in figure~\ref{figContoursDressed} was made using the matrix method~\citep{Kalnajs2} with a full basis, it captures the additional swing-amplification. Indeed, given the map of ${ N \text{div} \bm{\mathcal{F}}_{\rm tot} }$ obtained in figure~\ref{figContoursDressed}, one may estimate the typical time ${ \Delta \tau_{\rm BL} }$ required for such a flux to lead to the diffusion features observed in S12. The contours presented in figure $7$ of S12 are separated by an increment equal to ${ 0.1 \!\times\! F_{0}^{\rm max} }$, where ${ F_{0}^{\rm max} \!\simeq\! 0.12 }$ is the maximum of the normalized DF (via equation~\eqref{definition_Fstar}). In order to observe the resonant ridge, the value of the DF should typically change by an amount of the order ${ \Delta F_{0} \!\simeq\! 0.1 \!\times\! F_{0}^{\rm max} }$. From figure~\ref{figContoursDressed}, one can note that the maximum of the divergence of the diffusion flux is given by ${ |N \text{div} ( \bm{\mathcal{F}}_{\rm tot} ) |_{\rm max} \!\simeq\! 350 }$. Thanks to equation~\eqref{BL_div_Ftot}, one can immediately write the relation ${ \Delta F_{0} \!\simeq\! \Delta \tau_{\rm BL} |N \text{div} (\bm{\mathcal{F}}_{\rm tot})|_{\rm max} }$, where ${ \Delta \tau_{\rm BL} }$ is the time during which the Balescu-Lenard equation has to be evolved in order to develop a ridge. With the previous numerical values, one obtains ${ \Delta \tau_{\rm BL} \!\simeq\! 3 \!\times\! 10^{-5} }$. Comparing the numerically measured time ${ \Delta \tau_{\rm S12} }$ and the time ${ \Delta \tau_{\rm BL} }$ predicted from the Balescu-Lenard equation, one obtains
\begin{equation}
\frac{\Delta \tau_{\rm S12}}{\Delta \tau_{\rm BL}} \sim 1 \, .
\label{ratio_Delta_tau}
\end{equation}
As expected, the projection of the response over an unbiased basis
leads to over a hundredfold increase of the susceptibility of the disc and therefore to a 
very significant acceleration of  secular diffusion. Thanks to this  mechanism, we  now find a very good  agreement between the diffusion timescales observed in numerical simulations and the predictions from the Balescu-Lenard formalism.
This quantitative match is rewarding, both from the point of view of the accuracy of the integrator (symplecticity, timestep size, softening...), and from the relevance of the successive approximations 
underpinning 
the Balescu-Lenard formalism (timescale decoupling, truncation of the BBGKY hierarchy, neglect of the close encounter term...).

In Appendix~\ref{sec:swing-test}, we show that 
when considering either figure~\ref{figContoursBare}, for which the self-gravity of the system has been turned off,
 or figure~\ref{figbasistruncation} for which the loosely wound basis elements were not taken into account, one does not recover a 
narrow resonant ridge appearing on timescales compatible with S12 simulations. Therefore, the main source of  secular collisional diffusion oberved in S12 and recovered in figure~\ref{figContoursDressed} has to be the strong self-gravitating amplification of loosely wound perturbations, i.e.
a sequence of uncorrelated swing-amplified spirals sourced by finite${-N}$ effects is indeed the main driver of secular diffusion.
The WKB formalism from paper I identified correctly the family of orbits involved, 
but fell short in predicting how narrow the resonant ridge is and how strongly amplified the response is.

\section{Comparison to ${N-}$body simulations}
\label{sec:NB}

In paper I, we relied on simulations presented in \cite{Sellwood2012} to compare the divergence of the diffusion flux to the WKB prediction.
In order to probe the expected scalings with the number of particles or with the active fraction of the disc, we now  resort  to our own ${N-}$body simulations.

\subsection{${N-}$body integration}

The initial sampling of particles is critical when investigating the origin of secular evolution, as one must ensure that the disc is initially in a state of equilibrium.
The sampling strategy we implemented  is described in some detail in Appendix~\ref{sec:sampling}.

Once sampled, we evolve the initial conditions using a straightforward
particle-mesh
${N-}$body code with a single-timestep leapfrog integrator~\citep[e.g.,][\S3.4.1]{BinneyTremaine2008}.
We follow S12 and split the potential in which the particles move into
two parts: (i) an axisymmetric contribution $\psi_{\rm M}$ from the unperturbed Mestel disc, as in equation~\eqref{psi_Mestel_Sigma_Mestel} 
and (ii) a non-axisymmetric contribution ${ \psi_1(R,\phi) }$ that
develops as perturbations grow in the disc.
This splitting avoids difficulties in the treatment of the rigid
component of the potential that is not included in the DF,
due to the tapering functions and active fraction introduced in equation~\eqref{definition_Fstar}.
We calculate $\psi_1$ using cloud-in-cell interpolation~\citep[e.g.,][\S2.9.3]{BinneyTremaine2008} of the particles' masses
onto an ${ N_{\rm mesh} \!\times\! N_{\rm mesh} }$ mesh of square cells spaced ${ \Delta x }$ apart, then
filtering the resulting density field ${ \rho(x,y) }$ to isolate
the disc response (see below), before applying the usual Fourier-space
\textit{doubling up} procedure to obtain the potential~$\psi_1$ at the cell
vertices.
The contribution of $\psi_1$ to each particle's acceleration is
then obtained using the same cloud-in-cell interpolation scheme.

When computing the density mesh, we added a filtering scheme, to include only the ${ m_{\phi} \!=\! 2 }$ disc response,
similarly to what was considered in S12.  We isolate this
${ m_{\phi} \!=\! 2 }$ mode by calculating
\begin{equation}
 \rho_2(r)=\frac1{2\pi} \!\! \int \!\! \mathrm d\phi \, \rho(r \cos (\phi),r \sin (\phi) ) \,  e^{-2 i  \phi} ,
\label{eq:stupidring}
\end{equation}
immediately after the cloud-in-cell assignment of mass to the mesh at
each timestep, then imposing the new mesh mass distribution
\begin{equation}
  \rho(x_k,y_k) = \rho_2(r_k) \, e^{2  i\phi_k},
\end{equation}
with $(r_k,\phi_k)$ chosen according to
${ (x_k,y_k) \!=\! (r_k \cos (\phi_k) ,r_k \sin (\phi_k )) }$.
To obtain ${ \rho_2(r) }$ we use brute-force computation
of equation~\eqref{eq:stupidring} on a serie of $N_{\rm ring}$ radial rings with spacing ${ \Delta
r \!\ll\! \Delta x }$, using the trapezium rule with ${ N_{\phi} \!=\! 720 }$ points in~$\phi$ for
the angular integrals.
These models are designed to reproduce as closely as possible the
essential details of S12's simulations.
There are a couple of deliberate technical differences: S12 uses a
polar mesh to obtain $\psi_1$, whereas we use a cartesian mesh with
a ${ m_{\phi} \!=\! 2 }$ prefiltering of the density field; S12 has a block
timestep scheme instead of our simpler single-timestep one.

For the results presented here we used a timestep ${ \Delta
t \!=\! 10^{-3}R_{\rm i} /V_{0} }$ on a mesh that extends to ${ \pm R_{\rm max} \!\!=\! 20 \, R_{\rm i} }$ with ${ N_{\rm mesh} \!=\! 120 }$
cells, so that ${ \Delta x \!=\! R_{\rm i} / 3 }$. The filtering of the potential perturbations to the harmonic sector ${ m_{\phi} \!=\! 2 }$, was performed with ${ N_{\rm ring} \!=\! 1000 }$ radial rings, so that ${ \Delta r \!=\! 2 \, R_{\rm i} / 100 }$, and ${ N_{\phi} \!=\! 720 }$ points in the azimuthal direction. Finally, the computation of the potential from the density via Fourier transform, was performed with a softening length ${ \varepsilon \!=\! R_{\rm i}/6 }$, which is comparable to the value used in~\cite{Sellwood2012}, which considered a Plummer softening with ${ \varepsilon \!=\! R_{\rm i} / 8 }$.
The results are not significantly changed when we
halve the timestep or the mesh size.
In Appendix~\ref{sec:MatrixOK}, we detail the validation of our ${N-}$body code, by recovering known unstable modes of truncated Mestel discs~\citep{Zang1976,EvansRead1998II,SellwoodEvans2001}.

\subsection{Scaling with $N$}
\label{sec:Nscaling}

In order to  rid our measurements  of individual fluctuations, we run multiple simulations for the same number of particles and perform an ensemble average of different evolution realizations for the same number of particles.  It allows us to estimate only the \textit{mean evolution}, which is effectively what is described by the Balescu-Lenard formalism.

In order to study the scaling with $N$ of these numerical simulations, one has to extract from the simulations a quantity on which to test this scaling and compare it with the predictions from the Balescu-Lenard formalism. The statistical nature of the initial sampling presents an additional  difficulty. Indeed, because one only samples $N$ stars as described in Appendix~\ref{sec:sampling}, the initial effective DF fluctuates around the smooth background DF from equation~\eqref{definition_Fstar} as a  Poisson shot noise. These statistical fluctuations  originate from the \textit{initial} sampling and are not as such specific to the physical process captured by the Balescu-Lenard formalism, so that one should   carefully  disentangle  these two contributions. Hence we introduce the function ${ \tilde{h} (t , N) }$ defined as
\begin{equation}
\tilde{h} (t , N) = \left\langle h_{i} (t , N) \right\rangle \, ,
\label{definition_h_N}
\end{equation}
where the operator ${ \left\langle \, \cdot \,  \right\rangle }$ corresponds to the ensemble average, approximated here with the arithmetic average over the ${ p \!=\! 32 }$ different realizations of simulations for the same number of particles, indexed by $i$:  ${ \left\langle \, \cdot \,  \right\rangle \!=\! 1/p \!\sum_i  \left(\, \cdot \,\right) }$.
In equation~\eqref{definition_h_N} the function ${ h_{i} (t, N) }$ is a lag function which read
\begin{equation}
h_{i} (t ,N) = \!\! \int \!\! \mathrm{d} \bm{J} \, \left[ F_{i} (t ,\bm{J} , N) \!-\!  \left\langle F (t \!=\! 0 , \bm{J} , N) \right\rangle\right]^{2} \, ,
\label{definition_hi_N}
\end{equation}
where we defined as ${ F_{i} (t , \bm{J} , N) }$ the normalized DF of the $i^{\rm th}$ realization for a number $N$ of particles. Such a quantity intends to quantify the \textit{distance} between the initial mean DF ${ \left\langle F (t \!=\! 0)\right\rangle }$ and the evolved DF ${ F_{i} }$. We are interested in the early time behavior of the lag function $h$ from equation~\eqref{definition_h_N}, so that we may perform its Taylor expansion
\begin{equation}
\tilde{h} (t ,N) \simeq \tilde{h}_{0} (N) + \tilde{h}_{1} (N) \, t + \tilde{h}_{2} (N) \, \frac{t^{2}}{2} \, ,
\label{DL_h}
\end{equation}
where it is important to note that the coefficients $\tilde{h}_{0}$, $\tilde{h}_{1}$ and $\tilde{h}_{2}$ depend only on $N$ and are given by
\begin{equation}
\tilde{h}_{0} (N) = \tilde{h} (t \!=\! 0 , N) \; ; \; \tilde{h}_{1} (N) = \frac{\partial \tilde{h} }{\partial t} \bigg|_{t = 0} \; ; \; \tilde{h}_{2} (N) = \frac{\partial^2 \tilde{h}}{\partial t^2} \bigg|_{t = 0} \, .
\label{Taylor_h}
\end{equation}
Let us now  estimate each of these coefficients in turn. Thanks to equation~\eqref{definition_hi_N}, one can compute ${ \tilde{h}_{0} (N) }$ which reads
\begin{equation}
\tilde{h}_{0} (N) = \!\! \int \!\! \mathrm{d} \bm{J} \, \left\langle \left[ F \!-\! \left\langle F_{0} \right\rangle \right]^{2} \right\rangle \, ,
\label{calculation_h0}
\end{equation}
where we used the shortened notations ${ \left\langle F_{0} \right\rangle \!=\! \left\langle F (t \!=\! 0 , \bm{J} , N) \right\rangle }$ and ${ F \!=\! F (t \!=\! 0 , \bm{J} , N) }$. We note that this coefficient only depends on the  properties of the initial sampling, and not on its dynamics. Because discrete sampling obeys Poisson statistics, one can write
\begin{equation}
\tilde{h}_{0} (N) = \frac{\alpha_{0}}{N} \, ,
\label{behavior_h0}
\end{equation}
where ${ \alpha_{0} }$ is a constant independent of $N$. One may then compute ${ \tilde{h}_{1} (N) }$, which takes the form
\begin{equation}
\tilde{h}_{1} (N) = 2 \!\! \int \!\! \mathrm{d} \bm{J} \, \left\langle \left[ F \!-\! \left\langle F_{0} \right\rangle \right] F' \right\rangle \, ,
\label{calculation_h1}
\end{equation}
where we used the shortened notation ${ F ' \!=\! [ \partial F / \partial t ] (t \!=\! 0) }$. One should note that the terms appearing in equation~\eqref{calculation_h1} have two different physical contents. Indeed, the term ${ \left[ F \!-\! \left\langle F_{0} \right\rangle \right] }$ involves initial sampling, whereas ${ F' }$ is driven by the dynamics of the system. If we assume that the sampling and the system's dynamics are uncorrelated, one writes
\begin{equation}
\left\langle \left[ F \!-\! \left\langle F_{0} \right\rangle \right]  F'  \right\rangle
= \left\langle F \!-\! \left\langle F_{0} \right\rangle \right\rangle  \left\langle F' \right\rangle
= 0 \, .
\label{decorrelation_h1}
\end{equation}
As a consequence, one immediately obtains from equation~\eqref{calculation_h1} that ${ \tilde{h}_{1} (N) \!=\! 0 }$.
One can finally compute the coefficient ${ \tilde{h}_{2} (N) }$ which reads
\begin{equation}
\tilde{h}_{2} (N) = 2 \!\! \int \!\! \mathrm{d} \bm{J} \, \left\langle \big[ F' \big]^{2} \!+\! \big[ F \!-\! \left\langle F_{0} \right\rangle \!\big] \, F '' \right\rangle \, ,
\label{calculation_h2}
\end{equation}
where we used the shortened notation ${ F'' \!=\! [ \partial^{2} F / \partial t^{2} ] (t \!=\! 0) }$. Using the same argument as in equation~\eqref{decorrelation_h1}, we may get rid of the second term in the l.h.s of equation~\eqref{calculation_h2}. If we now also assume that the 
variance of ${ [F']^2 }$ is small compared to its expectation,
one can write ${ \big< [ F' ]^{2} \big>  \!=\! \left[ \left\langle F' \right\rangle \right]^{2} }$, 
so that 
equation~\eqref{calculation_h2} becomes
\begin{equation}
\tilde{h}_{2} (N) = 2 \!\! \int \!\! \mathrm{d} \bm{J} \, \left[ \left\langle F' \right\rangle \right]^{2} \, .
\label{calculation_h2_II}
\end{equation}
Now the dependence of the term ${  \left\langle F' \right\rangle }$ with $N$ follows from equation~\eqref{BL_with_N}, so that one can write
\begin{equation}
\tilde{h}_{2} (N) = \frac{\alpha_{2}}{N^{2}} \, ,
\label{behavior_h2}
\end{equation}
where $\alpha_{2}$ is an amplitude independent of $N$. 
This scaling 
is a prediction from the Balescu-Lenard formalism. If the secular evolution observed in S12 simulation was a \textit{Vlasov-only} evolution, i.e. a collisionless evolution, one would expect a scaling of $\tilde{h}_{2}$ such that ${ \partial \tilde{h}_{2} / \partial N \!=\! 0 }$.

\begin{figure}[!htbp]
\begin{center}
\epsfig{file=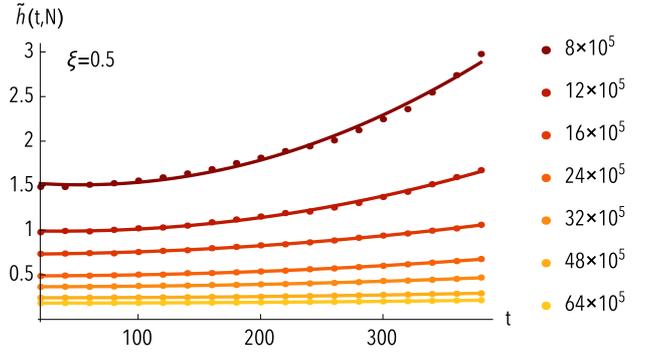,angle=-00,width=0.45\textwidth}
\caption{\small{Illustration of the behavior of the function ${ t \!\mapsto\! \tilde{h} (t, N) }$ defined in equation~\eqref{definition_h_N}, for an active fraction ${ \xi \!=\! 0.5 }$, when averaged on $32$ different realizations for particles numbers ${ N \!\in\! \{ \rm 8, \, 12, \, 16, \, 24, \, 32, \, 48, \, 64 \} \!\times\! 10^5 }$.  To compute ${ \tilde{h} (t,N) }$, we binned the action-space domain ${ (J_{\phi}, J_{r}) \!=\! [ 0 \, ; 2.5] \!\times\! [ 0 \, ; 0.2 ] }$ in ${ 100 \!\times\! 50 }$ regions. The values of ${\tilde{h} (t ,N)}$ have also been uniformly renormalized so as to clarify this representation. The dots corresponds to the snapshots of the simulations for which ${ \tilde{h} (t , N) }$ was computed, whereas the lines correspond to second-order fits. As expected, the smaller the number of particles, the \textit{noisier} the simulation and the larger ${ \tilde{h} (t,N) }$.
}}
\label{fig_scaling_run}
\end{center}
\end{figure}

\begin{figure}[!htbp]
\centering
\begin{tabular}{@{}cc@{}}
{\epsfig{file=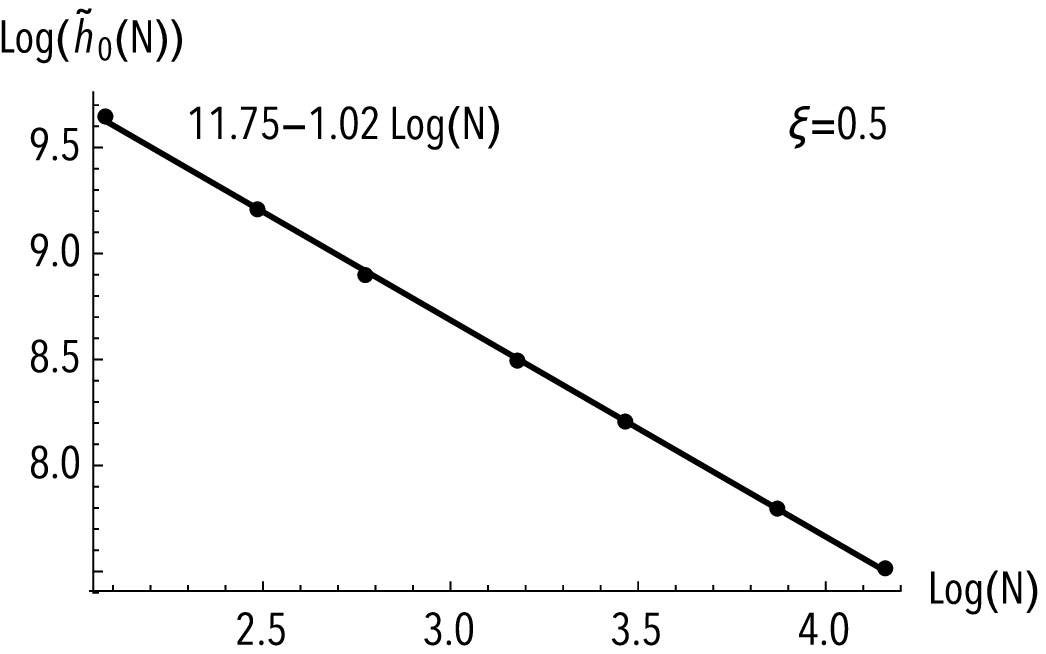,angle=-00,width=0.4\textwidth}} \\
{\epsfig{file=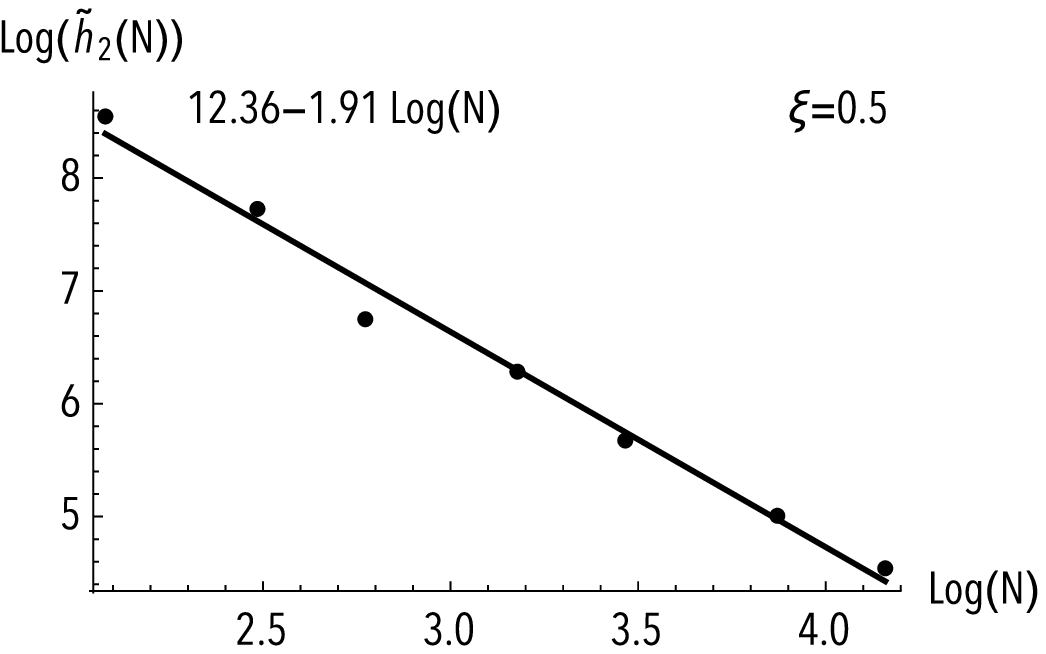,angle=-00,width=0.4\textwidth} }
\end{tabular}
\caption{\small{\textit{Top panel}: Illustration of the behavior of the function ${ \log (N) \!\mapsto\! \log(\tilde{h}_{0} (N)) }$, where $N$ has been rescaled by a factor ${ 10^{-5} }$ so as to simplify the representation. The dots correspond to computed values thanks to figure~\ref{fig_scaling_run}, while the line corresponds to a linear fit, which takes the form ${ \log (\tilde{h}_{0}(N)) \!\simeq\! 11.75 \!-\! 1.02 \log (N) }$. The coefficients ${ \tilde{h}_{0} (N) }$ have been uniformly renormalized so as to clarify this representation. \textit{Bottom panel}: Similar representation for the behavior of the function ${ \log(N) \!\mapsto\!  \log ( \tilde{h}_{2} (N) ) }$, whose linear fit takes the form ${ \log (\tilde{h}_{2} (N)) \!\simeq\! 12.36 \!-\! 1.91 \log(N) }$. Similarly, the coefficients ${ \tilde{h}_{2} (N) }$ have been uniformly renormalized so as to clarify this representation.
}}
\label{fig_scaling_h}
\end{figure}

One may now compare these predictions to the scalings obtained from ${N-}$body runs. We considered number of particles given by ${ N \!\in\! \{ \rm 8, \, 12, \, 16, \, 24, \, 32, \, 48, \, 64 \} } \!\times\! 10^5$, and for each of these values of $N$, we ran $32$ different simulations with different initial conditions while using the ${N-}$body code described in section~\ref{sec:NB}. For each value of N, one may study the function ${ t \!\mapsto\! \tilde{h} (t, N) }$, as illustrated in figure~\ref{fig_scaling_run}.
Once the behavior of the function ${ t \!\mapsto\! \tilde{h} (t,N) }$ is known, one can fit to these parabolas as in equation~\eqref{DL_h}, so as to determine the behavior of the functions $ N \!\mapsto\! \tilde{h}_{0} (N)$ and $ \tilde{h}_{2} (N) $. The dependence with $N$ of these coefficients is illustrated in figure~\ref{fig_scaling_h}.
From the top panel of figure~\ref{fig_scaling_h},  we recover the scaling of ${ \tilde{h}_{0} (N) }$ derived in equation~\eqref{behavior_h0} due to Poisson shot noise. The bottom panel of figure~\ref{fig_scaling_h}  displays the scaling ${ \tilde{h}_{2} (N) \!\propto\! N^{-1.91} }$. Given the finite number of simulations considered and the  uncertainties in the fits, this is in good agreement with the result presented in equation~\eqref{behavior_h2}.
This scaling  of ${ \tilde{h}_{2} (N) }$ with $N$ therefore confirms  the relevance of the Balescu-Lenard formalism in describing the secular evolution of S12 stable Mestel disc. Specifically, as explained below equation~\eqref{behavior_h2}, if the features observed in S12 simulation had only been the result of a collisionless mechanism, one would not have observed such a scaling of ${ \tilde{h}_{2} (N) }$ with $N$. 
This scaling confirms that the secular evolution of S12 stable Mestel disc is the result of a collisional evolution seeded by the discrete nature of the system and the effect of amplified distant resonant encounters.
Another probe of the collisional scaling, which allows to get rid of Poisson shot noise as present in equation~\eqref{behavior_h0}, is described in Appendix~\ref{sec:tthold}.

\subsection{Scaling with $\xi$}
\label{sec:xiscaling}

Since the novelty of the  Balescu-Lenard formalism is to capture the effect of gravitational polarization,
we now further compare qualitatively the prediction from section~\ref{sec:application} with the results obtained from numerical simulations, by studying the impact of the active fraction $\xi$ of the disc on the observed properties of the secular diffusion. Indeed, as detailed in section~\ref{sec:initialsetup}, the disc considered in S12 had an active fraction of ${ \xi \!=\! 0.5 }$, so that only one half of the potential was due to the active component. If one increases the active fraction of the disc, one will increase the strength of the self-gravitating amplification, and therefore accelerate the secular evolution of the disc, while still remaining in a regime of collisional evolution. Therefore the scaling of $\tilde{h}_{2}$ with $N$ given by equation~\eqref{behavior_h2} will remain the same, but the prefactor ${ \alpha_{2} (\xi) }$ will increase because the secular evolution will be amplified via a more efficient polarization. The dependence of $\alpha_{2}$ with $\xi$  can be both measured from ${N-}$body simulations but also predicted using  the Balescu-Lenard formalism via the  calculations presented in section~\ref{sec:initialD}. 
Let us consider the same sets of simulations as in section~\ref{sec:Nscaling}, so that the number of particles were given by ${ N \!\in\! \left\{ 8, \, 12, \, 16, \, 24, \, 32, \, 48, \, 64 \right\} \!\times\! 10^{5} }$, and for each of these values of $N$, 32 different simulations with ${ \xi \!=\! 0.6 }$ were performed, in order to carry out  ensemble averages.

\begin{figure}[!htbp]
\centering
\begin{tabular}{@{}cc@{}}
{\epsfig{file=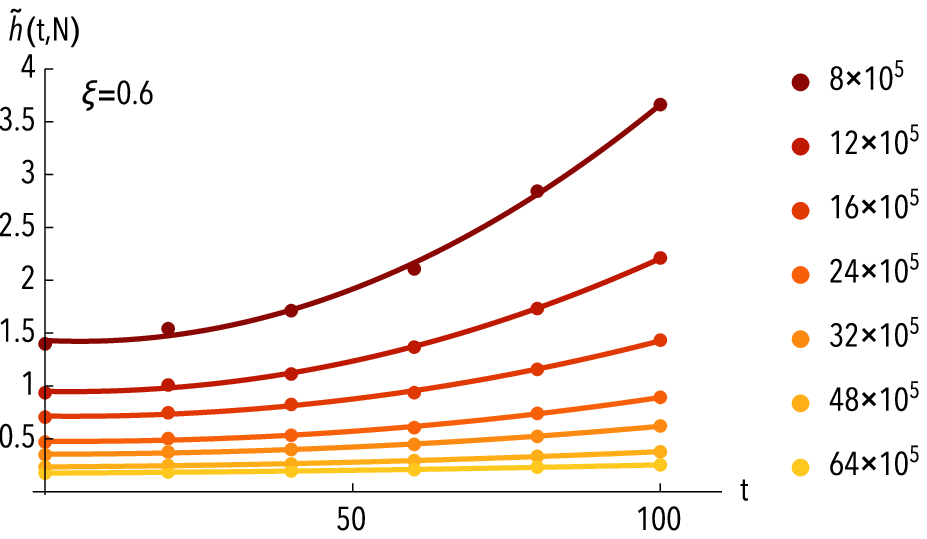,angle=-00,width=0.4\textwidth}} \\
{\epsfig{file=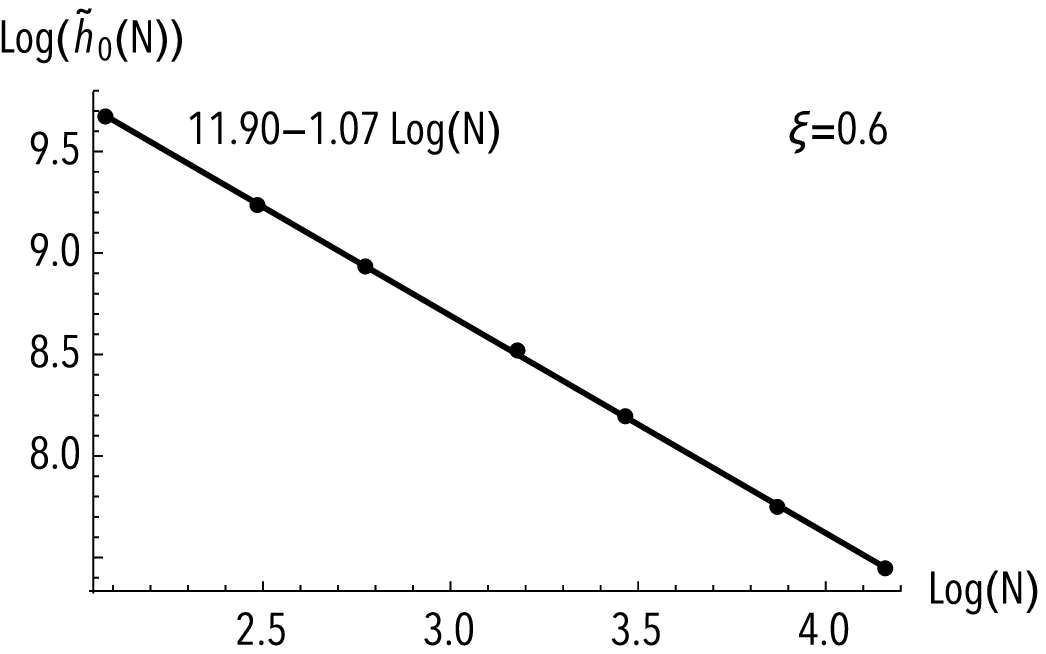,angle=-00,width=0.4\textwidth}} \\
{\epsfig{file=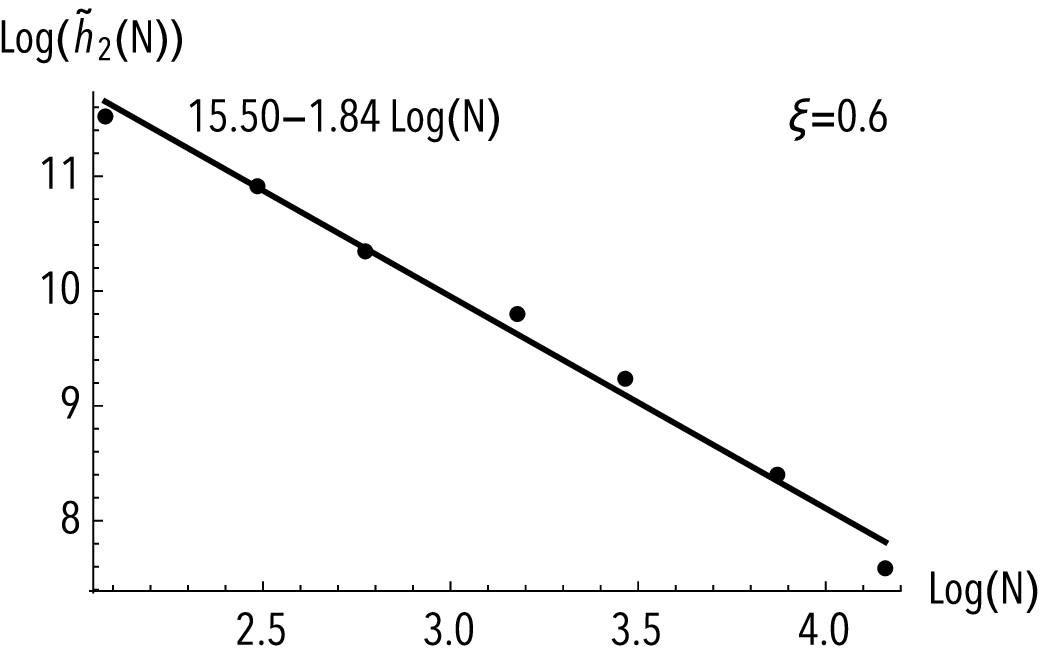,angle=-00,width=0.4\textwidth}}
\end{tabular}
\caption{\small{\textit{Top panel}: Illustration of the behavior of the function ${ t \!\mapsto\! \tilde{h} (t,N) }$ for an active fraction ${ \xi \!=\! 0.6 }$, using the same conventions as in figure~\ref{fig_scaling_run}. As expected, for an increased active fraction, the secular evolution of the system is fastened leading to a faster increase of the \textit{distance} ${ \tilde{h} (t,N) }$. \textit{Middle panel}: Behavior of the function ${ \log(N) \!\mapsto\! \log (\tilde{h_{0}} (N)) }$, for an active fraction ${ \xi \!=\! 0.6 }$, using the same conventions as in figure~\ref{fig_scaling_h}. Its linear fit takes the form ${ \log (\tilde{h}_{0} (N)) \!\simeq\! 11.90 \!-\! 1.07 \log(N) }$. One recovers the expected scaling of the Poisson shot noise sampling derived in equation~\eqref{behavior_h0}. \textit{Bottom panel}: Behavior of the function ${ \log(N) \!\mapsto\!  \log ( \tilde{h}_{2} (N) ) }$, for an active fraction ${ \xi \!=\! 0.6 }$, using the same conventions as in figure~\ref{fig_scaling_h}. Its linear fit takes the form ${ \log (\tilde{h}_{2} (N)) \!\simeq\! 15.48 \!-\! 1.84 \log(N) }$. One recovers the expected collisional scaling with $N$ obtained in equation~\eqref{behavior_h2}.
}}
\label{fig_scaling_xi_06}
\end{figure}

The equivalent of figures~\ref{fig_scaling_run} and~\ref{fig_scaling_h} for ${ \xi \!=\! 0.6 }$ is illustrated in figure~\ref{fig_scaling_xi_06}.
Even for ${ \xi \!=\! 0.6 }$, one finds that the function ${ t \!\mapsto\! \tilde{h}(t,N)}$ follows a parabola given by equation~\eqref{DL_h}. One also recovers the predicted scalings with $N$ of the functions ${ N \!\mapsto\! \tilde{h}_{0} (N) }$ and ${ N \!\mapsto\! \tilde{h}_{2} (N) }$, respectively representing the initial Poisson shot noise of the sampling and the collisional scaling of the Balescu-Lenard secular evolution. As expected, when the active fraction of the disc is increased the secular evolution is fastened. Thanks to these fits, one can study the dependence of the ratio ${ \alpha_{2} (\xi \!=\! 0.6) / \alpha_{2} (\xi \!=\! 0.5) }$, as defined in equation~\eqref{behavior_h2}, both from numerical simulations as described in figures~\ref{fig_scaling_h} and~\ref{fig_scaling_xi_06} and from the Balescu-Lenard equation using the matrix method described in section~\ref{sec:application}.

From the fits of ${ N \!\mapsto\! \tilde{h}_{2} (N) }$ from figures~\ref{fig_scaling_h} and~\ref{fig_scaling_xi_06}, one can write ${ \log (\tilde{h}_{2} (N)) \!\simeq\! 6.40 \!-\! 1.91 \, (\log(N) \!-\! 3.12) }$ for ${ \xi \!=\! 0.5 }$ and ${ \log(\tilde{h}_{2} (N)) \!\simeq\! 9.76 \!-\! 1.84 \, (\log(N) \!-\! 3.12) }$ for ${ \xi \!=\! 0.6 }$, where we shifted the intercept of the fits to correspond to the center of the considered region ${ \log(N) \!\in\! [\log(8) \,; \log(64)] }$. One therefore obtains the ratio
\begin{equation}
\frac{\alpha_{2} (0.6)}{\alpha_{2} (0.5)} \bigg|_{\rm NB} \!\!\!\! \simeq \exp \left[ 9.76 \!-\! 6.40 \right] \simeq 29 \, .
\label{ratio_alpha2_NB}
\end{equation}

\begin{figure}[!htbp]
\begin{center}
\epsfig{file=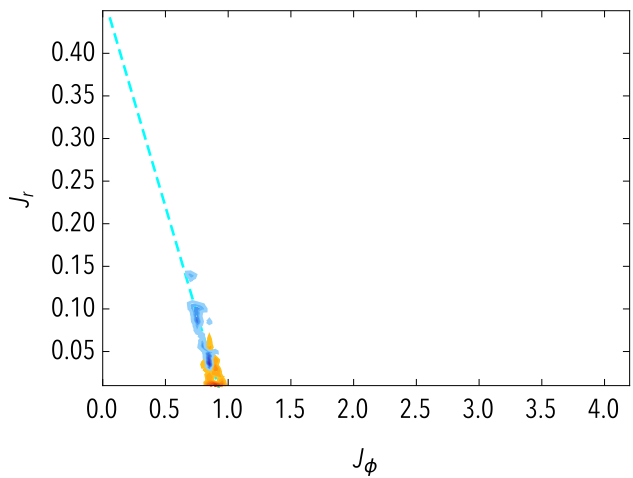,angle=-00,width=0.45\textwidth}
\caption{\small{Map of ${ N \text{div} (\bm{\mathcal{F}}_{\rm tot}) }$ with an active fraction ${ \xi \!=\! 0.6 }$ and using the same conventions as in figure~\ref{figContoursDressed}. The contours are spaced linearly between the minimum and the maximum of ${ N \text{div} (\bm{\mathcal{F}}_{\rm tot}) }$. The maximum value for the positive blue contours corresponds to ${ N \text{div} (\bm{\mathcal{F}}_{\rm tot}) \!\simeq\! 4200 }$, while the minimum value for the negative red contours corresponds to ${ N \text{div} (\bm{\mathcal{F}}_{\rm tot}) \!\simeq\! -3200 }$. As expected, one recovers that when the active fraction of the disc is increased, the susceptibility of the disc is increased, so that the norm of ${ N \text{div} (\bm{\mathcal{F}}_{\rm tot}) }$ gets larger and the secular diffusion is fastened. 
}}
\label{figContoursDressed_xi_06}
\end{center}
\end{figure}

One may now compare this ${N-}$body measurement to the same measurement performed via the Balescu-Lenard formalism.
Following equation~\eqref{calculation_h2_II}, one obtains that this ratio is given by
\begin{equation}
\frac{\alpha_{2} (\xi_{1})}{\alpha_{2} (\xi_{2})} = \frac{\displaystyle \int \!\! \mathrm{d} \bm{J} \, \big[ \text{div} (\bm{\mathcal{F}}_{\rm tot}^{\xi_{1}} ) \big]^{2} }{\displaystyle \int \!\! \mathrm{d} \bm{J} \, \big[ \text{div} (\bm{\mathcal{F}}_{\rm tot}^{\xi_{2}} ) \big]^{2}} \, ,
\label{ratio_alpha2_analytic}
\end{equation}
where $\bm{\mathcal{F}}_{\rm tot}^{\xi}$ stands for the secular diffusion flux at ${ t \!=\! 0^{+} }$ with an active fraction $\xi$. The value of ${ \alpha_{2} (\xi \!=\! 0.5) }$ can be determined via figure~\ref{figContoursDressed}, while the secular diffusion flux determined for ${ \xi \!=\! 0.6 }$ is illustrated in figure~\ref{figContoursDressed_xi_06}.
Thanks to the contours presented in figures~\ref{figContoursDressed} and~\ref{figContoursDressed_xi_06}, one can perform the same estimation as in equation~\eqref{ratio_alpha2_NB} starting from the Balescu-Lenard predictions. In order to focus on the contributions associated with the resonant ridge, the integrals on $\bm{J}$ in equation~\eqref{ratio_alpha2_analytic} were performed for ${J_{\phi} \!\in\! [ 0.5 \, ; \, 1.2 ]}$ and  ${ J_{r} \!\in\! [ 0.06 \, ; \, 0.15 ] }$. We measured
\begin{equation}
\frac{\alpha_{2} (0.6)}{\alpha_{2} (0.5)} \bigg|_{\rm BL} \!\!\!\! \simeq 42 \, .
\label{ratio_alpha2_Matrix}
\end{equation}
Despite the difficulty of this measurement which required to consider a much more sensitive disc with ${ \xi \!=\! 0.6 }$, the ratios of ${ \alpha_{2} }$ measured either via direct ${N-}$body simulations as in equation~\eqref{ratio_alpha2_NB} or via application of the Balescu-Lenard formalism in equation~\eqref{ratio_alpha2_Matrix} are within the same order of magnitude. As a consequence, one indeed checks that the Balescu-Lenard equation is able to correctly capture the relative effect of the disc susceptibility on the characteristics of the collisional secular diffusion. The strong consequence of modifying the active fraction, observed both in equations~\eqref{ratio_alpha2_NB} and~\eqref{ratio_alpha2_Matrix}, illustrates the relevance of the self-gravitating amplification in determining the typical timescale of secular diffusion of the system.

\subsection{Late-time evolution}

The predictions of the Balescu-Lenard secular diffusion flux presented in section~\ref{sec:application} were only applied for the initial time of evolution, i.e. for the estimation of ${ \bm{\mathcal{F}}_{\rm tot} (t \!=\! 0^{+}) }$. The ${N-}$body simulations presented in section~\ref{sec:NB} allowed us to verify the appropriate scaling of the response of the system with the number of particles for the initial time of evolution as illustrated in figure~\ref{fig_scaling_h}. Using the Balescu-Lenard formalism to probe the late secular evolution of the system would require to evolve iteratively equation~\eqref{definition_BL} over secular times. Such iterations are clearly beyond the scope of this first paper, however the use of ${N-}$body simulations allows us to start probing now such late times of evolution.

As discussed in section~\ref{sec:inhomogeneousBL}, the Balescu-Lenard equation describes the long-term evolution of a discrete self-gravitating inhomogeneous system. Such a collisional evolution is only relevant for stable systems, i.e. systems assumed to be stable in the  Vlasov sense. Because it has been obtained via a Taylor expansion of the dynamics at the order ${ 1/N }$ in the number of particles, it remains valid only for secular timescales of the order ${ N t_{D} }$, with $t_{D}$ the dynamical time.

On  such secular timescales, a Balescu-Lenard evolution can lead to two different outcomes. On the one hand, if the system remains stable during its entire evolution, the Balescu-Lenard equation will tend towards a ${1/N-}$stationary state\footnote{Boltzmann DF of the form ${ \exp [ - \beta \, H (\bm{J}) ] }$, when physically 
reachable, are obvious stationary states of the Balescu-Lenard equation.}. Once such a stationary state of evolution has been reached, the dynamics is then governed by the next order kinetic effects in ${ 1/N^{2} }$, which are not captured by the Balescu-Lenard equation.
On the other hand, the Balescu-Lenard collisional evolution may lead also to a destabilization of the system. Indeed, the long-term effects of the collisional diffusion, because they lead to an irreversible diffusion of the DF, may change its current state w.r.t. the collisionless (Vlasov) dynamics. After a slow and stable evolution sourced by collisional ${ 1/N }$ effects, the system may then become unstable with respect to collisionless dynamics, which becomes the main driver of its later-time evolution, as was suggested by \cite{Sellwood2012}. 
S12 observes an out-of-equilibrium  transition between the ${1/N}$ Balescu-Lenard collisional evolution and the collisionless Vlasov evolution. 

One can illustrate such a transition using the ${N-}$body simulations presented previously. In order to capture the change of \textit{regime} of evolution within the disc (collisional vs. collisionless), for a given value of the number of particles, we define the quantity ${ \Sigma_{2} (t , N) }$ as
\begin{equation}
\Sigma_{2} (t , N)
\!=\! \left\langle \int_{R_{\rm inf}}^{R_{\rm sup}} \!\!\!\!\!\!\!\! \mathrm{d} R \, R \, \mathrm{d} \phi \, \Sigma_{\rm star} (t, N, R , \phi) \, e^{- i 2 \phi}  \right\rangle
\!=\! \left\langle\mu\!\! \sum_{n} e^{- i 2 \phi_{n}} \right\rangle \, ,
\label{definition_Sigma2_BLtoVlasov}
\end{equation}
where as in equation~\eqref{definition_h_N}, the operator ${ \left\langle \, \cdot \,  \right\rangle }$ corresponds to the ensemble average, approximated here with the arithmetic average over the ${ p \!=\! 32 }$ different realizations of simulations for the same number of particles $N$. The radii considered are restricted to the range ${ R \!\in\! [ R_{\rm inf} \,;\, R_{\rm sup} ] \!=\! [1.2 \,;\, 5 ] }$, where the active surface density of the disc is little affected by the inner and outer tapers. Finally, to obtain the second equality in equation~\eqref{definition_Sigma2_BLtoVlasov}, as in equation~\eqref{Modes_star_surface}, we replaced the active surface density of the disc by a discrete sum over all the particles of the system, where the sum on $n$ is restricted to all the particle whose radius lies between $R_{\rm inf}$ and $R_{\rm sup}$, while their azimuthal phase was written as $\phi_{n}$. Such a quantity allows us to probe easily the presence of strong non-axisymmetric features within the disc.
\begin{figure}[!htbp]
\begin{center}
\epsfig{file=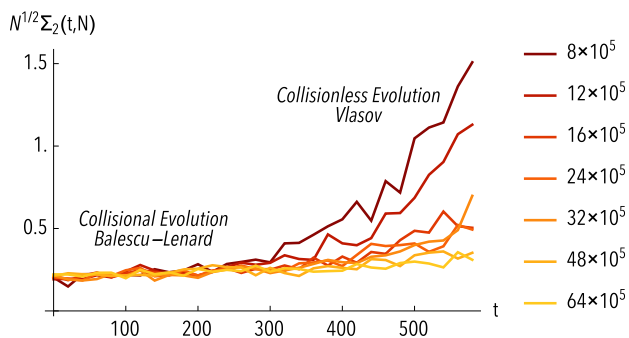,angle=-00,width=0.45\textwidth}
\caption{\small{Behavior of the function ${ t \!\mapsto\! \sqrt{N} \, \Sigma_{2} (t) }$ as defined in equation~\eqref{definition_Sigma2_BLtoVlasov}, for various values of the number of particles. The prefactor $\sqrt{N}$ has been added so as to \textit{mask} Poisson shot noise, allowing the initial values of ${ \sqrt{N} \, \Sigma_{2} }$ to be independent of $N$. It illustrates the bifurcation between the initial Balescu-Lenard collisional evolution, for which low values of $\Sigma_{2}$ are expected and the collisionless Vlasov evolution for which the system is no more axisymmetric leading to larger values of $\Sigma_{2}$. As expected, the larger the number of particles, the later the transition.
}}
\label{figBLtoVlasov}
\end{center}
\end{figure}
During the initial Balescu-Lenard collisional evolution of the system, one expects low values of $\Sigma_{2}$. Indeed, during this evolution, one relies on the phase averaging approximation, which assumes that ${ F \!=\! F (\bm{J} , t) }$, so that the DF of the system does not depend on the angles $\bm{\theta}$. During this collisional phase, $\Sigma_{2}$ still remains non-zero because the system develops  transient spiral waves, which  sustain  the secular evolution. On the long-term, this collisional evolution, through an irreversible diffusion of the DF, leads to a destabilization of the system. Eventually, the dynamical drivers of evolution are not any more discrete resonant collisionless effects but  exponentially growing dynamical instabilities. In this regime of collisionless unstable evolution, one expects much larger values of $\Sigma_{2}$, because of the appearance of strong non-axisymmetric bars within the disc. This bifurcation between these two regimes of diffusion is illustrated in figure~\ref{figBLtoVlasov}, through the behavior of the function ${ t \!\mapsto\! \Sigma_{2} (t,N) }$\footnote{A similar dynamical phase transition has been observed~\citep{CampaChavanis2008}  in a toy model of systems with long-range interactions called the Hamiltonian Mean Field (HMF) model. During the slow collisional evolution, because of finite${-N}$ effects, the distribution function of the system changes with time. In certain cases, the system may become dynamically (Vlasov) unstable and undergo a rapid phase transition from a homogeneous phase to an inhomogeneous phase. This phase transition can be monitored by the magnetization (see Fig. 1 of~\cite{CampaChavanis2008}) which is an order parameter playing a role similar to ${ \Sigma_{2} (t,N) }$.}. One can similarly observe this transition directly by looking at the active surface density ${ \Sigma_{\rm t} (R , \phi , t) }$ for these two different regimes. This is illustrated in figure~\ref{fig_Galaxy_Physical}, where one recovers that in the late time collisionless regime of evolution, the galaxy becomes strongly non-axisymmetric. 
S12 found that just after the disc becomes unstable, the pattern of the spiral response is consistent with the ILR frequency corresponding to the ridge\footnote{One could also check that the disc's distribution function corresponds at that stage to an unstable configuration, using the matrix method described in Appendix~\ref{sec:MatrixOK}.}.
\begin{figure}[!htbp]
\centering
\begin{tabular}{@{}cc@{}}
{\epsfig{file=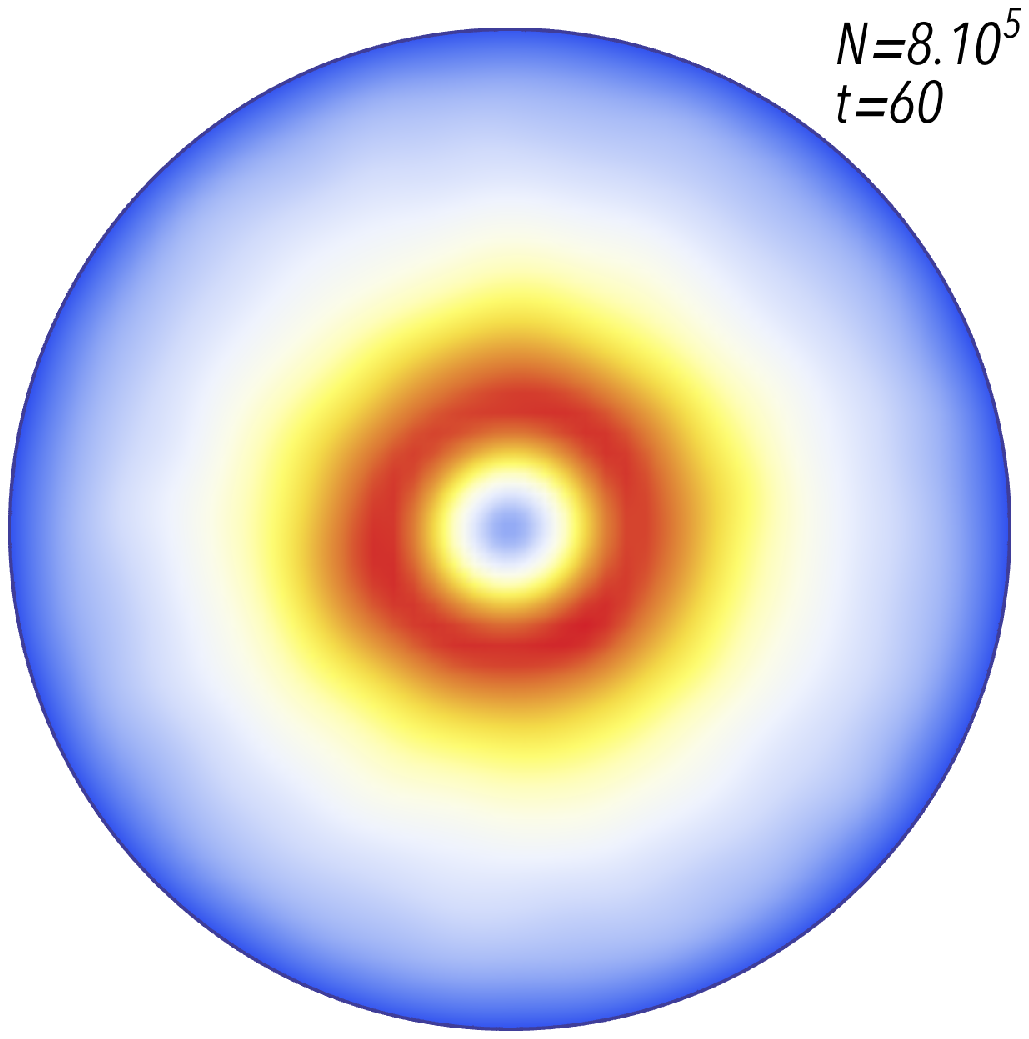,angle=-00,width=0.3\textwidth}} \\
{\epsfig{file=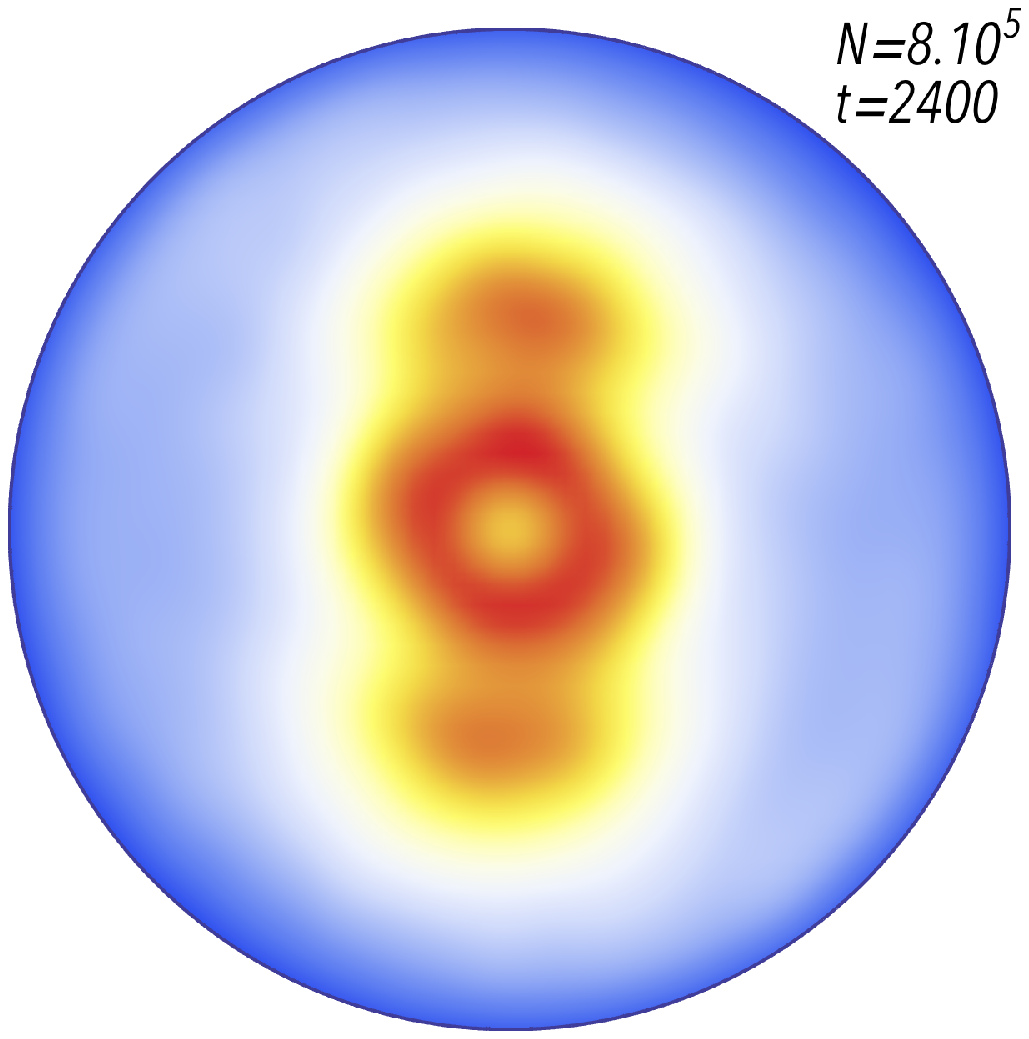,angle=-00,width=0.3\textwidth}}
\end{tabular}
\caption{\small{Illustration of the active surface density ${ \Sigma_{\rm t} }$ for a ${N-}$body run with ${ N \!=\! 8 \!\times\! 10^{5} }$, restriced to the range ${ R \!\leq\! 6 }$.
\textit{Top panel}: Active surface density $\Sigma_{\rm t}$ at an early time ${ t \!=\! 60 }$, for which the galaxy remains globally axisymmetric. In this regime, the dynamics of the system is collisional and governed by the Balescu-Lenard equation~\eqref{definition_BL}.
\textit{Bottom panel}: Active surface density $\Sigma_{\rm t}$ at a much later time ${ t \!=\! 2400 }$. The galaxy is then strongly non-axisymmetric. In this regime, the dynamics of the system is collisionless and governed by the Vlasov equation.
}}
\label{fig_Galaxy_Physical}
\end{figure}
Hence the phase transition observed in figure~\ref{fig_Galaxy_Physical} is driven by all the free energy available in a cold disc, which via spiral transients secularly heats the disc, but  \textit{only}
along a very tight resonant direction. This in turn  leads the disc towards an orbital instability,  transverse to the resonance
 \citep[via the direct azimuthal analogue to the two stream  instability in plasma physics,][]{Lynden-Bell1979,Pichon1994}. 
Qualitatively, one expects that the more massive and the narrower the ridge, the larger the number of  orbits trapped in ILR resonance
with little relative \textit{azimuthal dispersion}, 
and the earlier the instability
\citep{Penrose1960,Pichon1993}.

In closing, it is quite striking that an isolated galactic disc, fully stable in the mean field sense, will, given time, drive itself through two-point resonant correlations towards instability, demonstrating the extent to which such cold systems are truly secularly metastable.

\section{Conclusion}
\label{sec:conclusion}

Most astrophysical  discs  formed through dissipative processes  and
 have typically  evolved over many dynamical times.
Even in isolation, the  long range force of gravity allows their components  to  interact effectively through resonances, which given time may drive
them  secularly towards more likely equilibria. Such processes are captured by recent extensions of kinetic theories rewritten in angle-action variables
\citep{Heyvaerts2010,Chavanis2012}. Solving these equations provide astronomers with a unique opportunity to quantify the  induced  secular angular momentum redistribution within these discs~\citep{LyndenBell1972} over cosmic timescales.  
While challenging, the numerical computation of 
the corresponding diffusion and drift coefficients is as demonstrated within reach of a relatively straightforward extension of 
the so-called matrix method~\citep{Kalnajs2}, which computes the orbital response of self-gravitating discs using 
 quadratures and linear algebra.
 
Paper I presented asymptotic expressions in the tightly wound limit
and provided a qualitative insight into the physical processes at work during the secular diffusion of a self-gravitating discrete disc.  
Conversely, in this paper, we computed \textit{numerically} the drift and diffusion coefficients of the inhomogeneous Balescu-Lenard diffusion for such infinitely thin stellar discs.
The self-gravity of the disc was taken into account via the matrix method, validated on unstable Mestel discs.
We computed the divergence of the flux density in action space, ${ \text{div} ( \bm{\mathcal{F}}_{\rm tot} ) }$.
Swing amplification was shown to provide a significant boost for the diffusion timescale, which now matches the numerically measured one.
These computations are \textit{the first exact calculation} of the Balescu-Lenard diffusion and drift coefficients in the context of inhomogeneous multi-periodic systems.
They capture the essence of self-induced evolution (nature), which should 
compete with environmentally induced evolution (nurture).
We then compared these predictions to idealized numerical simulations of stable stationary and truncated Mestel discs sampled by pointwise particles, which were evolved for hundreds of dynamical times.
Using ensemble averages of our ${N-}$body runs,
we also identified a clear signature of the Balescu-Lenard process in the 
scaling of the diffusion features with $N$ and $\xi$, the fraction of the mass within the disc.
As originally identified by~\cite{GoldreichLyndenBell1965a,JulianToomre1966} in the context of their
linear response, the susceptibility of 
cold self-gravitating discs plays a critical role for their secular evolution as it is \textit{squared} in the Balescu-Lenard equation, 
which boosts considerably 
the effect of discreteness. Indeed, both the numerical experiments and our computation of the fluxes show that 
${ N_{\rm eff} \!\sim\! N/10^{4}}$, which is consistent with 
the predicted rescaling in ${ 1/ \mathcal{D}^{2} }$~\citep[it was shown forty years ago that for a Mestel disc ${ 1/ \mathcal{D} \!\sim\! 10^{2} }$, depending on the exact temperature of the disc,][]{Toomre1981}. 

Jointly with paper I we now have a qualitative and quantitative understanding of the 
initial secular orbital diffusion process induced by the discreteness of galactic discs.
Our qualitative understanding allows us to identify the role played by the square susceptibility 
in boosting the diffusion. 
Our quantitative agreement in both amplitude, position, width and scaling of the induced orbital signatures
strongly suggests that secular evolution is indeed driven by resonances as captured by the Balescu-Lenard formalism, 
and that it does \textit{not} depend on the 
initial phases of the disc (since the matching Balescu-Lenard fluxes are phase averaged).
It demonstrates that this equation initially reproduces the observed  evolution of self-gravitating discs driven by resonant two-point correlations
beyond the mean field approximation.
 
 The next step will be to evolve iteratively equation~\eqref{definition_BL}
 over a Hubble time, and compare with the result of ${N-}$body simulations.
 One should also model it jointly with an externally induced orbital diffusion~\citep{FouvryPichonPrunet2015} arising from e.g. 
a (possibly anisotropic) cosmic environment~\citep{Codis2012,Codis2015}
so as to assess which process dominates.
At the technical level, the   Balescu-Lenard formalism should be used to (in)validate ${N-}$body integrators accuracy over secular timescales.
There are indeed very few analytical predictions on which to calibrate ${N-}$body experiments  in this regime.
Such an exploration would also allow us to get a better grasp of the impact of the numerical parameters used in the ${N-}$body integration (such as timestep, mesh size or softening length) on the long-term dynamics of the system.

Beyond the application described in this paper,
the   Balescu-Lenard formalism may in the future also be numerically implemented to describe
 for instance   the secular diffusion of giant molecular clouds in galactic discs
(which in turn could play a role in migration-driven metallicity gradients and disc thickening),
 the secular migration of planetesimals in partially self-gravitating proto-planetary debris discs, 
 or even the long-term evolution of  population of stars and gas blobs  near the Galactic centre.
In 3D, assuming spherical symmetry, its implementation could be useful to describe spherical  systems dominated by 
radial orbits, or the secular evolution of tidal debris in our possibly flattened galactic halo using St\"ackel potentials.

\begin{acknowledgements}
JBF thanks the Institute of Astronomy, Cambridge, for hospitality
while this investigation was initiated. 
JBF and CP also thank the theoretical physics sub-department, Oxford, for hospitality and the CNRS-Oxford 
exchange program for funding. JBF, CP and PHC also thank the CNRS Inphyniti program for funding. CP thanks Clare and Churchill college, Cambridge,
the French embassy  and the community of ${\text{\url{http://mathematica.stackexchange.com}}}$ for their help. We thank Donald Lynden-Bell, James Binney, John Papaloizou, Walter Dehnen, Rebekka Bieri, Laura Monk, Gordon Ogilvie, Dmitry Pogosyan and
Simon Prunet for stimulating discussions, and Eric Pharabod for his help with Figure 4.
This work is partially supported by the Spin(e) grants ANR-13-BS05-0005 of the French \textit{Agence Nationale de la Recherche}
(${\text{\url{http://cosmicorigin.org}}}$)
and by the  LABEX Institut Lagrange de Paris (under reference ANR-10-LABX-63) which  is funded by  ANR-11-IDEX-0004-02.
\end{acknowledgements}

\bibliographystyle{aa}
\bibliography{references}

\appendix

\section{Kalnajs basis}
\label{sec:Kalnajsbasis}

We now detail the properties of the basis introduced in~\cite{Kalnajs2} to describe ${ 2D }$ discs\footnote{See also~\cite{Earn1995} for a similar rewriting of the basis normalizations.}. The basis will depend on two parameters: an index ${ k_{\rm Ka} \!\in\! \mathbb{N} }$ and a scale radius ${ r_{\rm Ka} \!\in\! \mathbb{R}^{+} }$. In all the upcoming formula of this section, in order to shorten the notations, we will write $r$ for the dimensionless quantity ${ r/r_{\rm Ka} }$. As introduced previously in equation~\eqref{definition_psi_p}, the basis elements will depend on two indices: the azimuthal number $\ell$ and the radial index $n$. One should note that we have ${ \ell \!\geq\! 0}$ and ${ n \!\geq\! 0 }$. The radial component of the potential elements are then of the form
\begin{equation}
\mathcal{U}_{n}^{\ell} (r) \!=\! - \frac{\sqrt{G}}{r_{\rm Ka}^{1\!/\!2}} \, \mathcal{P} (k_{\rm Ka} , \ell , n) \, r^{\ell} \!\sum_{i = 0}^{k} \!\sum_{j = 0}^{n} \!\alpha_{\rm Ka} (k_{\rm Ka} , \!\ell , \!n , \!i , \!j) \, r^{2 i + 2 j} \!.
\label{potential_2D_Kalnajs}
\end{equation}
The radial component of the density elements is given by
\begin{align}
\mathcal{D}_{n}^{\ell} (r) = & \frac{(-1)^{n}}{\sqrt{G} \, r_{\rm Ka}^{3\!/\!2}} \, \mathcal{S} (k_{\rm Ka} , \ell , n) \, (1 \!-\! r^{2})^{k_{\rm Ka} - 1/2} \, r^{\ell}   \nonumber
\\
& \times \sum_{j = 0}^{n} \beta_{\rm Ka} (k_{\rm Ka} , \ell , n , j) \, (1 \!-\! r^{2})^{j} \, .
\label{density_2D_Kalnajs}
\end{align}
In equations~\eqref{potential_2D_Kalnajs} and~\eqref{density_2D_Kalnajs}, the coefficients ${ \mathcal{P} (k , \ell , n) }$ and ${ \mathcal{S} (k , \ell , n) }$ are defined by
\begin{align}
\mathcal{P} (k ,\ell , n) \!=\! \Bigg\{\! & \frac{[2 k \!+\! \ell \!+\! 2 n \!+\! (1\!/2)] \Gamma [2 k \!+\! \ell \!+\! n \!+\! (1\!/2)]}{\Gamma [2 k \!+\! n \!+\! 1] \, \Gamma^{2} [\ell \!+\! 1] \, \Gamma[n \!+\! 1] } \nonumber
\\
& \times \Gamma[\ell \!+\! n \!+\! (1\!/2)]  \!\Bigg\}^{1\!/2} \! \, ,
\label{definition_P_Kalnajs}
\end{align}
and
\begin{align}
\mathcal{S} (k , & \ell , n) = \frac{\Gamma [k \!+\! 1]}{\pi \, \Gamma[2 k \!+\! 1] \, \Gamma [k \!+\! (1/2)]} \nonumber
\\
& \Bigg\{ \frac{[2k \!+\! \ell \!+\! 2 n \!+\! (1\!/2)] \, \Gamma [2 k \!+\! n \!+\! 1] \, \Gamma[2 k \!+\! \ell \!+\! n \!+\! (1\!/2)]}{\Gamma[\ell \!+\! n \!+\! (1/2)] \, \Gamma [n \!+\! 1]} \Bigg\}^{1/2} \!\! .
\label{definition_S_Kalnajs}
\end{align}
Finally, in equations~\eqref{potential_2D_Kalnajs} and~\eqref{density_2D_Kalnajs}, we have also introduced
\begin{align}
\alpha_{\rm Ka} (k , \ell , n , i , j) = & \frac{[-k]_{i} \, [\ell \!+\! (1\!/2)]_{i} \, [2 k \!+\! \ell \!+\! n \!+\! (1/2)]_{j}}{[\ell \!+\! 1]_{i} \, [1]_{i} \, [\ell \!+\! i \!+\! 1]_{j} [\ell \!+\! (1\!/2)]_{j} [1]_{j}} \nonumber
\\
& \times [i \!+\! \ell \!+\! (1\!/2)]_{j} \, [- n]_{j} \,  ,
\label{definition_alpha_Kalnajs}
\end{align}
and
\begin{equation}
\beta_{\rm Ka} (k , \ell , n , j) = \frac{[2 k \!+\! \ell \!+\! n \!+\! (1/2)]_{j} \, [k \!+\! 1]_{j} \, [- n]_{j}}{[2 k \!+\! 1]_{j} \, [k \!+\! (1/2)]_{j} \, [1]_{j} } \, .
\label{definition_beta_Kalnajs}
\end{equation}
In the two previous expressions, we introduced the rising Pochhammer symbol ${ [a]_{i} }$ defined as
\begin{equation}
[a]_{i} = 
\begin{cases}
\begin{aligned}
&1  & \text{if} \;\;\; i = 0 \, ,
\\
& a \, (a \!+\! 1) \, ... \, (a \!+\! n \!-\! 1) & \text{if} \;\;\; i > 0 \, .
\end{aligned}
\end{cases}
\label{definition_Pochhammer}
\end{equation}

\section{Calculation of $\aleph$}
\label{sec:aleph}

We now detail how the analytical function $\aleph$ from equation~\eqref{integration_subregion} may be computed. In order to ease the implementation of its computation, we rewrite $\aleph$ in an undimensionnalized way as follows
\begin{align}
\aleph (a_{g} , b_{g} , & c_{g} , a_{h} , b_{h} , c_{h} , \eta , \Delta r) \, = \nonumber
\\
& \!\! \int_{- \frac{\Delta r}{2}}^{\frac{\Delta r}{2}} \!\! \int_{- \frac{\Delta r}{2}}^{\frac{\Delta r}{2}} \!\! \mathrm{d} x_{p} \mathrm{d} x_{a} \, \frac{a_{g} \!+\! b_{g} x_{p} \!+\! c_{g} x_{a}}{a_{h} \!+\! b_{h} x_{p} \!+\! c_{h} x_{a} \!+\! i \eta} \nonumber
\\
& = \frac{a_{g}}{a_{h}} (\Delta r)^{2} \!\! \int_{- \frac{1}{2}}^{\frac{1}{2}} \!\! \int_{- \frac{1}{2}}^{\frac{1}{2}} \!\! \mathrm{d} x \mathrm{d} y \, \frac{1 \!+\! \frac{b_{g} \Delta r}{a_{g}} x \!+\! \frac{c_{g} \Delta r}{a_{g}} y }{1 \!+\! \frac{b_{h} \Delta r}{a_{h}} x \!+\! \frac{c_{h} \Delta r}{a_{h}} y \!+\! i \frac{\eta}{a_{h}} } \nonumber
\\
& = \frac{a_{g}}{a_{h}} (\Delta r)^{2} \, \aleph_{\rm D} \left[ \frac{b_{g} \Delta r}{a_{g}} , \frac{c_{g} \Delta r}{a_{g}} , \frac{b_{h} \Delta r}{a_{h}} , \frac{c_{h} \Delta r}{a_{h}} , \frac{\eta}{a_{h}} \right] \, ,
\label{calculation_aleph}
\end{align}
where we assumed that ${a_{g} , a_{h} \!\neq\! 0}$ and used the change of variables ${ x \!=\! x_{p} / \Delta r }$ and ${ y \!=\! x_{a} / \Delta r }$. We also defined the dimensionless function $\aleph_{\rm D}$ as
\begin{equation}
\aleph_{\rm D} (b, c, e, f, \eta) = \!\! \int_{- \frac{1}{2}}^{\frac{1}{2}} \!\! \int_{- \frac{1}{2}}^{\frac{1}{2}} \!\! \mathrm{d} x \mathrm{d} y \, \frac{1 \!+\! b x \!+\! c y}{1 \!+\! e x \!+\! f y \!+\! i \eta} \, .
\label{definition_aleph_D}
\end{equation}
To compute this integral, we may now exhibit a function ${ G (x , y) }$ such that
\begin{equation}
\frac{\partial^{2} G}{\partial x \partial y} = \frac{1 \!+\! b x \!+\! c y}{1 \!+\! e x \!+\! f y \!+\! i \eta} \, .
\label{property_G_aleph}
\end{equation}
One possible choice for $G$ is given by
\begin{align}
& G (x , y) =  \frac{1}{4 e^2 f^2} \log [ e^2 x^2 \!+\! 2 e (f x y\!+\!x)\!+\!f^2 y^2\!+\!2 f y\!+\!\eta ^2\!+\!1 ]  \nonumber
\\
& \bigg\{ b f  (e^2 x^2\!-\!(f y\!+\!i \eta \!+\!1)^2 ) \!+\!2 e f (e x\!+\!i \eta \!+\!1) \!-\!c e \nonumber
   (e x\!+\!i \eta \!+\!1)^2 \bigg\}
\\
 &  \!+\!\frac{i}{2 e^2 f^2} \bigg\{ \frac{\pi }{2}\!-\!\tan ^{-1} \!\bigg[ \frac{e x\!+\!f y\!+\!1}{\eta } \bigg] \bigg\} \nonumber
   \\
 &  \times \bigg\{ b f ( e^2 x^2\!-\!(f y\!+\!i \eta \!+\!1)^2 ) \!+\! 2 e f (e x\!+\!i \eta \!+\!1) \!-\!c e (e x\!+\!i \eta \!+\!1)^2 \bigg\} \nonumber
   \\
&    \!+\!\frac{y}{4 e^2 f} \bigg\{ f (\!-\!4 e\!+\!b (2 e x\!+\!f y\!+\!2 i \eta \!+\!2)) \nonumber
\\
& \!+\!c e (2 e x\!-\!f y\!+\!2 i \eta \!+\!2)\!+\!2 e f (c y\!+\!2) \log [e x\!+\!f y\!+\!i \eta \!+\!1] \bigg\}
\label{choice_G}
\end{align}
One should note in the previous expression the presence of a complex logarithm and a $\tan^{-1}$. However, because ${ e , f , \eta \!\in\! \mathbb{R} }$ and ${ \eta \!\neq\! 0 }$, one can easily show that the arguments of both of these functions never cross the usual branch-cut of these functions ${ \{ \text{Im}(z) \!=\! 0 \, ; \, \text{Re}(z) \!\leq\! 0 \} }$. As a consequence, the expression~\eqref{definition_aleph_D} can immediately be computed as
\begin{equation}
\aleph_{D} \!=\! G [ \tfrac{1}{2} , \tfrac{1}{2} ] \!-\! G [ \tfrac{1}{2} , - \tfrac{1}{2} ] \!-\! G [ - \tfrac{1}{2} , \tfrac{1}{2} ] \!+\! G [ - \tfrac{1}{2} , - \tfrac{1}{2} ] \, .
\label{computation_aleph_D}
\end{equation}

\section{Response Matrix and ${N-}$body validations}
\label{sec:MatrixOK}

\begin{figure}[!htbp]
\centering
\begin{tabular}{@{}cc@{}}
{\epsfig{file=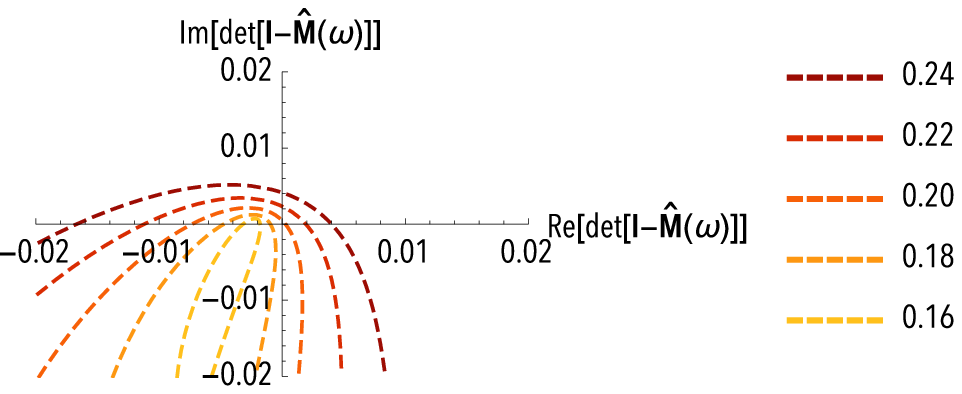,angle=-00,width=0.45\textwidth}} \\
{\epsfig{file=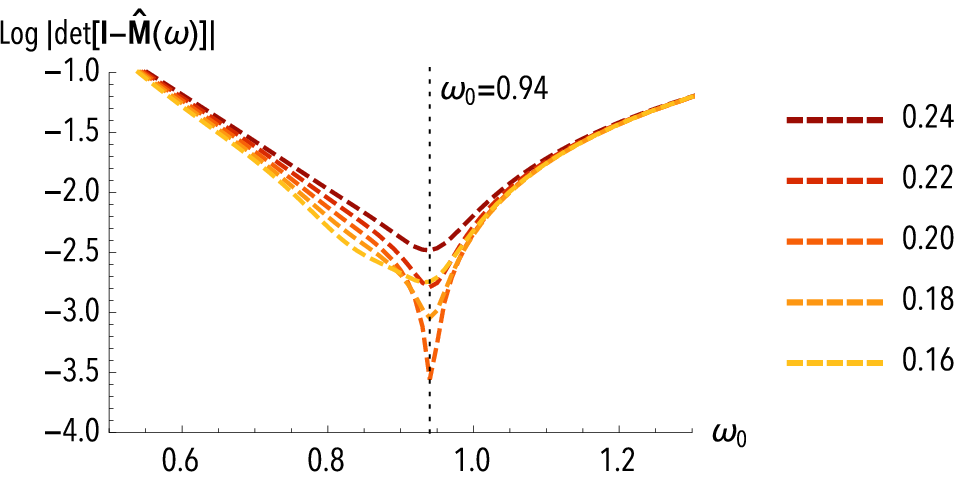,angle=-00,width=0.45\textwidth} }
\end{tabular}
\caption{\small{\textit{Top panel}: Zoomed Nyquist contours in the complex plane of ${ \omega_{0} \!\mapsto\! \det \big[ \mathbf{I} \!-\! \widehat{\mathbf{M}} (\omega_{0} , \eta) \big] }$ obtained via the matrix method for a truncated Mestel disc with ${ \nu_{\rm t} \!=\! 6 }$ and ${ q \!=\! 6 }$, looking for ${ m_{\phi} \!=\! 2 }$ modes. Each contour corresponds to a fixed value of $\eta$. For a growth rate of ${ \eta \!\simeq\! 0.20 }$, one can note that the contour crosses the origin, which corresponds to the presence of an unstable mode. \textit{Bottom panel}: Illustration of the function ${ \omega_{0} \!\mapsto\! \log  \big| \! \det \big[ \mathbf{I} \!-\! \widehat{\mathbf{M}}(\omega_{0},\eta) \big] \big| }$ for the same truncated Mestel disc. Each line corresponds to a fixed value of $\eta$. Such a representation allows to determine the pattern speed ${ \omega_{0} \!=\! m_{\phi} \Omega_{p} \!\simeq\! 0.94 }$ of the unstable mode.
}}
\label{figNyquist}
\end{figure}

The computation of the response matrix as described in section~\ref{sec:disccase} was validated by recovering the results of the pioneer work of~\cite{Zang1976}, extended in~\cite{EvansRead1998I,EvansRead1998II}, and recovered numerically in~\cite{SellwoodEvans2001}. These papers predicted the precession rate ${ \omega_{0} \!=\! m_{\phi} \Omega_{p} }$ and growth rate ${ \eta \!=\! s }$ of the unstable modes of a truncated Mestel disc similar to the stable one described in section~\ref{sec:initialsetup}. To build up an unstable disc similar to the ones considered in these previous works, one has to consider a fully active disc, so that ${ \xi \!=\! 1 }$. So as to have ${ Q \!=\! 1 }$~\citep{Toomre1964}, the velocity dispersion within the disc will be given by ${ q \!=\! 6 }$, where the parameter $q$ has been introduced in equation~\eqref{definition_q}. Finally, a last parameter one can tune in order to modify the properties of the disc is the truncation index of the inner tapering $\nu_{\rm t}$ defined in equation~\eqref{definition_tapering}. While looking only for ${ m_{\phi} \!=\! 2 }$ modes, we considered three different truncations indices given by ${ \nu_{\rm t} \!=\! 4, \, 6, \, 8 }$. To compute the response matrix, we used the same numerical parameters as described in section~\ref{sec:initialD}. Looking for unstable modes amounts to looking for complex frequencies ${ \omega \!=\! \omega_{0} \!+\! i \eta }$, such that the response matrix ${ \widehat{\mathbf{M}} (\omega_{0} , \eta) }$ from equation~\eqref{response_M_EL} possesses an eigenvalue equal to $1$. Such a complex frequency is then associated with an unstable mode of pattern speed $\omega_{0}$ and growth rate $\eta$. To determine the growth rate and pattern speed of the unstable modes, we relied on Nyquist contours similarly to the technique presented in~\cite{Pichon1997}. For a fixed value of $\eta$, one can study the continuous complex curve ${ \omega_{0} \!\mapsto\! \det \big[ \mathbf{I} \!-\! \widehat{\mathbf{M}} (\omega_{0} , \eta) \big] }$. Because for ${ \eta \!\to\! + \infty }$, one has ${ |\widehat{\mathbf{M}} (\omega_{0} , \eta)| \!\to\! 0 }$, the number of windings of this curve around the origin gives a lower bound on the number of unstable modes with a growth rate superior to ${\eta}$. By decreasing the value of $\eta$, one can then determine the largest value of $\eta$ admitting an unstable mode, and therefore the most unstable mode of the disc. The Nyquist contours obtained for the truncation index ${ \nu_{\rm t} \!=\! 6 }$ are illustrated in figure~\ref{figNyquist}, while the measurements are gathered in figure~\ref{figtabModes}.

\begin{figure}[!htbp]
\begin{center}
\epsfig{file=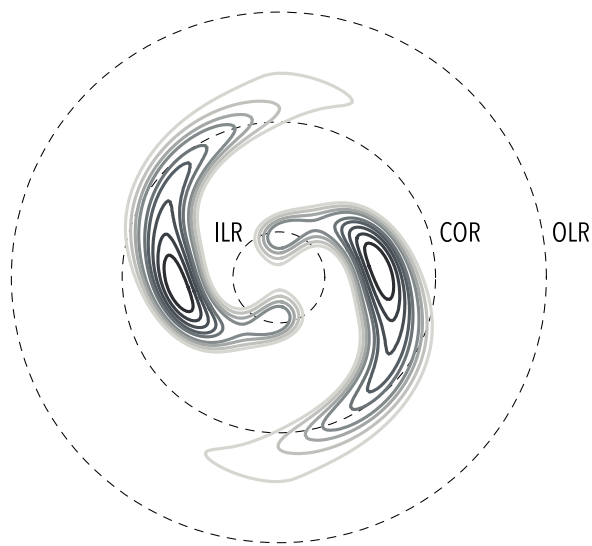,angle=-00,width=0.45\textwidth}
\caption{\small{Unstable mode for the truncated ${ \nu_{\rm t} \!=\! 4 }$ Mestel disc recovered via the matrix method presented in section~\ref{sec:disccase}. Only positive contour levels are shown and are spaced linearly between 10\% and 90\% of the maximum norm. The radii associated with the resonances ILR, COR and OLR have been represented, as given by ${ \omega_{0} \!=\! \bm{m} \!\cdot\! \bm{\Omega} (R_{\bm{m}}) }$, where the intrinsic frequencies ${ \bm{\Omega} (R) \!=\! (\Omega_{\phi} (R) , \, \kappa (R) ) }$ have to be computed within the epicyclic approximation. For a Mestel disc, they are given by ${ \Omega_{\phi} (R) \!=\! V_{0} / R }$ and ${ \kappa (R) \!=\! \sqrt{2} \, \Omega_{\phi} (R) }$.
}}
\label{figModeMestelMatrix}
\end{center}
\end{figure}

Once the characteristics ${ ( \omega_{0} , \eta ) }$ of the unstable modes have been determined, one can study in the physical space the shape of the mode. Indeed, for ${ \omega \!=\! \omega_{0} \!+\! i \eta }$, one can compute ${ \widehat{\mathbf{M}} (\omega_{0} , \eta) }$, and numerically diagonalize this matrix. One then considers its eigenvector ${ \bm{X}_{\rm mode} }$ (of size $n_{\rm max}$, where $n_{\rm max}$ is the number of basis elements considered) associated with the eigenvalue almost equal to $1$. The shape of the mode is then immediately given by
\begin{equation}
\Sigma_{\rm mode} (R , \phi) = \text{Re} \left[ \sum_{p} \bm{X}_{\rm mode}^{p} \, \Sigma^{(p)} (R , \phi) \right] \, , 
\label{shape_mode_Matrix_Method}
\end{equation}
where $\Sigma^{(p)}$ are the considered surface density basis elements.
The shape of the recovered unstable mode for the truncated ${ \nu_{\rm t} \!=\! 4 }$ Mestel disc is illustrated in figure~\ref{figModeMestelMatrix}.

\begin{figure}[!htbp]
\centering
\begin{tabular}{@{}cc@{}}
{\epsfig{file=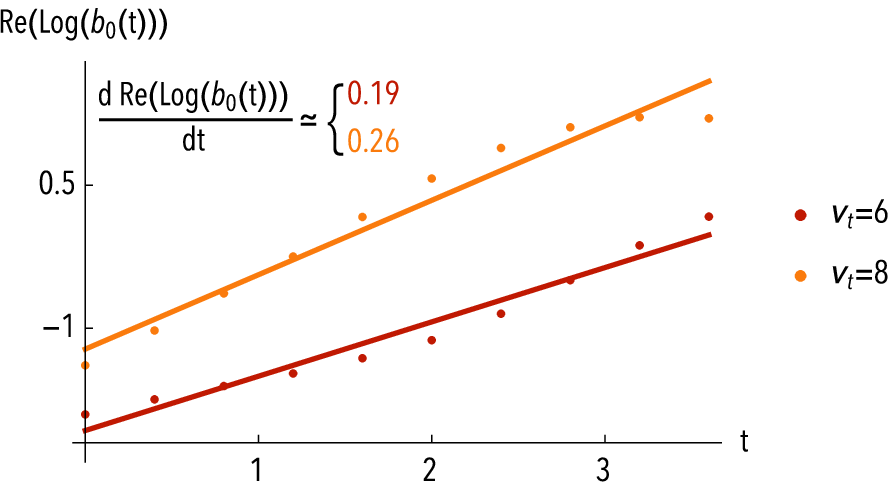,angle=-00,width=0.45\textwidth}} \\
{\epsfig{file=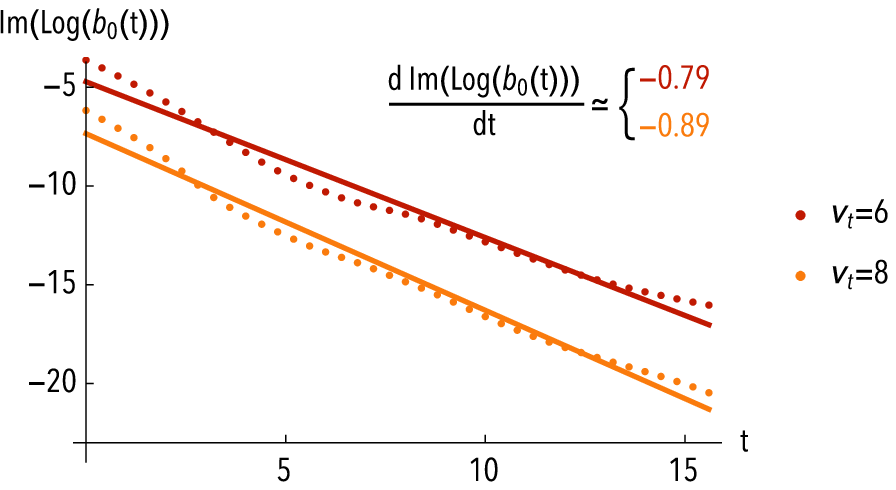,angle=-00,width=0.45\textwidth} }
\end{tabular}
\caption{\small{Measurement of the growth rate $\eta$ and pattern speed $\omega_{0}$ of the ${m_{\phi} \!=\! 2}$ unstable mode truncated Mestel discs with a random velocity given by ${q \!=\! 6}$ for various values of the truncation index ${ \nu_{\rm t} \!=\! 6, \, 8 }$. The basis coefficient plotted is associated with the basis element ${ (\ell , n) \!=\! (2,0) }$, using the same basis elements as for the matrix method in section~\ref{sec:initialD}.
}}
\label{figNBModes}
\end{figure}

\begin{figure}[!htbp]
\begin{center}
\epsfig{file=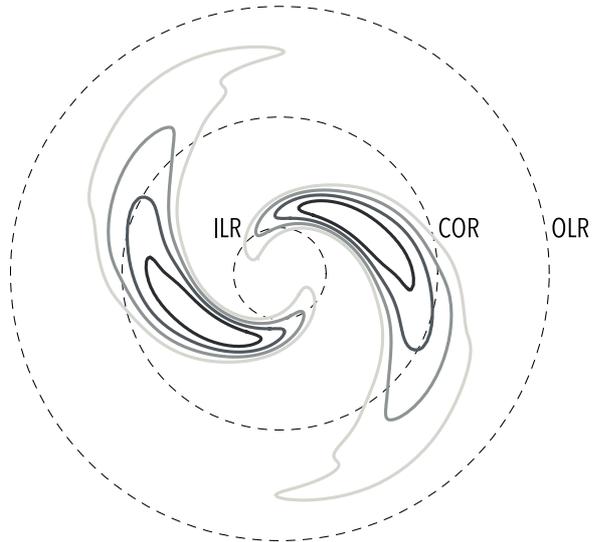,angle=-00,width=0.45\textwidth}
\caption{\small{Unstable mode for the truncated ${ \nu_{\rm t} \!=\! 4 }$ Mestel disc recovered via direct ${N-}$body simulations as presented in section~\ref{sec:NB}. Only positive contour levels are shown and are spaced linearly between 20\% and 80\% of the maximum norm. Similarly to figure~\ref{figModeMestelMatrix}, the radii associated with the resonances ILR, COR and OLR have been represented.
}}
\label{figModeMestelNB}
\end{center}
\end{figure}

The same unstable modes were also used to validate the ${N-}$body code presented in section~\ref{sec:NB}. To run these simulations, we used the same samping technique as described in section~\ref{sec:sampling}. In order not to be significantly impacted by the absence of a quiet start sampling~\citep{Sellwood1983}, for each value of $\nu_{\rm t}$, the measurements were performed with simulations of ${ 20M }$ particles. As observed in~\cite{SellwoodEvans2001}, the appropriate setting of the parameters of the ${N-}$body code are crucial to recover correctly the unstable modes of a disc. We considered a grid made with ${ N_{\rm mesh} \!=\! 120 }$ grid cells, while using a softening length equal to ${ \varepsilon \!=\! R_{\rm i} / 60 }$. As described, in section~\ref{sec:NB}, we similarly restricted the perturbing forces only to the harmonic sector ${ m_{\phi} \!=\! 2 }$, using ${ N_{\rm ring} \!=\! 2400 }$ radial rings, with ${ N_{\phi} \!=\! 720 }$ azimuthal points. In order to extract the properties of the mode present within the disc, one may proceed as follows. For each simulation snapshot, one can estimate the active surface density within the disc via
\begin{equation}
\Sigma_{\rm star} (\bm{x} , t) = \mu \sum_{i} \delta_{\rm D} (\bm{x} \!-\! \bm{x}_{i} (t)) \, ,
\label{Modes_star_surface}
\end{equation}
where the sum on $i$ is made on all the particles of the simulation and ${ \bm{x}_{i} (t) }$ is the position of the $i^{\rm th}$ particle at time $t$. Such a surface density can be decomposed on the basis elements from equation~\eqref{definition_basis}, under the form
\begin{equation}
\Sigma_{\rm star} (\bm{x} , t) = \sum_{p} b_{p} (t) \, \Sigma^{(p)} (\bm{x}) \, ,
\label{Modes_decomposition_star_surface}
\end{equation} 
where the sum on $p$ is made on all the basis elements $\Sigma^{(p)}$ considered. The effective basis elements used during our measurements are the same as the ones used in the matrix method from section~\ref{sec:initialD}. Thanks to the biorthogonality property from equation~\eqref{definition_basis}, the coefficients ${ b_{p} (t) }$ can be immediately determined as
\begin{equation}
b_{p} (t) 
= - \int \!\! \mathrm{d} \bm{x} \, \Sigma_{\rm star} (\bm{x} , t) \, \psi^{(p) *} (\bm{x})
= - \mu \sum_{i} \psi^{(p) *} (\bm{x}_{i} (t)) \, .
\label{Modes_calculation_bp}
\end{equation}
As we are looking for unstable modes within the disc, we expect to have ${ b_{p} (t) \!\propto\! \exp [ - i (\omega_{0} \!+\! i \eta) t ] }$, where ${ \omega_{0} \!=\! m_{\phi} \Omega_{p} }$ is the pattern speed of the mode and ${ s }$ its growth rate. As a consequence, one immediately obtains that
\begin{equation}
\frac{\mathrm{d} \, \text{Re} (\log (b_{p} (t)))}{\mathrm{d} t} = s \;\;\; ; \;\;\; \frac{\mathrm{d} \, \text{Im} (\log(b_{p} (t))) }{\mathrm{d} t} = - \omega_{0} \, ,
\label{Modes_Re_Im_bp}
\end{equation}
if one is sufficently careful with the branch-cut of the complex logarithm. Such a linear scaling with $t$ of ${ \text{Re} (\log (b_{p} (t))) }$ and ${ \text{Im} (\log(b_{p} (t)) ) }$ is therefeore the appropriate measurement procedure to use in order to estimate the growth rate and pattern speed of the unstable modes of these truncated Mestel discs. These measurements for the various values of the truncation index $\nu_{\rm t}$ are illustrated in figure~\ref{figNBModes}. 
Once the basis coefficients ${ b_{p} (t) }$ have been determined, one can study the shape of the recovered unstable modes in the physical space. Indeed, similarly to equation~\eqref{shape_mode_Matrix_Method}, the shape of the modes is given by
\begin{equation}
\Sigma_{\rm mode} (R , \phi , t) = \text{Re} \left[ \sum_{p} b_{p} (t) \, \Sigma^{(p)} (R , \phi) \right] \, .
\label{shape_mode_NB}
\end{equation}
In analogy with figure~\ref{figModeMestelMatrix}, for which the unstable modes have been obtained via the matrix method, figure~\ref{figModeMestelNB} illustrates the unstable mode of the same truncated ${ \nu_{\rm t} \!=\! 4 }$ Mestel disc.

As a conclusion, the growth rates and pattern speeds obtained either via the matrix method or direct ${N-}$body simulations are gathered in table~\ref{figtabModes}.
\begin{figure}[!htbp]
\begin{center}
  \begin{tabular}
  {|p{2.2cm}|p{0.6cm}p{0.6cm}|p{0.6cm}p{0.6cm}|p{0.6cm}p{0.6cm}|}
  \hline
\multicolumn{7}{|c|}{Unstable $m_{\phi} \!=\! 2$ modes of truncated Mestel discs, $q \!=\! 6.$}
\\ \hline
  & \multicolumn{2}{c} { $\nu_{\rm t} \!=\!4$ }
  & \multicolumn{2}{|c} { $\nu_{\rm t} \!=\! 6$ }
  & \multicolumn{2}{|c|} { $\nu_{\rm t} \!=\! 8$ }
  \\
  \textit{Method}  & $\omega_{0}$ & $\eta$ & $\omega_{0}$ & $\eta$ & $\omega_{0}$ & $\eta$\\ \hline
  Linear Theory
   &   0.88   &   0.13
   &   0.90   &   0.22
   &   0.92   &   0.27
\\ 
  Matrix Method
   &   0.93   &   0.11
   &   0.94   &   0.20
   &   0.95   &   0.24
\\ 
  ${N-}$body
   &   0.99   &   0.13
   &   0.79   &   0.19
   &   0.89   &   0.26
\\ 
\hline
\end{tabular}
\caption{\small{Measurements of the pattern speed ${ \omega_{0} \!=\! m_{\phi} \Omega_{p} }$ and growth rate ${ \eta \!=\!s }$ for unstable ${ m_{\phi} \!=\! 2 }$ modes of tapered Mestel discs. The velocity dispersion within these discs is characterised by ${ q \!=\! 6 }$, and the inner truncation power indices are given by ${ \nu_{\rm t} \!=\! 4, \, 6, \, 8 }$. The theoretical values were obtained from linear theory in~\cite{EvansRead1998II}. Our measurements were either performed via the response matrix method as in equation~\eqref{response_M_EL}, or via direct ${N-}$body simulations, using the ${N-}$body integrator described in section~\ref{sec:NB}. 
}}
\label{figtabModes}
\end{center}
\end{figure}
As observed in~\cite{SellwoodEvans2001}, the recovery of the unstable modes characteristics from direct ${N-}$body simulations when performed for truncated Mestel discs is a difficult task, for which convergence to the values predicted through linear theory may be difficult.

\section{Why swing matters?}
\label{sec:swing-test}

Let us investigate here briefly the importance of self-gravitation and the completeness of the projection basis in 
capturing the role of swing amplification.

\subsection{Turning off the self-gravitating amplification}
\label{sec:noamplification}

\begin{figure}[!htbp]
\begin{center}
\epsfig{file=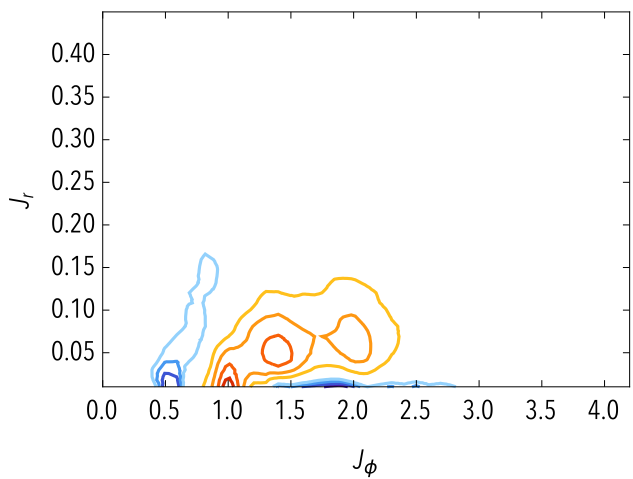,angle=-00,width=0.45\textwidth}
\caption{\small{Map of ${ N \text{div} (\bm{\mathcal{F}}_{\rm tot}^{\rm bare}) }$, corresponding to the bare secular diffusion flux, using the same conventions as in figure~\ref{figContoursDressed}. The contours are spaced linearly between the minimum and the maximum of ${ N \text{div} (\bm{\mathcal{F}}_{\rm tot}^{\rm bare}) }$. The maximum value for the positive blue contours corresponds to ${ N \text{div} (\bm{\mathcal{F}}_{\rm tot}^{\rm bare}) \!\simeq\! 0.30 }$, while the minimum value for the negative red contours is associated with ${ N \text{div} (\bm{\mathcal{F}}_{\rm tot}^{\rm bare}) \!\simeq\! -0.50 }$. This figure is qualitatively similar to the one obtained in figure $9$ of paper I.
}}
\label{figContoursBare}
\end{center}
\end{figure}

In order to investigate the role of the self-gravitating amplification, one may perform the same estimation as presented in figure~\ref{figContoursDressed}, while neglecting collective effects. 
When neglecting collective effects, i.e. when assuming that ${ \widehat{\mathbf{M}} \!\equiv\! 0 }$, one recovers the inhomogeneous Landau equation~\citep{Chavanis2013} which reads
\begin{align}
& \frac{\partial F}{\partial t} = \pi (2 \pi)^{d} \, \mu \, \frac{\partial}{\partial \bm{J}_{1}} \!\cdot\! \bigg[ \!\! \sum_{\bm{m}_{1} , \bm{m}_{2}} \!\! \bm{m}_{1} \!\! \int \!\! \mathrm{d} \bm{J}_{2} \, |A_{\bm{m}_{1} , \bm{m}_{2}} (\bm{J}_{1} , \bm{J}_{1})|^{2} \nonumber
\\
& \times  \delta_{\rm D} (\bm{m}_{1} \!\!\cdot\! \bm{\Omega}_{1} \!-\! \bm{m}_{2} \!\!\cdot\! \bm{\Omega}_{2})  \bigg(\! \bm{m}_{1} \!\!\cdot\! \frac{\partial }{\partial \bm{J}_{1}} \!-\! \bm{m}_{2} \!\!\cdot\! \frac{\partial }{\partial \bm{J}_{2}} \!\bigg) F (\bm{J}_{1} , t) F(\bm{J}_{2} , t) \bigg] .
\label{definition_Landau}
\end{align}
Equation~\eqref{definition_Landau} involves the bare susceptibility coefficients ${ |A_{\bm{m}_{1} , \bm{m}_{2} } (\bm{J}_{1} , \bm{J}_{2} )|^{2} }$, which can be equivalently defined (see Appendix B of paper I) by
\begin{align}
& A_{\bm{m}_{1} , \bm{m}_{2}} (\bm{J}_{1} , \bm{J}_{2}) = - \sum_{p} \psi_{\bm{m}_{1}}^{(p)} (\bm{J}_{1}) \, \psi_{\bm{m}_{2}}^{(q) *} (\bm{J}_{2}) \nonumber
\\
& \hskip -0.3cm \;\;\;\;\;\; = \frac{1}{(2 \pi)^{4}} \!\! \int \!\! \mathrm{d} \bm{\theta}_{1} \mathrm{d} \bm{\theta}_{2} \, u ( | \bm{x} (\bm{\theta}_{1} , \bm{J}_{1}) \!-\! \bm{x} (\bm{\theta}_{2} , \bm{J}_{2}) | ) \, e^{i (\bm{m}_{1} \cdot \bm{\theta}_{1} - \bm{m}_{2} \cdot \bm{\theta}_{2})} \, , 
\label{definition_bare_A}
\end{align}
where ${ u(\bm{x}) }$ is the binary potential of interaction potential given by ${ u(\bm{x}) \!=\! - G / |\bm{x}| }$ for gravity.
 This estimation therefore does not require to estimate the response matrix from equation~\eqref{response_M_EL}, but one still has to perform integrations along the resonant lines as in equation~\eqref{rewriting_drift_diff_II}. 
 Let us provide a first numerical implementation of this equation in the context of galactic dynamics.
 The contours of ${ N \text{div} ( \bm{\mathcal{F}}_{\rm tot}^{\rm bare} ) }$ are illustrated in figure~\ref{figContoursBare}.
Comparing the maps of the dressed diffusion flux ${ N \text{div} ( \bm{\mathcal{F}}_{\rm tot} ) }$ from figure~\ref{figContoursDressed} and the bare diffusion flux ${ N \text{div} (\bm{\mathcal{F}}_{\rm tot}^{\rm bare} )}$, allows to assess the strength of the self-gravitating amplification. As expected, when turning off the self-gravity of the system, one reduces significantly the susceptibility of the system and therefore slows down its secular evolution, by a factor of about ${ 1000 }$. One may also remark that while the secular appearance of a resonant ridge in the dressed diffusion from figure~\ref{figContoursDressed} was obvious, the shape of the contours obtained in the bare figure~\ref{figContoursBare} do not emphasize as clearly the appearance of such a narrow resonant ridge. One can still remark that the structure of the bare contours obtained in figure~\ref{figContoursBare} is similar to what was obtained in figure $9$ of paper I, through the WKB limit of the Balescu-Lenard equation.
One can finally note that the amplitudes of the bare divergence contours obtained previously are similar to the WKB values obtained in paper I.
As a consequence, the comparison of figures~\ref{figContoursDressed} and~\ref{figContoursBare} emphasizes that the strong self-gravitating amplification of loosely wound perturbations is indeed responsible for the appearance of a narrow ridge, while also ensuring that this appearance is sufficiently rapid, as observed in the diffusion timescales comparison from equation~\eqref{ratio_Delta_tau}.

\subsection{Turning off loosely-wound contributions}
\label{sec:backtoWKB}

\begin{figure}[!htbp]
\begin{center}
\epsfig{file=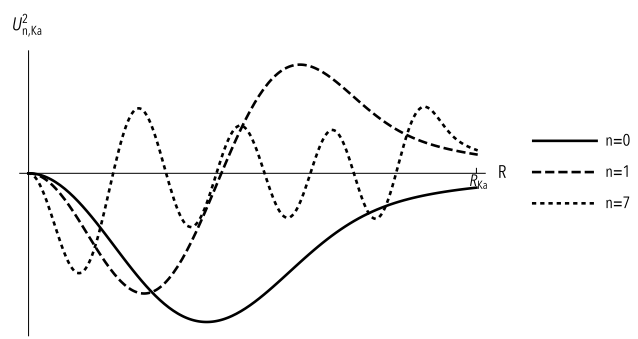,angle=-00,width=0.45\textwidth}
\caption{\small{Illustration of the radial basis elements of the ${ \ell \!=\! 2 }$ Kalnajs basis elements for $k_{\rm Ka} \!=\! 7$, defined in Appendix~\ref{sec:Kalnajsbasis}, which were used in the estimation of the Balescu-Lenard diffusion flux in section~\ref{sec:initialD}. As the radial index $n$ increases, the basis elements get more and more wound.
}}
\label{figBasisKalnajs}
\end{center}
\end{figure}

\begin{figure}[!htbp]
\begin{center}
\epsfig{file=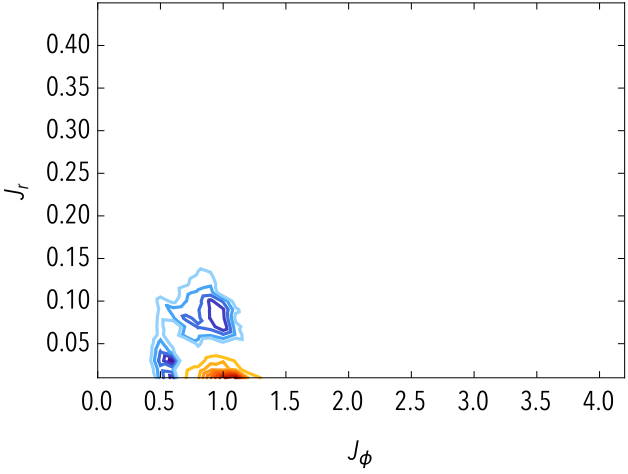,angle=-00,width=0.45\textwidth}
\caption{\small{Map of ${ N \text{div} (\bm{\mathcal{F}}_{\rm tot}^{\rm WKB}) }$, corresponding to the dressed secular diffusion flux, using the same conventions as in figure~\ref{figContoursDressed}. In order to limit ourselves only to tightly wound contributions, the basis elements associated with the radial basis index ${ n \!\in\! \{ 0, 1 \} }$ have not been taken into account. The contours are spaced linearly between the minimum and the maximum of ${ N \text{div} (\bm{\mathcal{F}}_{\rm tot}^{\rm WKB}) }$. The maximum value for the positive blue contours corresponds to ${ N \text{div} (\bm{\mathcal{F}}_{\rm tot}^{\rm WKB}) \!\simeq\! 0.7 }$, while the minimum for the negative red contours is associated with ${ N \text{div} (\bm{\mathcal{F}}_{\rm tot}^{\rm WKB}) \!\simeq\! -4.5 }$. This figure is to be compared to figure $9$ of paper I.
}}
\label{figbasistruncation}
\end{center}
\end{figure}

As emphasized in the Introduction, the WKB limit of the Balescu-Lenard equation presented in~\cite{Fouvry2015} was not able to capture the mechanism of swing amplification, which involves unwinding perturbations. By considering a complete and global basis as in equation~\eqref{definition_psi_p}, we have shown in figure~\ref{figContoursDressed} how the missing amplification from~\cite{Fouvry2015} could be recovered. Using the numerical method of estimation of the secular diffusion flux as presented in section~\ref{sec:disccase}, one can try to recover the results obtained within the WKB formalism by carefully choosing the considered basis elements generically introduced in equation~\eqref{definition_psi_p} and chosen to be by Kalnajs basis elements as detailed in Appendix~\ref{sec:Kalnajsbasis}. We recall that each basis element depends on two indices: an azimuthal index ${ \ell }$ and a radial one $n$. Because in S12's simulation perturbations were restricted to the harmonic sector ${ m_{\phi} \!=\! 2 }$, one only has to consider basis elements associated with ${ \ell \!=\! 2 }$. Moreover, as illustrated in figure~\ref{figBasisKalnajs}, 
% Indeed, 
the larger $n$ the radial index, the faster the radial variation of the basis elements and therefore the more tightly wound the basis elements. So as to get rid of the loosely-wound basis elements which are the ones which can get swing-amplified, we perform a truncation of the radial indices considered. Therefore, we define the secular diffusion flux ${ N \text{div} ( \bm{\mathcal{F}}_{\rm tot}^{\rm WKB} ) }$ computed in the same way than ${ N \text{div} (\bm{\mathcal{F}}_{\rm tot} ) }$ as presented in section~\ref{sec:initialD}, except that the basis elements are such that ${ n_{\rm cut} \!\leq\! n \!\leq\! n_{\rm max} }$, with ${ n_{\rm cut} \!=\! 2 }$ and ${ n_{\rm max} \!=\! 8 }$. By keeping only the tightly wound basis elements, one can therefore consider the same contribution as the one considered in the WKB limit presented in paper I. The contours of ${ N \text{div}  (\bm{\mathcal{F}}_{\rm tot}^{\rm WKB} ) }$ are illustrated in figure~\ref{figbasistruncation}.
One can note that the values of the contours obtained in the map of ${ N \text{div} (\bm{\mathcal{F}}_{\rm tot}^{\rm WKB}) }$ illustrated in figure~\ref{figbasistruncation} are in the same order of magnitude as the ones which were presented in figure $9$ of paper I in the WKB limit. The presence of positive blue contours of ${ N \text{div} (\bm{\mathcal{F}}_{\rm tot}^{\rm WKB}) }$ is also in agreement with a secular heating of the disc (i.e. an increase of $J_{r}$). However, these contours do not display a narrow resonant ridge as was observed in S12 simulation or in figure~\ref{figContoursDressed}.

\section{Sampling of the DF}
\label{sec:sampling}

In order to use the ${N-}$body integrator described in section~\ref{sec:NB}, one has to sample the particles according to the DF given by equation~\eqref{definition_Fstar}. We introduce the probability distribution function $F_{\rm sp}$, normalized to $1$ and thanks to which the sampling is performed. This probability DF $F_{\rm sp}$ is directly proportional to the active distribution function $F_{\rm star}$ from equation~\eqref{definition_Fstar}, so that we may write
\begin{equation}
F_{\rm sp} (E , L) = C_{\rm sp} \, L^{q}  \exp [ -E / \sigma_{r}^{2} ] \, T_{\rm inner} (L) \, T_{\rm outer} (L) \, , 
\label{definition_F_sampling}
\end{equation}
where $C_{\rm sp}$ is a normalization constant which will be determined in the upcoming calculations. Because the mapping ${ (E,L) \!\mapsto\! (J_{r} , J_{\phi}) }$ from equation~\eqref{definition_actions} is not a trivial one, we will not perform the sampling of the stars in the action space ${ (J_{r} , J_{\phi}) }$, but rather in the ${ (E,L)-}$space. Moreover, one should pay attention to the fact that the DF $F_{\rm sp}$ from equation~\eqref{definition_F_sampling} is a probability distribution function in the ${ (\bm{x} , \bm{v})-}$space, so that ${ \mathrm{d}^{2} \bm{x} \mathrm{d}^{2} \bm{v} F_{\rm sp} (\bm{x} , \bm{v}) }$ is proportional to the number of particles in the infinitesimal volume ${ \mathrm{d}^{2} \bm{x} \mathrm{d}^{2} \bm{v} }$ around the position ${ (\bm{x} , \bm{v}) }$. As we want to sample the particles in the ${ (E,L)-}$ space, we introduce the function ${ h_{\rm sp} (E,L) }$ such that ${ \mathrm{d} E \mathrm{d} L \, h_{\rm sp} (E,L) }$ is proportional to the number of particles in the volume ${ \mathrm{d} E \mathrm{d} L }$ around the location ${ (E,L) }$. One can now determine ${ h_{\rm sp} (E,L) }$ as a function of ${ F (E,L) }$. Indeed, we have
\begin{align}
h_{\rm sp} (E',L') & = \!\! \int \!\! \mathrm{d} \bm{x} \mathrm{d} \bm{v} \, \delta_{\rm D} (E' \!-\! E ) \, \delta_{\rm D} (L' \!-\! L) \, F_{\rm sp} (E , L) \nonumber
\\
& = 2 \pi \!\! \int \!\! \mathrm{d} r \, r \!\! \int \!\! \mathrm{d} v_{\rm r} \mathrm{d} v_{\rm t} \, \delta_{\rm D} (E' \!-\! E) \, \delta_{\rm D} (L' \!-\! L) \, F_{\rm sp} (E ,L) \nonumber
\\
& = 2 \pi \!\! \int \!\! \mathrm{d} r \!\! \int \!\! \mathrm{d} v_{\rm r} \, \delta_{\rm D} (E' \!-\! E) \, F_{\rm sp} (E , L') \, ,
\label{link_h_F_sampling}
\end{align}
using the fact that the tangential velocity satisfies ${ v_{\rm t} \!=\! L / r }$. The last step is then to perform the change of variable ${ v_{\rm r} \!\to\! E }$. One has ${ v_{\rm r}^{2} \!=\! 2 (E \!-\! \psi_{\rm M} (r)) \!-\! L^{2} / r^{2}}$, so that ${ \mathrm{d} v_{\rm r} \!=\! \mathrm{d} E / \!\sqrt{2 (E \!-\! \psi_{\rm M} (r)) \!-\! L^{2} / r^{2}} }$. Because the radial velocity can be both positive and negative, equation~\eqref{link_h_F_sampling} takes the form
\begin{align}
h_{\rm sp} (E' , L ') & = 4 \pi \!\! \int \!\! \mathrm{d} r \!\! \int \!\! \mathrm{d} E \, \frac{\delta_{\rm D} (E' \!-\! E) \, F_{\rm sp} (E , L')}{\sqrt{2 (E \!-\! \psi_{\rm M} (r)) \!-\! L'^{2} / r^{2}}} \nonumber
\\
& = \frac{4 \pi^{2}}{\Omega_{1} (E' , L')} F_{\rm sp} (E' , L') \, ,
\label{link_h_F_sampling_II}
\end{align}
where we used the definition~\eqref{definition_Omega1} of the radial intrinsic frequency $\Omega_{1}$. One can then correctly normalize the probability distribution $h_{\rm sp}$ and determine the value of the constant ${ C_{\rm sp} }$ from equation~\eqref{definition_F_sampling}. One should pay attention to the fact that in addition to the tapering functions $T_{\rm inner}$ and $T_{\rm outer}$ from equations~\eqref{definition_tapering}, we also assume that no stars have orbits that extend beyond $R_{\rm max}$. As a consequence, the allowed region in the ${ (E,L)-}$space has to satisfy two constraints. First of all, the angular momentum $L_{\rm star}$ has to satisfy
\begin{equation}
L_{\rm min} \!\!=\! 0 \leq L_{\rm star} \leq R_{\rm max} V_{0} \!=\! L_{\rm max} \, .
\label{constraint_Lstar}
\end{equation}
Then, for a given value of $L_{\rm star}$, one can show that the energy of the star $E_{\rm star}$ must satisfy the constraint
\begin{equation}
E_{\rm min} (L_{\rm star}) \!=\! \psi_{\rm M} \bigg[ \frac{L_{\rm star}}{V_{0}} \bigg] \!+\! \frac{V_{0}^{2}}{2}  \!\leq\!  E_{\rm star}  \!\leq\!  \frac{L_{\rm star}^{2}}{2 R_{\rm max}^{2}} \!=\! E_{\rm max} (L_{\rm star}) \, .
\label{constraint_Estar}
\end{equation}
These two constraints allow to completely characterize the ${ (E,L)-}$space on which the sampling ${ (E,L) }$ will have to be performed. One finally has to satisfy the constraint
\begin{equation}
\int_{L_{\rm min}}^{L_{\rm star}} \!\!\!\! \mathrm{d} L \!\! \int_{E_{\rm min} (L)}^{E_{\rm max} (L)} \!\!\!\!\!\! \mathrm{d} E \, h_{\rm sp} (E,L) = 1 \, .
\label{constraint_C_sampling}
\end{equation}
Given the parameters presented after equation~\eqref{definition_Fstar}, one can numerically determine the value of the constant $C_{\rm sp}$ which reads
\begin{equation}
C_{\rm sp} \simeq 1.4723 \times 10^{-15} \, .
\label{numerical_C_sampling}
\end{equation}
We may now proceed to the sampling of the coordinates of the particles. Up to the sign of its radial velocity, one star is characterized by the set ${ \{ E_{\rm star} , L_{\rm star} , R_{\rm star} , \phi_{\rm star} \} }$. Given that the initial state is axisymmetric, the azimuthal angle of the star can be uniformly sampled between $0$ and ${ 2 \pi }$. The next step is then to successively sample ${ (L_{\rm star} , E_{\rm star}) }$ and finally $R_{\rm star}$, using successive rejection samplings as we will now detail.

The heart of the rejection sampling is as follows. Let us assume that we want to generate sampling values from a function ${ f(x) }$, from which it is difficult to sample. However, we assume that we have at our disposal another distribution function ${ g (x) }$ from which the sampling is simple, and such that there exists a bound ${ M \!>\! 1 }$ satisfying ${ f (x) \!<\! M \, g (x) }$. The smaller $M$, the more efficient the sampling. One then has to proceed as follows: sample both a proposition $x$ from $g$ and $u$ uniformly between ${  [0 \,;1]}$. One then applies the selection
\begin{equation}
\alpha = \frac{f (x)}{M g(x)} \; : \; 
\begin{cases}
\displaystyle u < \alpha  \; \Longrightarrow \; x \text{ is kept.}
\\
\displaystyle u \geq \alpha \; \Longrightarrow \; x \text{ is rejected.}
\end{cases}
\label{presentation_rejection_sampling}
\end{equation}
In order to have an efficient sampling, one should try to consider a function $g$ \textit{close} to $f$.

We may now directly sample ${ (E,L) }$ thanks to this algorithm. The \textit{true} sampling function is ${ f_{(E,L)} \!=\! h_{\rm sp} }$ from equation~\eqref{link_h_F_sampling}. The \textit{simple} sampling function is ${ g_{(E,L)} \!\propto\! 1 }$, defined on the domain characterized by the constraints from equation~\eqref{constraint_Lstar} and~\eqref{constraint_Estar}. When performing a rejection sampling with such an uniform $g_{(E,L)}$, in order to determine the bound $M_{(E,L)}$, one only has to determine an uniform bound for $f_{(E,L)}$. With the numerical values introduced after equation~\eqref{definition_Fstar}, one can check that $f_{(E,L)}$ is such that
\begin{equation}
f_{(E,L)} (E,L) \leq 1.4 \, .
\label{uniform_bound_f_sampling}
\end{equation} 
The final element required to be able to perform the rejection sampling with $f_{(E,L)}$ is to be able to draw uniformly candidate ${(E,L)}$ in the domains defined by the constraints from equation~\eqref{constraint_Lstar} and~\eqref{constraint_Estar}, which is equivalent as sampling candidates ${ (E,L) }$ from the uniform probability distribution function $g_{(E,L)}$. To perform this uniform sampling, since the constraints from equation~\eqref{constraint_Estar} are expressed for a given value of $L_{\rm star}$, it is more natural to first draw $L_{\rm star}$ and then $E_{\rm star}$. The probability distribution according to which $L_{\rm star}$ has to be drawn is of the form ${ f_{L} \!\propto\! (E_{\rm max} (L) \!-\! E_{\rm min} (L)) }$. When correctly normalized, it reads
\begin{equation}
f_{L} (L) = \frac{3}{2} \frac{1}{R_{\rm max} V_{0}} \left[ \frac{L^{2}}{2 R_{\rm max}^{2} V_{0}^{2}} \!-\! \frac{1}{2}  \!-\! \log \bigg[ \frac{L}{R_{\rm max} V_{0}} \bigg] \right] \, ,
\label{pdf_Lstar}
\end{equation}
To sample $L$ from $f_{L}$, we will use another rejection sampling by introducing the additional simple probability distribution function $g_{L}$ defined as
\begin{equation}
g_{L} (L) = - \frac{1}{R_{\rm max} V_{0}} \log \bigg[ \frac{L}{R_{\rm max} V_{0}} \bigg] \, .
\label{pdf_g_Lstar}
\end{equation}
It is straightforward to check that ${ f_{L} \!<\! (3/2) \, g_{L} }$, so that we may use the bound ${ M_{L} \!=\! 3/2 }$ to perform the rejection sampling of $L_{\rm star}$. The final remark is to note that sampling $L$ from $g_{L}$ is simple since its cumulative distribution function $G_{L} \!=\! \int_{L_{\rm min}}^{L} \!\!\! \mathrm{d} L'  g_{L} (L') $ can be inverted so as to read
\begin{equation}
G_{L}^{-1} (u) = R_{\rm max} V_{0} \exp \bigg[ 1 \!+\! W_{-1} \bigg(\! - \frac{u}{\text{e}} \bigg) \bigg] \, ,
\label{inverse_cumulative_g_Lstar}
\end{equation}
where ${ W_{-1} }$ is the lower branch of the Lambert function ${ W (x) }$ for ${ x \!\in\! [-1/\text{e} \,; 0] }$. With all these elements, the rejection sampling of $L_{\rm star}$ following $f_{L}$ from equation~\eqref{pdf_Lstar} can be performed.

Once $L_{\rm star}$ has been drawn, it only remains to sample uniformly $E_{\rm star}$ on the interval ${ E_{\rm star} \!\in\! [E_{\rm min} (L_{\rm star}) \, ; E_{\rm max} (L_{\rm star}) ] }$, as given by equation~\eqref{constraint_Lstar}. Thanks to these uniformly drawn candidates ${ (E_{\rm star} , L_{\rm star}) }$ and the uniform bound from equation~\eqref{uniform_bound_f_sampling}, one can perform the rejection sampling from the probability distribution $f_{(E,L)}$.

For $E_{\rm star}$ and $L_{\rm star}$ succesfully sampled, one may then sample the radius $R_{\rm star}$ using a similar rejection sampling. The radius has to be sampled according to the probability distribution $f_{R}$ given by
\begin{equation}
f_{R} (r) =  \frac{\Omega_{1} / \pi}{\sqrt{2 (E \!-\! \psi_{\rm M} (r)) \!-\! L^{2} / r^{2}}} \, ,
\label{definition_pr_sampling}
\end{equation}
so that one has ${ f_{R} \!\propto\! v_{\rm r} }$. However, one should note that for ${ r \!\to\! r_{p/a} }$, one has ${ f_{R} (r) \!\to\! + \infty }$, so that the rejection sampling cannot be used without considering a probability DF $g$ which also diverges for ${ r \!\to\! r_{p/a} }$. In order to get rid of these divergences, instead of sampling the variable $r$, we will sample the angle ${ u \!\in\! [-\pi/2\,; \pi/2] }$, where we have defined the mapping ${ r \!\mapsto\! u (r) }$ as
\begin{equation}
r (u) = \frac{r_{p} \!+\! r_{a}}{2} \!+\! \frac{r_{a} \!-\! r_{p}}{2} \, \sin(u) \, ,
\label{definition_mapping_u_sampling}
\end{equation}
so that one naturally has ${ r (- \pi /2) \!=\! r_{p} }$ and ${ r (\pi/2) \!=\! r_{a} }$. The probability distribution function from which $u$ has to be sampled is immediately given by
\begin{equation}
f_{u} (u) = \frac{r_{a} \!-\! r_{p}}{2} \cos (u) \, p_{r} (r(u)) \, .
\label{definition_pu_sampling}
\end{equation}
Using the fact that the maximum of $f_{u}$ is reached for ${ u \!=\! \pi / 2 }$, one can then sample $u$ from $f_{u}$ using a rejection sampling with a uniform control probability distribution function ${ g_{u} (u) \!=\! 1/\pi }$. Once $u$ is known, it only remains to compute ${ R_{\rm star} \!=\! r (u_{\rm star}) }$, so that the sampling of all the required quantities for one star has been performed.

The final step of the sampling of the particles is to determine the physical coordinates of the particles ${ (\bm{x} , \bm{v}) }$ associated with the set ${ \{ E_{\rm star} , L_{\rm star} , R_{\rm star} , \phi_{\rm star} \} }$. These physical coordinates are the ones which will be given to the ${N-}$body integrator. We draw uniformly the sign of the radial velocity ${ \varepsilon_{\rm r} \!\in\! \{ -1 , 1 \} }$. Because we are considering a disc made only of prograde stars, one immediately obtains that the radial and tangential velocities $v_{\rm r}$ and $v_{\rm t}$ are given by
\begin{equation}
\begin{cases}
\displaystyle v_{\rm r} = \varepsilon_{\rm r} \sqrt{2 (E \!-\! \psi_{\rm M} (R_{\rm star})) \!-\! L_{\rm star}^{2} / R_{\rm part}^{2}} \, ,
\\
\displaystyle v_{\rm t} = L_{\rm star} / R_{\rm star} \, .
\end{cases}
\label{radial_tangential_veloticies_sampling}
\end{equation}
The final step of the transformation to the ${ (\bm{x} , \bm{v})-}$coordinates is then straightforward, since one naturally has
\begin{equation}
\begin{cases}
\begin{aligned}
\displaystyle x & = && \!\!\! R_{\rm star} \cos (\phi_{\rm star}) \, ,
\\
\displaystyle y & = && \!\!\! R_{\rm star} \sin (\phi_{\rm star}) \, ,
\\
\displaystyle v_{x} & = && \!\!\! v_{\rm r} \cos (\phi_{\rm star}) \!-\! v_{\rm t} \sin (\phi_{\rm star}) \, ,
\\
\displaystyle v_{y} & = && \!\!\! v_{\rm r} \sin (\phi_{\rm star}) \!+\! v_{\rm t} \cos (\phi_{\rm star}) \, .  
\end{aligned}
\end{cases}
\label{coordinates_transformation_sampling}
\end{equation}

One should note that the sampling procedure described previously does not correspond to a quiet start procedure~\citep{Sellwood1983}, which would allow a reduction of the initial shot noise within the disc, as briefly discussed in section~\ref{sec:MatrixOK}, with regard to the validation of the ${N-}$body code.

\section{Another test of the scaling with $N$}
\label{sec:tthold}

\begin{figure}[!htbp]
\centering
\begin{tabular}{@{}cc@{}}
{\epsfig{file=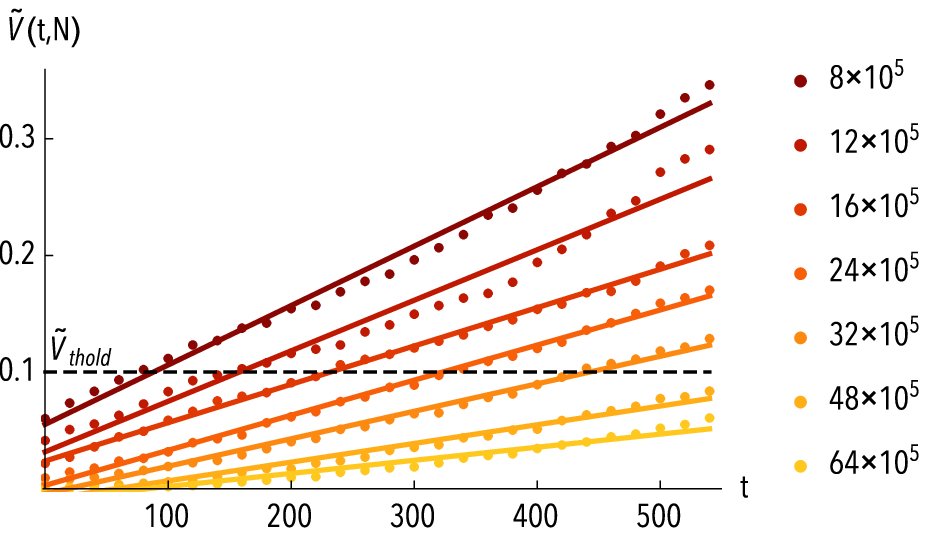,angle=-00,width=0.45\textwidth}} \\
{\epsfig{file=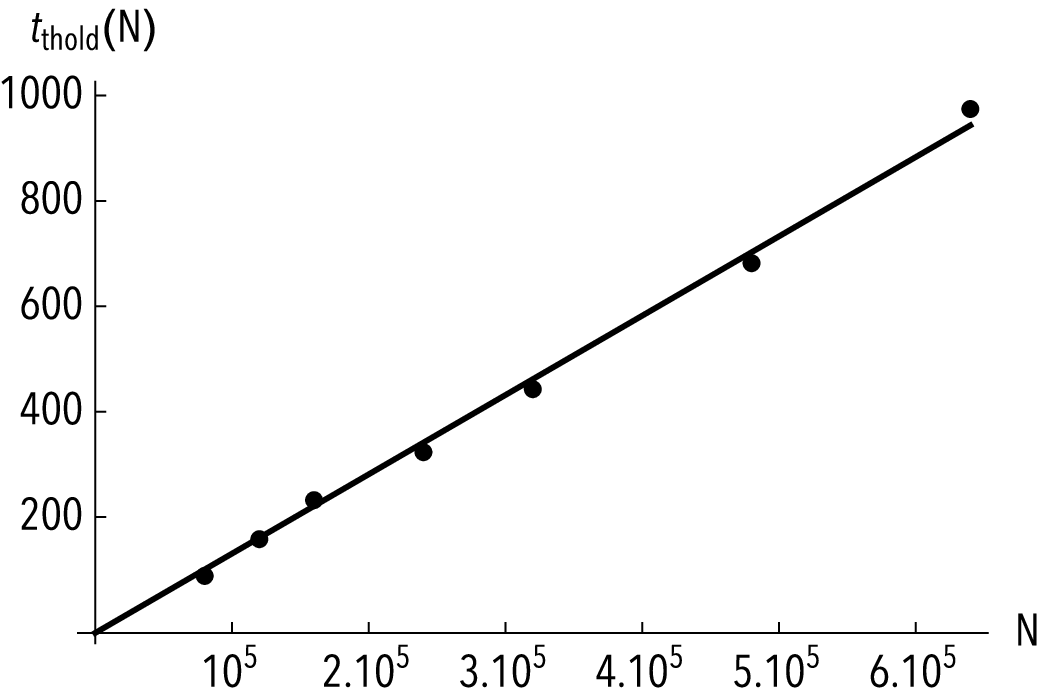,angle=-00,width=0.45\textwidth} }
\end{tabular}
\caption{\small{\textit{Top panel}: Illustration of the behavior of the function ${ t \!\mapsto\! \tilde{V} (t , N) }$ from equation~\eqref{definition_V}, when averaged on ${32}$ different realizations for particles numbers ${ N \!\in\! \{ 8,\, 12,\, 16,\, 24,\, 32,\, 48,\, 64 \} \!\times\! 10^{5} }$, along with the associated linear fits. To compute ${ \tilde{V} (t,N) }$, we used the same binning of the action-space ${ (J_{\phi}, J_{r}) }$ as in figure~\ref{fig_scaling_run}. As obtained in equation~\eqref{scaling_V}, one recovers that for a fixed value of $N$, the function ${ t \!\mapsto\! \tilde{V} (t , N) }$ is linear. The horizontal dashed line illustrates the threshold value $\tilde{V}_{\rm thold}$ for which the threshold time $t_{\rm thold}$ is determined.
\textit{Bottom panel}: Illustration of the behavior of the function ${ N \!\mapsto\! t_{\rm thold} (N) }$. As derived in equation~\eqref{behavior_thold}, one recovers a linear dependence of $t_{\rm thold}$ with $N$.
}}
\label{figVthold}
\end{figure}

One difficulty with the measurement of the scaling with $N$ presented in section~\ref{sec:Nscaling} is that one has to disentangle the contributions from the initial sampling Poisson shot noise present through $h_{0}$ from equation~\eqref{behavior_h0} and the effects due collisional Balescu-Lenard diffusion scaling through $h_{2}$ from equation~\eqref{behavior_h0}. Indeed, Poisson shot noise leads to fluctuations of the system DF about its mean value. In order not to be sensitive to such fluctuations, one could only consider fluctuations sufficiently large, i.e. fluctuations caused by an effective secular diffusion rather than caused by inevitable Poisson fluctuations. As a consequence, by restricting ourselves only to \textit{large} fluctuations, we can get rid of Poisson's effects. We therefore define the function ${ \tilde{V} (t,N) }$ as
\begin{equation}
\tilde{V} (t, N) = \!\! \int \!\! \mathrm{d} \bm{J} \, \chi \left[ \left\langle F (t , \bm{J} , N) \right\rangle \!-\! \left\langle F (t\!=\! 0 , \bm{J} , N) \right\rangle  \!<\! C_{\tilde{V}} \right] \, ,
\label{definition_V}
\end{equation}
where we introduced a threshold ${ C_{\tilde{V}} \!<\! 0 }$. Here ${ \chi \left[ \,\cdot\, \right] }$ is a characteristic function equal to $1$ if ${ (\left\langle F (t , \bm{J} , N) \right\rangle \!-\! \left\langle F (t\!=\! 0 , \bm{J} , N) \right\rangle  \!<\! C_{\tilde{V}}) }$, and $0$ otherwise. As a consequence, ${ \tilde{V} (t , N) }$ measures the volume in action space of the regions (depleted from particles, since ${ C_{\tilde{V}} \!<\! 0 }$) for which the mean DF has changed by more than $C_{\tilde{V}}$. For a sufficiently large value of the threshold $C_{\tilde{V}}$, such a construction allows not to be polluted by Poisson sampling shot noise.
For the initial times, as in equation~\eqref{BL_with_N}, it is straightforward to study the scaling of ${ \tilde{V} (t,N) }$ with $t$ and $N$. Indeed, one can write
\begin{equation}
\left\langle F (t , \bm{J} , N) \right\rangle \!-\! \left\langle F (t\!=\! 0 , \bm{J} , N) \right\rangle  \!\simeq\!  \Delta \tau \, \text{div} (\bm{\mathcal{F}}_{\rm tot})  \!\simeq\!  \frac{t}{N} \, \text{div} (\bm{\mathcal{F}}_{\rm tot}) \, .
\label{calculation_scaling_V}
\end{equation}
Introducing ${ \tilde{V}_{0} \!=\! \int \! \mathrm{d} \bm{J} \, \chi \left[ \text{div} (\bm{\mathcal{F}}_{\rm tot}) \!<\! C_{\tilde{V}} \right] }$, one can rewrite equation~\eqref{calculation_scaling_V} under the form
\begin{equation}
\tilde{V} (t , N) = \frac{t}{N} \tilde{V}_{0} \, .
\label{scaling_V}
\end{equation}
Therefore, for a fixed value of $N$, one expects to observe a linear time dependence of the function ${ \tilde{V} (t, N) }$, as illustrated in figure~\ref{figVthold}. In order to test the scaling of equation~\eqref{scaling_V} with $N$, one may proceed as follows. Introducing a threshold value ${ \tilde{V}_{\rm thold} }$, for each value of $N$, we define the associated threshold time ${ t_{\rm thold} (N) }$ as
\begin{equation}
\tilde{V} (t_{\rm thold} (N) , N) = \tilde{V}_{\rm thold} \, .
\label{definition_tthold}
\end{equation}
Thanks to the scalings from equation~\eqref{scaling_V}, one immediately obtains that
\begin{equation}
t_{\rm thold} (N) \simeq N \frac{\tilde{V}_{\rm thold}}{\tilde{V}_{0}} \, .
\label{behavior_thold}
\end{equation}
Such a linear scaling of ${ t_{\rm thold} (N) }$ with $N$ is a prediction from the Balescu-Lenard formalism and is nicely recovered in figure~\ref{figVthold}.

\section{Distributed code description}
\label{sec:codes}

For the sake of reproducibility, which has been lacking in the context of 
the linear response of stellar systems,
we distribute the linear matrix response code we 
wrote for this paper 
both as a {\sc Mathematica} package 
(\path{http://www.iap.fr/users/pichon/matrix-method/code/matrix-method.m}), 
and a notebook
(\path{http://www.iap.fr/users/pichon/matrix-method/code/matrix-method.nb}).
The functions therein allow for:
\begin{itemize}
\item[$\bullet$]  the determination as a function of ${ (r_{p} , r_{a}) }$ of the orbits quantites: $E$, $L$, $J_{r}$, $\Omega_{1}$ and $\Omega_{2}$.
\item[$\bullet$]  the construction of the ${2D}$ basis from~\cite{Kalnajs2}.
\item[$\bullet$]  the computation of the Fourier transform w.r.t. the angles, i.e. the computation of ${ \mathcal{W}_{\ell^{p} m_{2} n^{p}}^{m_{1}} (\bm{J}) }$ from equation~\eqref{expression_W}.
\item[$\bullet$]  the calculation of the ${ 2D }$ response matrix via equation~\eqref{response_M_sum}.
\end{itemize}
It has been tested for the isochrone and the Mestel disc.
\vfill

\end{document}